\documentclass[amsmath,
 amssymb,
 aps,
 nofootinbib,
 preprintnumbers
]{revtex4-1}
\usepackage[utf8]{inputenc}
\usepackage{simplewick}
\usepackage{footnote}
\usepackage{graphicx}
\usepackage{dcolumn}
\usepackage{bm}
\usepackage{stmaryrd}
\usepackage{amsmath}
\usepackage{physics}
\usepackage{amssymb}
\usepackage{mathtools}
\usepackage{bbm}
\usepackage{amsfonts}
\usepackage{xcolor}
\usepackage{footmisc}
\usepackage{comment}
\usepackage[normalem]{ulem}
\usepackage{slashed}
\usepackage{multirow}
\usepackage{appendix}
\usepackage{float}
\usepackage{braket}
\usepackage[linktocpage=true,bookmarksnumbered=true,colorlinks=true,plainpages,linkcolor=blue,citecolor=red]{hyperref}
\usepackage{cancel}

\begin{document}
\preprint{ADP-24-T1256,CPPC-2024-10}
\title{ Limits on Kaluza-Klein Portal Dark Matter Models}

\author{R. Sekhar Chivukula$^{a}$}
\author{Joshua A. Gill$^{b,d}$}
\author{Kirtimaan A. Mohan$^{c}$}
\author{George Sanamyan$^{b}$}
\author{Dipan~Sengupta$^{d}$}
\author{Elizabeth H. Simmons$^{a}$}
\author{Xing Wang$^{a,e}$}

\affiliation{$^{a}$Department of Physics and Astronomy, University of California, San Diego, 9500 Gilman Drive, La Jolla, CA-92093, USA
}
\affiliation{$^{b}$ARC Centre of Excellence for Dark Matter Particle Physics, Department of Physics, University of Adelaide, South Australia 5005, Australia}
\affiliation{$^{c}$Department of Physics and Astronomy, 
Michigan State University\\
567 Wilson Road, East Lansing, MI-48824, USA
}

\affiliation{$^d$Sydney Consortium for Particle Physics and Cosmology, School of Physics,\\ The University of New South Wales, Sydney NSW 2052, Australia}

\affiliation{$^e$Dipartimento di Matematica e Fisica, Università degli studi Roma Tre,\\ Via della Vasca Navale 84, I-00146, Rome, Italy}

\begin{abstract}
We revisit the phenomenology of dark-matter (DM) scenarios within radius-stabilized Randall-Sundrum models. Specifically, we consider models where the dark matter candidates are Standard Model (SM) singlets confined to the TeV brane and interact with the SM via spin-2 and spin-0 gravitational Kaluza-Klein (KK) modes. We compute the thermal relic density of DM particles in these models by applying recent work showing that scattering amplitudes of massive spin-2 KK states involve an intricate cancellation between various diagrams. Considering the resulting DM abundance, collider searches, and the absence of a signal in direct DM detection experiments, we show that spin-2 KK portal DM models are highly constrained. In particular, we confirm that within the usual thermal freeze-out scenario, scalar dark matter models are essentially ruled out. In contrast, we show that fermion and vector dark matter models are viable in a region of parameter space in which dark matter annihilation through a KK graviton is resonant.
Specifically, vector models are viable for dark matter masses ranging from 1.1 TeV to 5.5 TeV  for theories in which the scale of couplings of the KK modes is of order 40 TeV or lower. Fermion dark matter models are viable for a similar mass region, but only for KK coupling scales of order 20 TeV. 
In this work, we provide a complete description of the calculations needed to arrive at these results and, in an appendix, a discussion of new KK-graviton couplings needed for the computations, which have not previously been discussed in the literature.  Here,        we focus on models in which the radion is light, and the back-reaction of the radion stabilization dynamics on the gravitational background can be neglected. The phenomenology of a model with a heavy radion and the consideration of the effects of the radion stabilization dynamics on the DM abundance are being addressed in forthcoming work. 

\end{abstract}

\maketitle

\tableofcontents

\vfill\eject

\section{Introduction}
Theories of extra-dimensions are well-motivated extensions of the standard model. Introduced initially by Kaluza and Klein~\cite{Kaluza:1921tu,Klein:1926tv} as a means to unify electromagnetism and gravity, such theories have emerged as a solution to the electroweak hierarchy problem~\cite{Antoniadis:1990ew,ArkaniHamed:1998rs,Randall:1999ee,Randall:1999vf}. These include models compactified over a flat extra dimension~\cite{Antoniadis:1990ew,ArkaniHamed:1998rs}, as well as those with a warped extra dimension based on a slice of Anti-de-Sitter (AdS) space, known as Randall-Sundrum (RS) models~\cite{Randall:1999ee,Randall:1999vf}. Theories of compact extra dimensions are expected to be viewed as low-energy effective field theories originating from string theory, with significant work devoted to the ADS/CFT correspondence~\cite{Maldacena:1997re,Aharony:1999ti}, especially in RS models~\cite{Arkani-Hamed:2000ijo}.
Extra dimensional models have found applications in flavor physics (see, for example, \cite{Arkani-Hamed:1999pwe,Agashe:2004cp}), studies of the electroweak phase transition~\cite{Creminelli:2001th,Nardini:2007me}, cosmological considerations~\cite{Csaki:1999jh,Cline:1999ts}, and grand unification~\cite{Agashe:2002pr}, as well as in scenarios intended to provide viable dark matter candidates~\cite{Servant:2002aq,Servant:2002hb}\footnote{Extensive reviews covering a variety of these developments can be found in, for example,~\cite{Rattazzi:2003ea,Csaki:2004ay,Gabadadze:2003ii,Quevedo:2010ui}.}.

In recent years, there has been a surge of interest in dark matter models within extra-dimensional theories~\cite{Csaki:2021gfm,Gonzalo:2022jac,Garny:2015sjg,Folgado:2019sgz,Lee:2013bua,Brax:2019koq,Koutroulis:2024wjl,Fichet:2022ixi,Fichet:2022xol}.
Specifically, models where the extra-dimensional gravitational theory provides a portal between the dark sector and the standard model (SM) have been the subject of study. These include models based on an effective theory with a single massive spin-2 mediator or a full Kaluza-Klein (KK) theory~\cite{Rueter:2017nbk,Garny:2015sjg,Folgado:2019sgz,Lee:2013bua}. In these models, scattering amplitudes and squared matrix elements involving massive spin-2 and spin-0 resonances originating from the compactified gravity sector of the extra-dimensional model play a key role in estimating various relevant observables.  
These observables include velocity-averaged cross-sections $\langle\sigma v\rangle$, required to ascertain the observed DM relic density $\Omega_{\rm DM}$ of the Universe, spin-independent (SI) and spin-dependent (SD) direct detection rates, and various signatures at high-energy colliders like the LHC.  

Calculations involving massive spin-2 resonances originating from extra-dimensions have been shown to be plagued by amplitudes that appear to grow (anomalously) rapidly with the scattering energy $\sqrt{s}$, seemingly signaling a breakdown of the effective field theory at a scale much lower than the intrinsic scale of the compact extra dimension. The source of this anomalous growth can be traced back to the helicity-0 and helicity-1 modes of the massive spin-2 resonances in the final states. For instance, a power-counting estimate of the cross-section for DM annihilation into the helicity-0 states grows as $\mathcal{O}\left(s^{3}/(\Lambda^2 M_{KK}^{4})\right)$, where $\sqrt{s}$ is the center of mass energy, $M_{KK}$ is the mass of the Kaluza-Klein particle, and $\Lambda$ is a characteristic gravitational scale of the underlying theory~\cite{Folgado:2019sgz,Garny:2015sjg,Lee:2013bua,Bernal:2020fvw}. The consequences of this assumed anomalous growth for the calculation of DM observables are significant, as the rapid growth would predict that at large scattering energy $\sqrt{s}$, the DM thermal relic density could be satisfied for small values of the effective coupling away from the resonant funnel region~\cite{Folgado:2019sgz,Garny:2015sjg}. It would also predict that the DM annihilation cross-section (or the scattering cross-section for direct detection experiments) would violate perturbative unitarity at scales much lower than the effective field theory scale \footnote{For compactified flat extra-dimensions the scale $\Lambda$ is simply the Planck mass $M_{Pl}$. In contrast, for RS models the scale is $\Lambda_{\pi}=M_{Pl}e^{-k r_{c}\pi}$, where $k$ is the curvature and $r_{c}$ the compactification radius.}. 

In previous work, the authors of this paper and collaborators have performed detailed and increasingly more refined calculations of effective spin-2 KK mode effective field theory of compactified extra-dimensional models:

\begin{itemize}
    \item  First, it was shown that while amplitudes of individual diagrams involving spin-2 KK modes (as well as the spin-0 fluctuation, the radion) grew as fast as $\mathcal{O}\left(s^{5}/(\Lambda^2M_{KK}^{8})\right)$ in unitary gauge, intricate cancellations between different diagrams, which are ultimately due to the underlying symmetries of the higher dimensional theory, ensure that the final amplitude grows no faster than $\mathcal{O}(s/\Lambda^2)$~\cite{SekharChivukula:2019yul,SekharChivukula:2019qih,Chivukula:2020hvi,Foren:2020egq,Chivukula:2022kju}. 

    \item Next, this analysis of the spin-2 effective KK theory was extended to models where the compactification radius was stabilized ({\it i.e.}, models in which the spin-0 radion mode obtains a mass) via the Goldberger-Wise (GW) mechanism~\cite{Goldberger:1999uk,Goldberger:1999un}.  In this case, the authors and their collaborators showed that while the naive introduction of a radion mass-term by hand reintroduces a bad high energy growth in the scattering amplitudes, the dynamics, including the entire GW scalar sector, ensure that the full spin-2 scattering amplitudes again grow no faster than $\mathcal{O}(s/\Lambda^2)$~\cite{Chivukula:2021xod,Chivukula:2022tla}.

    \item Subsequently, it was shown that the $\mathcal{O}(s/\Lambda^2)$  high-energy behavior of the scattering amplitudes in the effective KK theory is transparent in `t-Hooft-Feynman gauge and that this can be understood using an equivalence theorem which ensures that the scattering amplitudes of helicity-0 and helicity-1 spin-2 Kaluza-Klein states equal those of the corresponding Goldstone bosons~\cite{Chivukula:2023qrt}. The properties of the effective KK theory are, hence, directly the result of the underlying diffeomorphism symmetry of the full five-dimensional gravitational theory. 

    \item Finally, the investigation showed that the cancellation properties of the gravitational KK sector described above extend to the coupling of the gravitational sector to the matter sector, whether matter lies in the bulk or is localized to a brane. Specifically, in the full KK-matter amplitudes, the leading  ${\cal O}(s^3)$ terms in the high energy expansions cancel among themselves and the ${\cal O}(s^2)$ terms do likewise. This behavior of the scattering amplitudes can also be understood via the equivalence theorem~\cite{Chivukula:2023sua}.\footnote{An analogous calculation for scalar brane-matter in an unstabilized RS1 model was performed in~\cite{deGiorgi:2020qlg,deGiorgi:2023mdy}, following the investigations described in the previous paragraph. It was also shown in~\cite{Gill:2023kyz} that for a single massive spin-2 model emission Ward identities ensure that there are is no anomalous growth, whether in Fierz-Pauli theories or extra dimensions.}  
\end{itemize}

\begin{figure}[t]
    \centering
\includegraphics[width=0.60\textwidth]{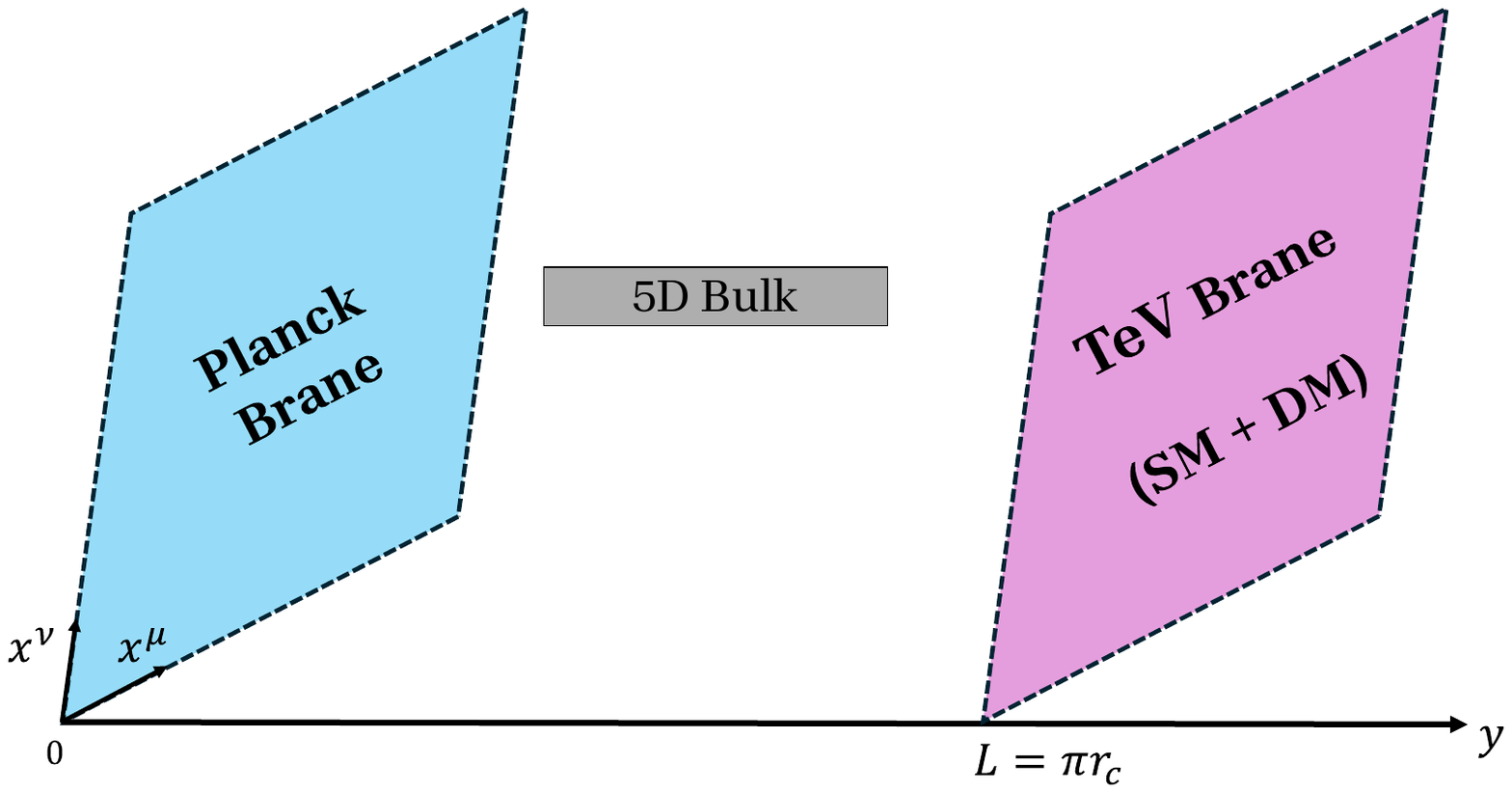}

    \caption{Pictorial Representation of the Randall-Sundrum 1 Model with a dark sector localized on the TeV brane.}
    \label{fig:RS1}
\end{figure}

In this work, we apply our understanding of the scattering amplitudes in RS KK theories to provide a comprehensive analysis of the thermal freeze-out scenario in KK-portal DM models, schematically depicted in Fig.~\ref{fig:RS1}. Initial steps towards a consistent calculation for the KK-portal dark matter model were performed in~\cite{deGiorgi:2021xvm} for brane-localized scalar DM. In this work, we provide a comprehensive analysis of the spin-2 KK portal model by including the following, 
\begin{itemize}
    \item We extend the analysis of~\cite{Rueter:2017nbk} and\cite{deGiorgi:2021xvm}   to include both brane-localized fermion and vector dark matter candidates and consider all of the annihilation channels. 
    \item We include the dynamics of the Goldberger-Wise~\cite{Goldberger:1999uk,Goldberger:1999un} stabilized RS model in which the background scalar and metric fields can be determined exactly~\cite{DeWolfe:1999cp}. In this paper, we focus on the case of a light radion, for which the back-reaction of the stabilization sector can be neglected and for which the effects of this sector on DM relic abundance are small. In a forthcoming paper, we describe the phenomenology of a model with a heavy radion and show that a massive radion and the Goldberger-Wise scalar states can be relevant for the relic abundance of dark matter. 
    \item  We extend our previous scattering amplitude analyses to include final states comprising either two spin-zero modes or one spin-2 KK mode plus one spin-0 mode; these have not previously been discussed in the literature. 
    
    \item We provide a unified picture of the status of the KK-portal dark matter models using the latest collider and direct detection constraints.
\end{itemize}

We find that a combination of relic abundance computations with collider bounds on the spin-2 KK resonances from diphoton searches at the LHC~\cite{ATLAS:2021uiz} place constraints on the KK portal models, as illustrated in Fig.~\ref{fig:FFscan} and Fig.~\ref{fig:VVscan}.  We show that DM candidates within this scenario remain viable for masses of 1.1~TeV and 5.5~TeV, so long as the scale $\Lambda_\pi$ of the gravitational theory is of order 80 (25) TeV or lower for vector (fermion) DM candidates. In the case of DM candidates heavier than 2.3 TeV, dark matter resonant annihilation occurs for a Kaluza-Klein graviton at level 2 or 3 -- leading to the possibility of a lighter KK resonance not directly coupling to dark matter being uncovered at by the high-luminosity LHC. We further show that these scenarios can also be probed by direct detection experiments if the radion mass is of order 1 GeV, with some regions being excluded by the 2024 Lux-Zepellin (LZ)~\cite{LZCollaboration:2024lux} experiment, and others potentially observable in the future.

In this work, we focus on models in which the radion is light (with a mass of order 100 GeV or lower), and hence, the back-reaction of the radion stabilization dynamics on the gravitational background can be neglected. We will show in Sec.~\ref{sec:DDconstraints} that the light radion impacts direct detection constraints severely,
 The phenomenology of a model with a heavy radion and the consideration of the effects of the radion stabilization dynamics on the DM abundance will be addressed in a forthcoming paper~\cite{Chivukula:2024}. 

The rest of the paper is organized as follows. The next section, Sec.~\ref{sec:model}, provides the details of the model being considered, including a specification of the Dewolfe-Freedman-Gubser-Karch (DFGK)~\cite{DeWolfe:1999cp} stabilized RS model, and describes the corresponding KK decompositions and the masses and couplings of the spin-2 and spin-0 KK modes. In Sec.~\ref{sec:cross-sections}, we describe the calculation of the various annihilation cross-sections needed for the relic abundance computation. To perform these calculations, we need the decay widths 
of the spin-2 and spin-0 KK modes to SM particles (as described in appendices \ref{sec:smdecaywidth} and \ref{sec:decaywidthradion}) and to other spin-2 and spin-0 KK modes (as described in appendix \ref{sec:RSdecaywidth}). The entire computation performed here also requires understanding new interactions and coupling structures for the KK modes not previously explored in the literature. These are explained in appendix \ref{sec:newcouplings}, and various numerical tests are presented to show that the amplitudes for these new processes have the appropriate high-energy behavior (and thereby provide a check of the accuracy of these new results).
We review and update the collider and direct detection constraints on  KK-portal dark matter in sections \ref{sec:ColliderConstraints} and \ref{sec:DDconstraints}, respectively. The allowed region of parameter space for KK portal DM models is explored in Sec.~\ref{sec:results}, and we outline our conclusions in Sec.~\ref{sec:conclusion}.

\section{The Model}

\label{sec:model}

This section specifies our notation and describes the model we analyze.

\subsection{5D Gravitational and Scalar Fields}
\label{sec:gsector}

Following~\cite{Chivukula:2021xod,Chivukula:2022tla}, we write the Lagrangian for the stabilized RS model as
\begin{equation}
    S = \int_V d^4 x~ d y \left( \mathcal{L}_{\text{EH}} + \mathcal{L}_{\phi \phi} + \mathcal{L}_{\text{pot}} + \mathcal{L}_{\text{GHY}} + \Delta \mathcal{L} \right),
    \label{eq:stablag}
\end{equation}
where $\mathcal{L}_{\text{EH}}$ is the usual $5$D Einstein-Hilbert term, $\mathcal{L}_{\phi \phi}$  and $\mathcal{L}_\text{pot}$ are the kinetic and potential energy terms of the GW bulk scalar field $\hat{\phi}(x,y)$, $\mathcal{L}_\text{GHY}$ is the Gibbons-Hawking-York term~\cite{York:1972sj,Gibbons:1976ue} required for a well-posed variational problem in the gravitational spin-2 KK and the spin-0 GW sector, and $\Delta \mathcal{L}$ is a total derivative term added to the action to cancel out terms linear in the field fluctuations as well as to eliminate the mixing between the tensor and scalar $5$D fields.

In unitary gauge, the $5$D metric for an RS-like background is parametrized as~\cite{Chivukula:2022tla} 
\begin{equation}
    G_{M N} = \begin{pmatrix} w g_{\mu \nu} & 0 \\ 0 & -v^2 \end{pmatrix} \text{ and } G^{M N} = \begin{pmatrix} g^{\mu \nu} / w & 0 \\ 0 & - 1 / v^2 \end{pmatrix},\, 
    \label{eqn:metricparametrisation}
\end{equation}
in terms of coordinates $x^M=(x^\mu,y)$, where $y \in (-\pi r_c,+\pi r_c]$ parametrizes an orbifolded extra-dimension (where $y$ and $-y$ are identified); for convenience we define $\varphi \equiv y/r_c$, and use $y$ or $\varphi$ interchangeably as the coordinate of the fifth dimension.
Here 
\begin{align}
    g_{\mu\nu}\equiv \eta_{\mu\nu} + \kappa \hat{h}_{\mu\nu}(x_{\mu},y)~,  \label{eq:metricparam1}
\end{align}
where $\eta_{\mu \nu}$ is the usual $4$D ``mostly-minus" metric and $\hat{h}_{\mu\nu}(x,y)$ parametrizes metric fluctuations. The quantities $w$ and $v$ are defined as, 
\begin{eqnarray}
    w & = & \exp\left[-2\left(A(y)+\frac{ e^{2A(y)}}{2\sqrt{6}} \kappa\,\hat{r}(x,y)\right)\right] \nonumber \\
    v&=&  \left(1 + \frac{ e^{2A(y)}}{\sqrt{6}} \kappa\,\hat{r}(x,y)\right)~,
    \label{eq:metricparam2}
\end{eqnarray}
where the function $A(y)$ specifies the background RS-like geometry\footnote{For RS1~\cite{Randall:1999ee}, in which the extra dimension is unstabilized, $A(y) = k|y|$ as sourced by brane and bulk cosmological constants.} and $\hat{r}(x,y)$ parametrizes scalar metric fluctuations around this background.
We normalize the Lagrangian such that the 5D gravitational coupling $\kappa$ is related to the 5D Planck mass $M_5$ according to $\kappa^2=4/M_5^{3}$.\footnote{This specific form of the metric in Eq.~(\ref{eqn:metricparametrisation}) with the definitions in Eqs.~(\ref{eq:metricparam1}) and (\ref{eq:metricparam2}) was used in our previous works 
\cite{SekharChivukula:2019yul,SekharChivukula:2019qih,Chivukula:2020hvi}
for the unstabilized RS1 metric, following~\cite{Charmousis:1999rg}, to ensure that the quadratic gravitational Lagrangian is brought to a canonical form, free of any mixing between the scalar and tensor parts of the metric.}

The GW stabilization mechanism~\cite{Goldberger:1999uk,Goldberger:1999un} relies on the existence of a non-trivial background scalar field configuration $\phi_0(y)$ such that the size of the extra dimension is stabilized by a competition between the kinetic (gradient) and potential energies of this field, which is ``pinned" to different values at the two ends (branes) of the orbifold. We expand the bulk scalar $\hat{\phi}(x,y)$ as
\begin{align}  
    \hat{\phi} (x,y) &\equiv \frac{1}{\kappa}\phi \equiv \frac{1}{\kappa}\left[\phi_{0}(y) + \hat{f}(x,y)\right] ~,\label{eq:gandPhi}
\end{align}
where scalar fluctuations are encoded in $\hat{f}(x,y)$.\footnote{The factors of $\kappa$ (in units of energy$^{-3/2}$) included in Eq.~\eqref{eq:gandPhi} are defined so that $\phi_{0}$ and $\hat{f}$ are dimensionless in natural units.} 

The background scalar and metric field configurations specified by $\phi_{0}(y)$ and $A(y)$ must be found by solving the coupled gravity/bulk-scalar field equations~\cite{DeWolfe:1999cp}, and requires that we specify the form of the GW potential in ${\cal L}_{pot}$. Of note is that there is non-trivial mixing between the bulk scalar field and the scalar metric fluctuations in the presence of
a non-trivial scalar background. The analysis of this system can be simplified by imposing the  gauge fixing constraint~\cite{Boos:2005dc,Kofman:2004tk},
\begin{align}
    (\partial_{\varphi}\phi_{0})\,\hat{f}(x,y) \equiv \kappa \sqrt{6}\, e^{2A(\varphi)}\,\big(\partial_{\varphi}\hat{r}\big)~. \label{SRS1GaugeCondition}
\end{align}
Having imposed this condition, there is only one set of physical scalar-fluctuations, and all fluctuations in the model are contained in the five-dimensional fields $\hat{h}_{\mu\nu}(x,y)$ and $\hat{r}(x,y)$. 
In the next subsection, we describe the effective 4D KK modes that arise from these 5D fields.

\subsection{Tensor and Scalar KK Modes}
\label{sec:kkmodes}

For any arbitrary potential that stabilizes the RS1 geometry, the quadratic fluctuation terms of the full Lagrangian in Eq.~(\ref{eq:stablag}), once compactified with proper boundary conditions, will generate a pair of Sturm-Liouville (SL) equation that can be solved to yield the Kaluza-Klein spectrum of the gravitational (spin-2) sector and the scalar sector. A full description of the procedure to bring the KK mode Lagrangian to a quadratic form can be found in~\cite{Chivukula:2021xod,Chivukula:2022tla}, and we quote the results below.\footnote{For convenience, we have chosen a mode normalization here that differs slightly from that in~\cite{Chivukula:2022tla} and absorbed a factor of $1/\sqrt{\pi}$ in the definitions of $\psi_n(\varphi)$ and $\gamma_i(\varphi)$.}

For the tensor fluctuations, which give the masses and the wavefunctions of the spin-2 KK modes, the tensor field $\hat{h}_{\mu\nu}(x,y)$ can be decomposed into a tower of 4D KK states $\hat{h}^{(n)}_{\mu\nu}(x)$ with $\varphi \equiv y/r_{c}$,
\begin{align}
    \hat{h}_{\mu\nu}(x,y) = \dfrac{1}{\sqrt{ r_{c}}} \sum_{n=0}^{+\infty} \hat{h}^{(n)}_{\mu\nu}(x)\, \psi_{n}(\varphi)~,
    \label{eq:spin2mode-equation}
\end{align}
where, as described earlier, $r_c$ is the radius of the compact extra dimension. Here $\psi_n (\varphi)$ is the 5D wavefunction of the $n^{\text{th}}$ mode that satisfies the  SL differential equation:
\begin{align}
    \partial_{\varphi}[e^{-4A}\,\partial_{\varphi}\psi_{n}]=-\mu_{n}^{2} e^{-2A}\psi_{n}.
     \label{eq:spin2-SLeqn}
\end{align}
 These wavefunctions satisfy the Neumann boundary conditions where $(\partial_{\varphi}\psi_{n}) = 0$ at $\varphi \in \{0,\pi\}$. 
The eigenvalues $\mu_n = m_{n} r_{c}$ give the masses $m_{n}$ of the $n^{\text{th}}$ spin-2 KK mode. The wavefunctions are normalized according to 
\begin{align}
    \int_{-\pi}^{+\pi} d\varphi\hspace{10 pt} e^{-2A}\,\psi_{m}\,\psi_{n} &= \delta_{m,n} \ ,
    \label{eq:spin2-mode-norm}
\end{align}
and satisfy the completeness relation
\begin{align}
    \delta(\varphi_{2}-\varphi_{1}) =  \, e^{-2A}\sum_{j=0}^{+\infty} \psi_{j}(\varphi_{1})\, \psi_{j}(\varphi_{2})~.
    \label{eq:spin2-completeness}
\end{align}
This form of this SL problem for the spin-2 KK sector is {\it identical} to the unstabilized case. The difference is encoded in the new background geometry with a modified warp factor $A(y)$. 

The Neumann boundary conditions $(\partial_{\varphi}\psi_{n}) = 0$ implies that there is always a massless 4D graviton mode (with a wavefunction $\psi_0$ which is constant in $\varphi$) in the spin-2 KK sector. From the form of the spin-2 mode expansion in Eq.~(\ref{eq:spin2mode-equation}) and the constant graviton wavefunction, we immediately find the relationship between the 5D Planck mass and the observed 4D mass $M_{Pl}$
 \begin{equation}
     \dfrac{1}{M^2_{Pl}} = \dfrac{\abs{\psi_0}^2}{r_c M^3_5}~,
 \end{equation}
and hence, using the normalization condition in Eq.~(\ref{eq:spin2-mode-norm}),
\begin{equation}
    M^2_{Pl} = \dfrac{r_c M^3_5}{\abs{\psi_0}^2}=r_c M^3_5\int_{-\pi}^{+\pi} d\varphi\, e^{-2A(\varphi)}~.
\label{eq:4DMPl}
\end{equation}

For the spin-0 sector, in which the metric fluctuation and the bulk scalar mix proportional to background scalar field $\phi_{0}(\varphi)$,  the KK decomposition  of the 5D scalar field $\hat{r}(x,y)$ into a tower of spin-0 KK modes proceeds by introducing extra-dimensional wavefunctions $\gamma_{i}(\varphi)$ and a tower of 4D scalar fields $\hat{r}^{(i)}(x)$ parameterized as follows:
\begin{align}
    \hat{r}(x,y) = \dfrac{1}{\sqrt{ r_{c}}} \sum_{i=0}^{+\infty} \hat{r}^{(i)}(x)\, \gamma_{i}(\varphi)~.
    \label{eq:spin0-mode-expansion}
\end{align}
The mode equation that brings the 5D scalar Lagrangian to canonical form (including the effects of the mixing between the GW and gravitational sectors) is given by~\cite{Boos:2005dc,Kofman:2004tk},
\begin{align}
    \partial_{\varphi}\bigg[\dfrac{e^{2A}}{(\phi_{0}^{\prime})^{2}} (\partial_{\varphi}\gamma_{i})\bigg] - \dfrac{e^{2A}}{6}\gamma_{i} = -\mu_{(i)}^{2}\,\dfrac{e^{4A}}{(\phi_{0}^{\prime})^{2}} \, \gamma_{i} \, \Bigg\{ 1 + \frac{2\,\delta(\varphi)}{\Big[ 2 \ddot{V}_{1} r_{c} - \frac{\phi _{0}^{\prime\prime}}{\phi_{0}^{\prime}} \Big]} +  \frac{2\,\delta(\varphi - \pi )}{\Big[ 2 \ddot{V}_{2} r_{c} + \frac{\phi _{0}^{\prime\prime}}{\phi_{0}^{\prime}} \Big]} \Bigg\}~,\label{eq:spin0-SLeqn}
\end{align}
where $\phi^{\prime}_{0} \equiv (\partial_{\varphi}\phi_{0})$ and the eigenvalues $\mu_{(n)} = m_{(n)} r_{c}$ give the masses $m_{(n)}$ of the $n^{\text{th}}$ scalar KK mode.  The Dirac delta-function terms enforce the following (orbifold) boundary conditions:
\begin{align}
    (\partial_{\varphi}\gamma_{i})\bigg|_{\varphi = 0} &= -\bigg[2\ddot{V}_{1}r_{c} - \dfrac{\phi_{0}^{\prime\prime}}{\phi_{0}^{\prime}} \bigg]^{-1}\,\mu_{(i)}^{2}\,e^{2A}\,\gamma_{i} \bigg|_{\varphi = 0}~,\nonumber\\
    (\partial_{\varphi}\gamma_{i})\bigg|_{\varphi = \pi} &= +\bigg[2\ddot{V}_{2}\,r_{c} + \dfrac{\phi_{0}^{\prime\prime}}{\phi_{0}^{\prime}}\bigg]^{-1}\,\mu_{(i)}^{2}\,e^{2A}\,\gamma_{i} \bigg|_{\varphi = \pi}~,
    \label{eq:scalarBCtext}
\end{align}
where $\ddot{V}_{1,2}$ are the second functional derivatives of the brane potentials evaluated at the background-field configuration. Note that these boundary conditions reduce to Neumann form in the ``stiff-wall" limit, $\ddot{V}_{1,2} \to +\infty$, a limit which we will use in the subsection~\ref{sec:DFGK-model} in the context of the DFGK~\cite{DeWolfe:1999cp} method to construct analytic background solutions. The mixing between the gravitational and bulk scalar sectors also generates an
unconventional normalization of the scalar wavefunctions to
bring the scalar kinetic energy terms to canonical form~\cite{Chivukula:2022tla},
\begin{align}
    \delta_{mn} 
    & = 6\int_{-\pi}^{+\pi} d\varphi\hspace{5 pt} \bigg[\dfrac{e^{2A}}{(\phi_{0}^{\prime})^{2}} \gamma_{m}^{\,\prime} \, \gamma_{n}^{\,\prime} + \dfrac{e^{2A}}{6} \gamma_{m}\gamma_{n} \bigg]~. \label{eqScalar:Norm1}
\end{align}
 The scalar wavefunction completeness relation is then given by, 
\begin{equation}
    \delta(\varphi_2-\varphi_1) =  \dfrac{6 e^{4A(\varphi_1)}}{(\phi_{0}^{\prime}(\varphi_1))^{2}} \, \Bigg\{ 1 + \frac{2\,\delta(\varphi_1)}{\Big[ 2 \ddot{V}_{1} r_{c} - \frac{\phi _{0}^{\prime\prime}}{\phi_{0}^{\prime}} \Big]} +  \frac{2\,\delta(\varphi_1 - \pi )}{\Big[ 2 \ddot{V}_{2} r_{c} + \frac{\phi _{0}^{\prime\prime}}{\phi_{0}^{\prime}} \Big]}\Bigg\}~
    \sum_{j=0}^{+\infty} \mu_{(j)}^{2} \gamma_{j}(\varphi_{1})\, \gamma_{j}(\varphi_{2})~.
    \label{eq:spin0-completeness}
\end{equation}
In the scalar sector, due to the non-constant expectation value of the background scalar field, the lightest spin-0 state (identified as the radion with a wavefunction $\gamma_{0}$) acquires a mass $\mu_{(0)}>0$. 

Before explaining how we find $A(\varphi)$ and $\phi_0(\varphi)$, which satisfy the coupled gravity/scalar field equations, we specify the form of the Lagrangian for the brane-localized DM and SM fields.

\subsection{The Brane-Localized Dark and Standard Model Sectors}
\label{sec:DMLagrangian}

We write the Lagrangian of the brane localized matter (both dark matter and standard model matter) on the TeV brane  interacting with 
the spin-2 KK sector as~\cite{Chivukula:2023sua} \footnote{We follow the notations introduced in~\cite{Chivukula:2023sua}. While this was written in conformal coordinates in~\cite{Chivukula:2023sua}, the conversion to the coordinate choice used in this paper is straightforward.}, 

 \begin{equation}
    \mathcal{L}_{\rm brane~matter} = \mathcal{L}_{S,\text{brane}} + \mathcal{L}_{V,\text{brane}} + \mathcal{L}_{\chi,\text{brane}},
\end{equation}
where the Lagrangian for brane localized scalars, vector fields and fermions (which are taken here to include both the standard model fields and a putative DM candidate) are respectively of the form
\begin{align}
    \mathcal{L}_{S,\text{brane}} &= \int_{-\pi}^{\pi}d\varphi \sqrt{-\bar{G}} \left[ \frac{1}{2} \bar{G}^{\mu\nu} \partial_\mu \hat{S} \partial_\nu \hat{S} - \frac{1}{2} M_S^2 \hat{S}^2 \right] e^{2 A \left( \phi \right)} \delta \left( \phi - \pi \right),\label{eqn:Lphibrane} \\
    \mathcal{L}_{V,\text{brane}} &= \int_{-\pi}^{\pi}d\varphi \sqrt{-\bar{G}} \left[ - \frac{1}{4} \bar{G}^{\mu\rho} \bar{G}^{\nu
    \sigma} F_{\mu \nu} F_{\rho \sigma} + \frac{1}{2} M_{V}^2 \bar{G}^{\mu\nu} \hat{V}_\mu \hat{V}_\nu \right] \delta \left( \phi - \pi \right), \label{eqn:LAbrane} \\
    \mathcal{L}_{\mathcal{\chi}, \text{brane}} &= \int_{-\pi}^{\pi}d\varphi \sqrt{-\bar{G}} \left[ \bar{\hat{\mathcal{\chi}}} i e^{\mu}_{\bar{\alpha}} \gamma^{\bar{\alpha}} D_{\mu} \hat{\mathcal{\chi}} - M_\mathcal{\chi} \bar{\hat{\mathcal{\chi}}} \hat{\mathcal{\chi}} \right] e^{3 A \left( \phi \right)}  \delta \left( \phi - \pi \right)~. \label{eqn:LPsibrane}
\end{align}
Here $\bar{G}^{\mu\nu}$ and its determinant are the induced metric on the TeV brane
\begin{equation}
    \bar{G}_{\mu\nu} = \left[ wg_{\mu\nu}\right]_{\varphi=\pi}~.
\end{equation}
In addition, the vector field strength is $F_{M N} = \nabla_M \hat{V}_N - \nabla_N \hat{V}_M$, and the fermion covariant derivative is defined as 
\begin{equation}
    D_{\mu} \hat{\mathcal{\chi}} = \partial_\mu \hat{\mathcal{\chi}} + \frac{1}{2} \Omega^{\bar{\alpha} \bar{\beta}}_\mu \sigma_{\bar{\alpha} \bar{\beta}} \hat{\mathcal{\chi}}, 
\end{equation}
where $\sigma_{\bar{\alpha} \bar{\beta}} = \left[ \gamma_{\bar{\alpha}}, \gamma_{\bar{\beta}} \right] / 4$, with $\gamma_{\bar{\alpha}}$ being the gamma matrices defined over the tetrad $e^{\mu \bar{\alpha}}$. For simplicity, we take the scalar dark matter candidate to be real, while the fermion is assumed to be Dirac.

Integrating Eqs.~(\ref{eqn:Lphibrane})-(\ref{eqn:LPsibrane}) over the fifth dimension, as enforced by the delta function, the quadratic part of the action  has the following form 
\begin{align}
    S_{S S} &= \int d^4 x \left[ \frac{1}{2} \partial_\mu \hat{S} \partial^\mu \hat{S} - \frac{1}{2} M_S^2 e^{- 2  A\left( \pi \right)} \hat{S}^2 \right], \\
    S_{V V} &= \int d^4 x \left[ - \frac{1}{4} F^{\mu \nu} F_{\mu \nu} + \frac{1}{2} M_V^2 e^{- 2 A \left( \pi \right)} \hat{V}^\mu \hat{V}_\mu \right], \\
    S_{\mathcal{\chi} \mathcal{\chi}} &= \int d^4 x \left[ i \bar{\hat{\mathcal{\chi}}} \slashed{\partial} \hat{\mathcal{\chi}} - M_\mathcal{\chi} e^{-A \left( \pi \right)} \bar{\hat{\mathcal{\chi}}} \hat{\mathcal{\chi}} \right], 
\end{align}
where the fields are evaluated on the brane, $\varphi = \pi$. In what follows, we use reparametrized mass terms of the scalar, fermion and vector fields, which are
\begin{eqnarray}
     m_{{S}} & = & e^{-A(\pi)}M_{{S}}, \\
      m_{\chi} & = & e^{-A(\pi)}M_{\chi}, \\
      m_{{V}} & = & e^{-A(\pi)}M_{{V}}.
\end{eqnarray}
Unlike the situation with bulk fields, there are no interactions of the brane-localized fields that contain an explicit derivative in the fifth dimension.  The 3- and 4-point interactions of the KK sector and matter are written out in {section IV} and {appendix C} of~\cite{Chivukula:2023sua}\looseness=-1.

\subsection{The DFGK Model}
\label{sec:DFGK-model}

While the general properties of the KK gravity (and gravity-matter) scattering amplitudes do not depend on the specific form of the GW potential and the background geometry, the calculation of the DM relic abundance will require knowing the KK masses and couplings in order to generate the relevant cross sections -- and the masses and couplings are dependent on the solving the spin-2 and spin-0 Sturm-Liouville systems described in subsection~\ref{sec:kkmodes} above. Hence, we will need a method of finding a consistent set of background $A(\varphi)$ and $\phi_0(\varphi)$ fields for a GW stabilized RS-like geometry -- and from these, we will (numerically) solve for the KK wavefunctions and masses, and compute the needed couplings.

To find consistent background solutions we will use the strategy employed in the DFGK model~\cite{DeWolfe:1999cp}, with the introduction of a superpotential-inspired function $W[\hat{\phi}]$ that can be used to derive a GW potential for which the background equations can be easily solved. In this formulation, the scalar bulk and brane potentials are parameterized (in dimensionless form) as:
\begin{align}
    &\hspace{-60 pt}V r_{c}^{2} = \dfrac{1}{8}\bigg(\dfrac{dW}{d\hat{\phi}}\bigg)^{2} - \dfrac{W^2}{24}~,\\
   \varphi\equiv 0: \, \,  V_{1}r_{c} = +\dfrac{W}{2} + \beta_{1}^{2}\bigg[\hat{\phi}(\varphi) -\phi_{1}\bigg]^2~,\hspace{20 pt}&\hspace{20 pt}\varphi\equiv\pi:\, \, V_{2}r_{c} = -\dfrac{W}{2} + \beta_{2}^{2}\bigg[\hat{\phi}(\varphi)-\phi_{2}\bigg]^2~,
   \label{eq:branepotentials}
\end{align}
 and we take the ``stiff-wall" limit: $\beta_{1,2}\to +\infty$, so that $\phi_{1} \equiv \hat{\phi}(0)$ and $\phi_{2} \equiv \hat{\phi}(\pi)$.
 The background scalar and Einstein equations can then be analytically solved to give,
\begin{align}
       (\partial_{\varphi} A) = \frac{W}{12}\bigg|_{\hat{\phi}=\phi_{0}}\text{ sign}(\varphi)~,\hspace{20 pt}&\hspace{20 pt}(\partial_{\varphi} \phi_{0}) = \dfrac{dW}{d\hat{\phi}}\bigg|_{\hat{\phi}=\phi_{0}}\text{ sign}(\varphi)~.\label{AAndPhiInTermsOfW}
\end{align}

The DFGK analysis \cite{DeWolfe:1999cp} then introduces a convenient $W[\hat{\phi}]$ with the following specific form:
\begin{align}
    W[\hat{\phi}(\varphi)] = 12 kr_{c} - \dfrac{1}{2}\hat{\phi}(\varphi)^{2}\,ur_c~.
\end{align}
Plugging this into Eq.~(\ref{AAndPhiInTermsOfW}), we find solutions for the bulk scalar vacuum and the warp factor:
\begin{align}
    \phi_0(\varphi) &= \phi_{1}e^{-ur_c|\varphi|}~, \label{eq:scalarbackgroundi}\\
    A(\varphi) &= kr_{c}|\varphi| + \dfrac{1}{48}\phi_{1}^{2}\bigg[e^{-2ur_c|\varphi|} - 1\bigg]\ ,
    \label{eq:AnPhi}
\end{align}
where the parameters $u$, $\phi_{1}$, and $\phi_{2}$ are related according to
\begin{equation}
    ur_c=\dfrac{1}{\pi}\log\dfrac{\phi_1}{\phi_2}~.
\end{equation}
Given these  $\phi_{0}(\varphi)$ and $A(\varphi)$, we solve the numerically for the mass spectrum and wavefunctions of the spin-2 KK sector and the spin-0 GW scalar sector. 

One convenient feature of this choice of GW potential is that, in the limit $u \to 0$, it reproduces the unstabilized RS1 model with $A(y)=kr_c|\varphi|$ (and for which $\hat{\phi}(\varphi) \equiv \phi_1=\phi_2$ is constant). In addition, for the light radion (small $u$), the effect of the back-reaction of the GW dynamics on the RS geometry is small, and we can use the perturbative DFGK model described in~\cite{Chivukula:2022tla}. In the next section, we will describe the masses and couplings of the KK modes for the light radion limit. A study of the model with a heavy radion will be related in a subsequent work~\cite{Chivukula:2024}.

\subsection{Masses and Matter Couplings in the Light Radion Limit}
\label{sec:LightRadion}

In this section, we will exploit the light radion limit to explain the general features of the KK masses and their couplings with brane-localized matter. We note, however, that all of the numerical analyses presented later incorporate numerical solutions whose accuracy goes beyond the perturbative analysis given in this section.
Expanding the solutions given in Eqs.~(\ref{eq:scalarbackgroundi}) and (\ref{eq:AnPhi})  for small $u$, and solving the Sturm-Liouville equation for the radion (the lightest spin-0 KK mode with wavefunction $\gamma_{(0)}(\varphi)$) perturbatively, we find that the warp factor and the radion mass can be expressed~\cite{Chivukula:2022tla} in terms of three parameters $\{ \tilde{k}, r_c, \epsilon \}$ as
\begin{align}
    A{\left( \varphi \right)} &= \tilde{k} r_c \abs{\varphi}+{\cal O}(\epsilon^2 \abs{\varphi}^2), \\
    m_r^2 &= \frac{8 \epsilon^2}{r_c^2 \left(  1 + e^{2 \pi \tilde{k} r_c } \right)} + \order{ \epsilon^3},
\end{align}
where the two parameters $\tilde{k}$ and $\epsilon$ are related to the original three potential parameters $\{k, u,\phi_1 \}$ by
\begin{align}
    \tilde{k} &= k - \frac{\phi_1^2 u}{24}, \\
    \epsilon &= \frac{\phi_1 u r_c}{\sqrt{24}}. 
\end{align}

To leading order in this limit, from Eq.~(\ref{eq:4DMPl}) we find the usual RS relation between the 5D and 4D Planck masses
\begin{equation}
    M^2_{Pl} = \dfrac{M^3_5}{\tilde{k}}\left( 1-e^{-2\tilde{k}\pi r_c}\right)~.
\end{equation}
For large ${\tilde k} r_c$, we also find the usual masses of the spin-2 KK modes in as~\cite{Chivukula:2022tla} 
\begin{equation}
    m_{n}= x_{n} \tilde{k}e^{-\tilde{k}\pi r_{c}}\ ,
\end{equation}
where $x_{n}$ are zeroes of the Bessel function of the first kind.
The spin-2 KK modes couple to the energy-momentum tensor of the TeV-brane localized matter through the induced 4D metric on the brane, $\bar{G}_{\mu\nu}=[w g_{\mu\nu}]_{\varphi=\pi}$
\begin{align}
{\cal L}_{\rm spin-2~couplings} =     \dfrac{1}{\sqrt{r_c} M^{3/2}_5} \sum_n {\hat h}^{(n)}_{\mu\nu}(x) \psi_n(\varphi=\pi) T^{\mu\nu} ~,
\end{align}
and we find immediately, from Eq.~(\ref{eq:4DMPl}), that the graviton couples with strength proportional to $M^{-1}_{Pl}$. The coupling of spin-2 KK modes can be rewritten as 
\begin{align}
 {\cal L}_{\rm spin-2~couplings} =     \dfrac{1}{\psi_1(\varphi=\pi) \Lambda_\pi} \sum_n {\hat h}^{(n)}_{\mu\nu}(x) \psi_n(\varphi=\pi) T^{\mu\nu} ~,
\end{align}
where $\Lambda_\pi$ characterizes the coupling to the first spin-2 KK mode. In the large $\tilde{k}r_c$ limit, $\Lambda_\pi$ is given by
\begin{align}
    \Lambda_\pi& = \dfrac{\psi_0}{\psi_1(\varphi=\pi)} M_{Pl} \approx e^{-\tilde{k}r_c \pi} M_{Pl}~.
\end{align}
Thus, in the light radion limit,  we may trade the parameters $\{ \tilde{k}, r_c, \epsilon \}$ in favor of three physical parameters $\{ \Lambda_\pi, m_1, m_r \}$ in the gravitational sector.

For the spin-0 GW sector, as we noted in~\cite{Chivukula:2022tla}, the 
masses of the GW scalars for large $\tilde{k}r_c$ can be expressed as, 
\begin{equation}
    m_{(n)}= z_{n} \tilde{k}e^{-\tilde{k}\pi r_{c}}
\end{equation}
 where $z_{n}$ are the roots of the Bessel functions of the second kind. From Eq.~(\ref{eq:metricparam2}), we see that the scalar fields couple to the trace of the TeV-brane matter energy-momentum tensor,
 \begin{align}
     {\cal L}_{\rm spin-0~couplings} & \propto - \dfrac{e^{2A(\varphi=\pi)}}{\sqrt{r_c} M^{3/2}_5} \sum_i {\hat r}^{(i)}(x) \gamma_i(\varphi=\pi) T^\mu_\mu = -\dfrac{e^{2A(\varphi=\pi)}}{\psi_1(\varphi=\pi) \Lambda_\pi} \sum_i {\hat r}^{(i)}(x) \gamma_i(\varphi=\pi) T^\mu_\mu ~.
 \end{align}

\begin{figure}[t]
    \centering
\includegraphics[width=0.7\linewidth]{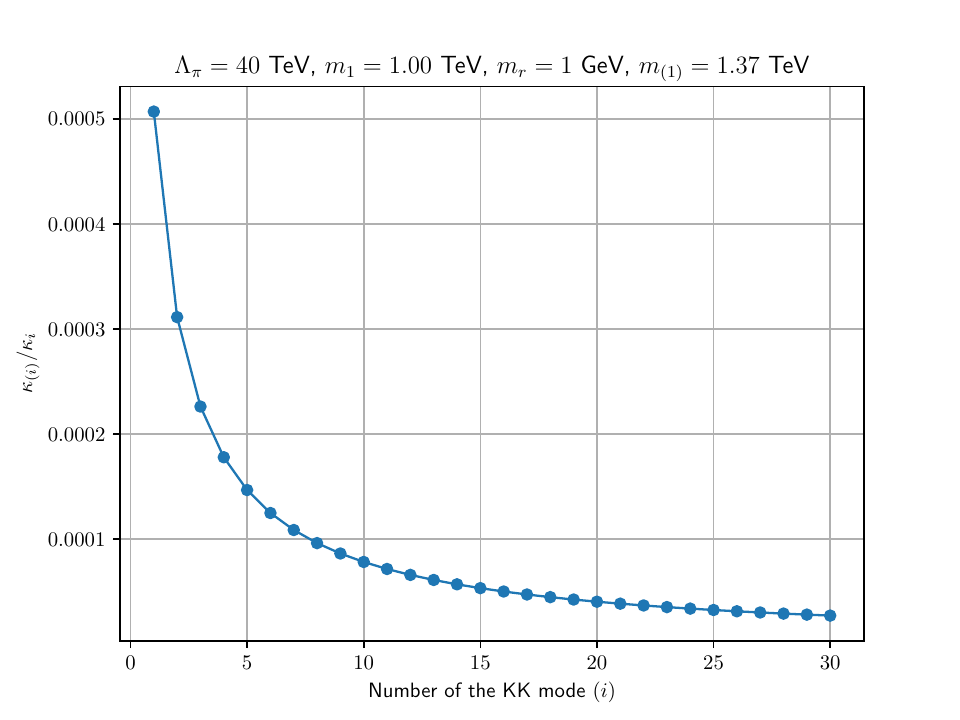}
    \caption{ The ratio of the  GW scalar couplings ($\kappa_{(i)}$) to the spin-2 KK mode couplings ($\kappa_{i}$), starting at mode number i=1. See Eq. \ref{eqn:DefinedKappaCouplings}, and the corresponding discussion in the text. \looseness=-1
     \label{fig:couplplots}}
\end{figure}

For brevity of notation, in the following sections, we will use these definitions for couplings of the spin-2 and spin-0 KK modes to TeV-brane matter, \looseness=-1
\begin{equation}
    \kappa_m \equiv \frac{\psi_m \left( \pi \right)}{\psi_1 \left( \pi \right)} \frac{1}{\Lambda_{\pi}} \text{ and } \kappa_{\left( m \right)} \equiv \frac{1}{\Lambda_{\pi}} \frac{\gamma_m \left( \pi \right)}{\psi_1 \left( \pi \right)} e^{2 A\left( \varphi=\pi \right)}. \label{eqn:DefinedKappaCouplings}
\end{equation}
From the form of couplings it is not obvious that $\kappa_{(m)}$ is suppressed relative to $\kappa_{m}$. However, there is an exponential suppression of $\gamma_{m}(\pi)e^{2A(\pi)}$, relative to $\psi_{m}(\pi)$, which leads to the spin-0 couplings being orders of magnitude smaller than the spin-2 couplings. More physically, since the mixing between the tower of scalar states and the gravitational sector vanishes in the limit of a massless radion, $\mu_{(0)}\to 0$, when the background scalar field configuration becomes trivial (flat), the scalar couplings are suppressed in the light radion limit. The suppression of the couplings $\kappa_{(i)}$  is illustrated in  Fig. \ref{fig:couplplots}, where we plot the ratio of the GW scalar couplings  to the spin-2 KK couplings $\kappa_{(i)} / \kappa_{i}$, corresponding to Eq. \ref{eqn:DefinedKappaCouplings}.

Thus, for given values of $\{ \Lambda_\pi, m_1, m_r \}$, we can determine the masses, wavefunctions, and couplings of all KK modes. Using this input, we can compute the DM annihilation cross sections and consider the thermal relic abundance of these particles; that is the subject of the next section.

\section{Dark Matter Annihilation Cross Sections}

\label{sec:cross-sections}

\begin{figure}[t]
    \centering
    \includegraphics[width=\textwidth]{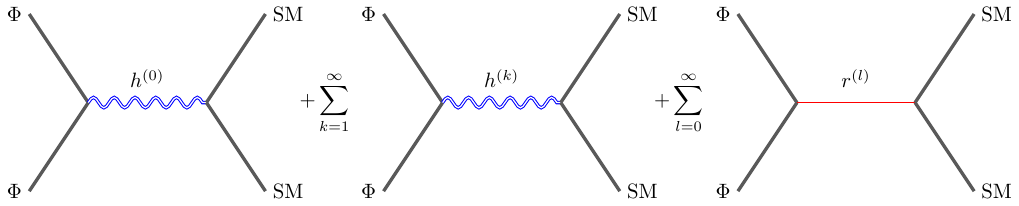}
\caption{ The diagrams for brane dark matter annihilating into Standard Model modes via the exchange of the massless graviton ($h^{(0)}$), spin-2 KK modes ($h^{(k)}$), or scalar sector modes ($r^{(l)}$). $\Phi$ denotes all possible types of brane dark matter: scalars $S$, fermions $\chi$, and vectors $V$; SM, likewise, denotes all Standard Model modes. \label{fig:KKPortalDiag}} 
\end{figure}

\begin{figure}
    \includegraphics[width=\textwidth]{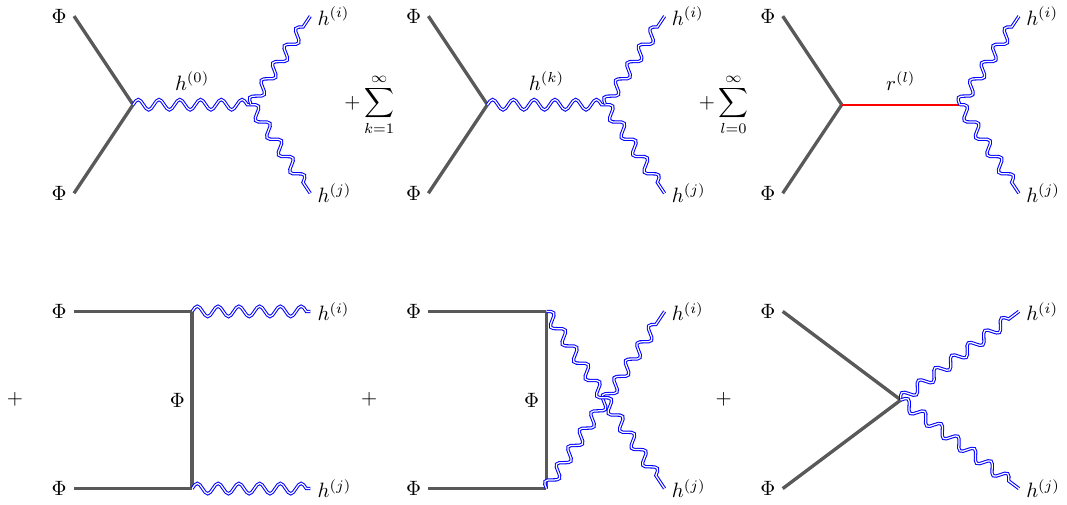}
     \caption{Brane localized 
 dark matter ($\Phi = \left(  S, V, \chi \right)$, as in Fig. \ref{fig:KKPortalDiag}) annihilating to spin-2 KK modes ($h^{(k)}$). Here $r^{\left(i\right)}$ represents the the $i$th spin-$0$ (radion or GW scalar) KK mode.} 
    \label{fig:branescalar}
\end{figure}

\begin{figure}[t]
    \centering
    \includegraphics[width=\textwidth]{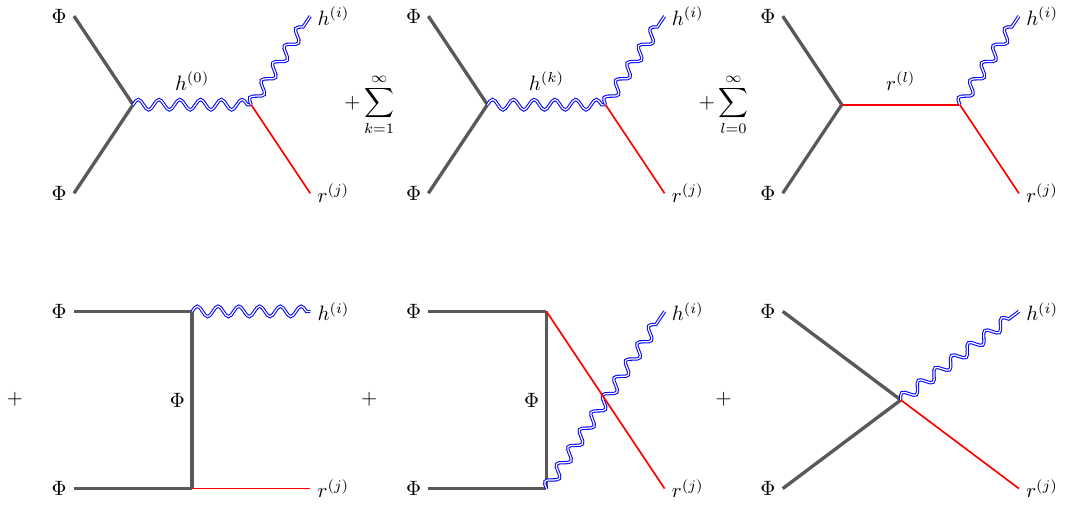}
    \caption{ Diagrams for brane dark matter annihilating to a mixed final state: one spin-$2$ and one spin-$0$ KK mode. As in Fig. \ref{fig:KKPortalDiag} ($\Phi = \left(  S, V, \chi \right)$) represents the brane dark matter particle (scalar, vector boson), while $h^{\left(i\right)}$ denotes the $i$th spin-$2$ KK mode and $r^{\left(i\right)}$ denotes the $i$th spin-$0$ KK mode (only relevant for the brane stabilized RS model). \label{fig:ProcessInQuestion2}}
\end{figure}

\begin{figure}[t]
    \centering
    \includegraphics[width=\textwidth]{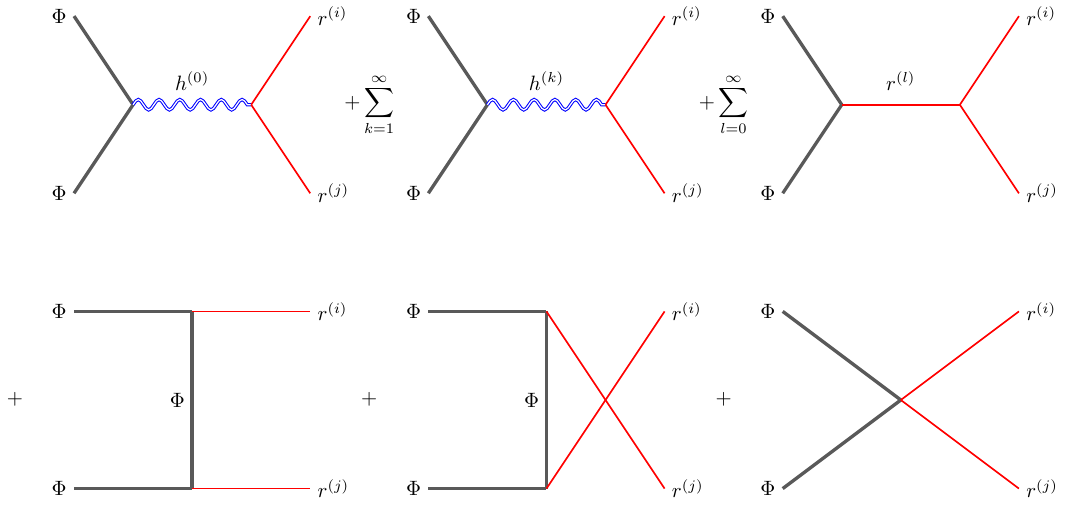}
    \caption{ Diagrams for brane dark matter annihilating to two spin-$0$ KK modes. As in Fig. \ref{fig:KKPortalDiag} ($\Phi = \left(  S, V, \chi \right)$) represents the brane dark matter particle (scalar, vector boson), while $h^{\left(i\right)}$ denotes the $i$th spin-$2$ KK mode and $r^{\left(i\right)}$ denotes the $i$th spin-$0$ KK mode. \looseness=-1\label{fig:ProcessInQuestion3}}
\end{figure}

In order to compute the relic abundance of the freeze-out process, we require the velocity-averaged cross-sections governing the thermal equilibrium abundance of dark matter in the early universe. In the Kaluza-Klein portal dark matter models we are considering, we need the amplitudes for 3 distinct types of processes,
\begin{itemize}
\item Annihilation of the dark sector particles to SM via the KK portal propagators, which includes the spin-2 KK modes, the massive radion, and the GW scalars, described in Fig.~\ref{fig:KKPortalDiag}.
\item Annihilation of dark sector particles into the spin-2 KK sector, described in Fig.~\ref{fig:branescalar} 
\item Annihilation to one or two GW scalar sector modes, including the now massive radion, described in Fig.~\ref{fig:ProcessInQuestion2}~and~\ref{fig:ProcessInQuestion3}.   
\end{itemize}
The potentially problematic high-energy scattering behavior is intrinsic to the diagrams involving spin-2 KK mode self-couplings, and hence in the last two classes of contributions illustrated in Figs.~\ref{fig:branescalar},~\ref{fig:ProcessInQuestion2},~and~\ref{fig:ProcessInQuestion3} -- but are absent from the $s$-channel spin-2 diagrams in Fig.  \ref{fig:KKPortalDiag} since the energy-momentum tensors of the DM and SM sectors are conserved (to this order in perturbation theory).

We describe each of these annihilation channels in the subsections that follow.


\subsection{Annihilation to SM particles via the KK portal}
\label{sec:SMann}

For the KK portal DM scenario, the spin-2 and the spin-0 KK modes mediate the only interactions linking the SM and DM sectors. The couplings of the gravity sector are, as described above, proportional to the energy-momentum tensors of these TeV-brane localized sectors. The Feynman diagrams for dark matter annihilation to SM particles are depicted in Fig.~\ref{fig:KKPortalDiag}. The first two diagrams describe the dark species ($\Phi=(S,V,\chi)$) annihilating via the massless graviton $h^{\left( 0 \right)}$, and the tower of spin-2 KK modes $h^{\left( k \right)}$. The third diagram depicts the annihilation via the scalar sector $r^{\left( l \right)}$, of which the $l=0$ state describes the massive radion, while the $l\neq 0$ state describes the GW scalars. All final state SM species are included: the Higgs, the SM fermions, the massive weak gauge bosons, and the photon and the gluons. \looseness=-1

As an example, for scalar DM annihilating to the Higgs final state, the matrix element of the processes in Fig.~\ref{fig:KKPortalDiag} are given by
\begin{equation}
\aligned
    \mathcal{M}^{S S} =&~ \frac{i}{24} \left[ \frac{3 \kappa_0}{s} \left( s \left(4 \left( m_{h}^2 + m_S^2 \right) + s \right) + \left( 4 m_{h}^2 - s \right) \left( s - 4 m_S^2 \right) \cos 2 \theta \right) \right. \\ 
    &~ + \left. \left( 4 m_{h}^2 - s \right) \left( s - 4 m_S^2 \right) \sum_{k=1}^\infty \frac{\kappa_k^2}{s - m_k^2 + i m_k \Gamma_{h^{\left( k \right)}}} - 4 \left( 2 m_{h}^2 + s \right) \left( 2 m_S^2 + s \right) \sum_{n=0}^\infty \frac{\kappa_{\left( n \right)}^2}{s - m_{(n)}^2 + i m_{(n)} \Gamma_{r^{\left( n \right)}}} \right],
    \label{eqn:HiggsChannelScalarDMMatrixElement}
\endaligned
\end{equation}
where $\sqrt{s}$ and $\theta$ are the center of mass energy-squared and scattering angle. Here $m_k$ and $m_{(n)}$ are masses of the intermediate spin-$2$ and spin-$0$ KK modes, respectively, and the $\kappa_k$ and $\kappa_{(n)}$ are the corresponding couplings as defined in Eq.~(\ref{eqn:DefinedKappaCouplings}). Furthermore, the $\Gamma_{h^{\left( k \right)}}$ and $\Gamma_{r^{\left( n \right)}}$ are the decay widths of the KK modes, expressions for which are provided in appendix~\ref{sec:smdecaywidth}.\footnote{The expressions we find agree with those provided in~\cite{Han:1998sg}}. Similar expressions are found for brane scalar annihilation to SM vectors and fermions.

 The decay width of the $k$th spin-$2$ KK mode $\Gamma_{h^{\left( k \right)}}$ depends on the quantity $\kappa_k^2 m_k^3$ as can be seen from appendix~\ref{sec:decayspin2}, and therefore the matrix element in Eq.~(\ref{eqn:HiggsChannelScalarDMMatrixElement}) when evaluated on the $k$th spin-$2$ resonance ($s = m_k^2$) is independent of the coupling constant $\kappa_k$ and the mass $m_k$. Hence, due to the phase-space pre-factor, the corresponding cross-section evaluated at the $k$th spin-$2$ resonance scales as $\simeq 1 / m_k^2$.

\begin{figure}[t]
    \centering
    \includegraphics[width=0.49\linewidth]{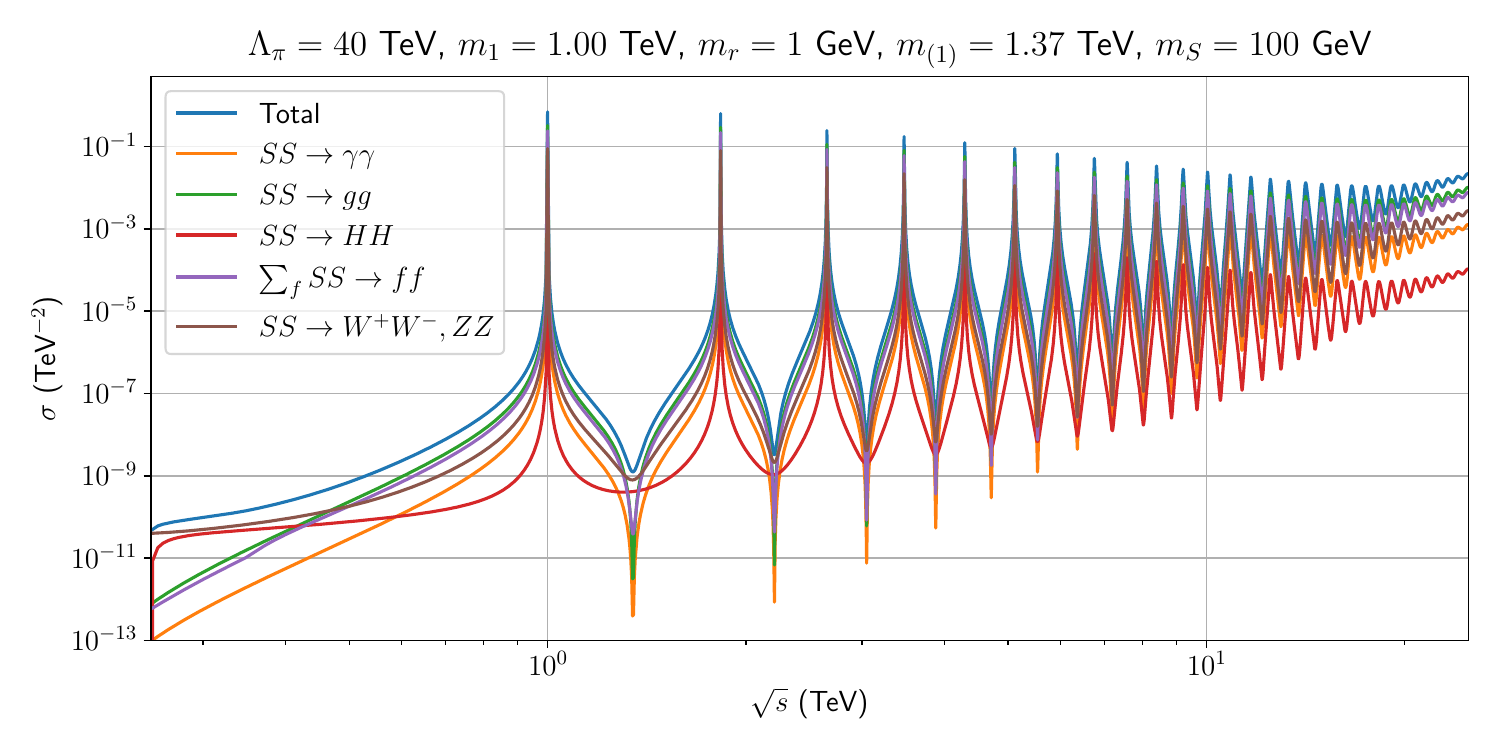}
    \includegraphics[width=0.49\linewidth]{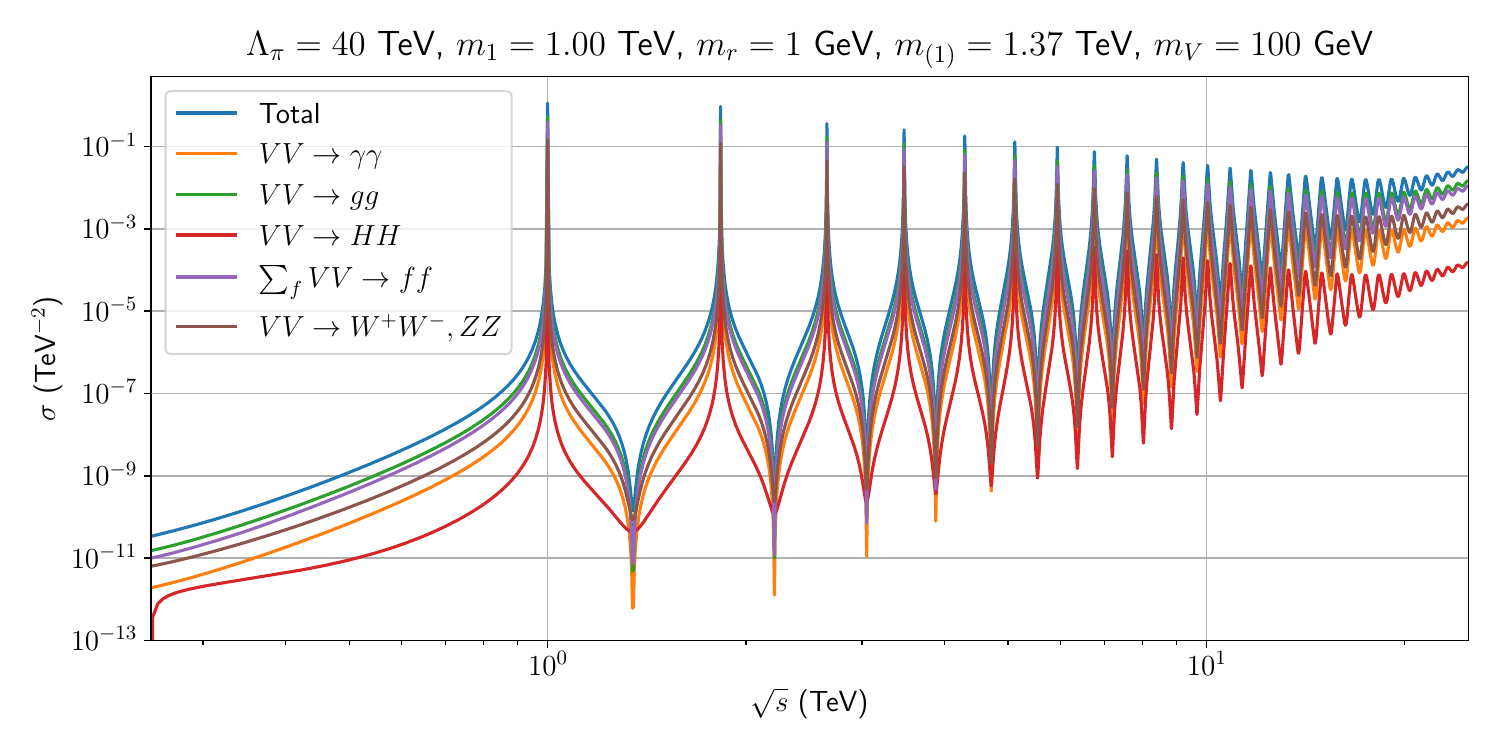} 
    \caption{Individual contributions to the cross-section for the scalar dark matter process $S S \rightarrow \rm SM~SM$ (left panel) and vector process $V~V\to \rm SM~SM $ (right panel) for a dark matter mass of $100$ GeV. The cross-section is obtained via a sum over a truncated tower of $25$ internal KK states. Here $\gamma$ corresponds to a photon, $g$ to a gluon, $H$ to Higgs, $f$ is a placeholder for the SM fermion with $\sum_f$ denoting the sum over all SM fermions. \label{fig:SigmaSSKKPortal} }
\end{figure}

For a given choice of model parameters $\{\Lambda_{\pi},m_{1},m_{r}\}$ we numerically solve for the KK mode masses, wavefunctions, and couplings (which we do without recourse to the light radion or large $\tilde{k} r_c$ approximations). Then, we compute the scalar DM annihilation cross-sections for any scalar DM mass. In the left panel of Fig.~\ref{fig:SigmaSSKKPortal}, we plot the behavior of the cross-section of the dark matter particle annihilation $ S~S\to \rm SM~SM$ for $\rm \Lambda_{\pi}= 40~TeV$, $\rm m_{1}=1~ TeV$, and $\rm m_{r}= 1~GeV$ (from which we find that $\rm m_{(1)}\simeq~1.4~TeV$, where $\rm m_{(1)}$ denotes the mass of the first GW scalar). We plot both the total annihilation cross-section and those to particular SM final states.
We analogously compute the annihilation of vector or fermion dark matter particles, with an example of vector DM annihilation $ V~V~\to \rm SM~SM $ plotted in the right panel of Fig.~\ref{fig:SigmaSSKKPortal}.

We observe that the scalar and vector annihilation cross-sections are very similar and, as might be expected, maximally enhanced on the KK resonances.  Furthermore, the vector annihilation cross-section resonance peak heights are slightly larger than the corresponding scalar peak heights. Correspondingly, we find that fermion DM annihilation results in a cross-section (which is not shown here) with resonance peak heights between those found for scalars and vectors. \looseness=-1

In addition, while the GW scalar resonances are also present, their contribution to the total cross-sections is significantly smaller. This is a consequence of the light radion parameter space investigated here. The GW scalars have no direct coupling to DM; rather, the GW scalar contribution to the DM annihilation cross-section comes entirely from the ``mixing" of the GW scalars with the gravity sector, a mixing that vanishes in the limit of zero radion mass.\looseness=-1

\subsection{Annihilation to spin-2 KK graviton}

The second set of processes we are interested in is DM annihilations to spin-2 KK gravitons, depicted in Fig.~\ref{fig:branescalar}. 
These processes are calculated in~\cite{Chivukula:2023sua} in unitary gauge\footnote{These processes were also analyzed in~\cite{Chivukula:2023qrt} using 't-Hooft-Feynman gauge and the Goldstone boson equivalence theorem.} and, as a check of the numerical unitary-gauge computations in this work, we verified the cancellation of the bad high-energy behavior between the different diagrams. We briefly review the behavior of these scattering amplitudes below and then provide illustrative examples of the behavior of the corresponding cross-sections.

Consider the scattering of a pair of brane-localized matter fields $\bar{\Phi}$ into a pair of longitudinally polarized KK gravitons, \begin{align}
    {\Phi}_{\lambda}\bar{\Phi}_{\bar{\lambda}} \rightarrow h_{L}^{(i)}h_{L}^{(j)}, 
\end{align}
where the ${\Phi}$ represent incoming brane matter fields with ${\Phi} = {S},\chi,{V}$; here $\lambda$, $\bar{\lambda}$ denote their helicities.
The computation of these contributions involves the self-couplings of the spin-2 KK modes (including the graviton) and the couplings between a spin-0 mode (the radion or GW scalars) and two KK modes. The form of the self-couplings of the gravity KK sector are the same as those in the unstabilized RS1 model~\cite{SekharChivukula:2019yul,SekharChivukula:2019qih,Chivukula:2020hvi,Foren:2020egq}. The inclusion of the GW sector introduces a new set of coupling structures, which were discussed in~\cite{Chivukula:2021xod,Chivukula:2022tla}.

There are two regimes of interest in analyzing the behavior of this class of amplitudes and their contribution to the relic density. First, the resonance or ``funnel" region, with $m_{i},m_{(r_{l})}\simeq 2 m_{\bar{\Phi}}$, where the intermediate KK graviton or the (massive) scalar propagator goes on resonance. Accurate computations in this regime require numerical computations of sufficient accuracy for all of the spin-2 and spin-0 mode masses, wavefunctions, and couplings.

The second region of interest is the high energy $\sqrt{s}\to \infty$ limit, where individual diagrams in the amplitude would naively seem to grow as $\mathcal{M}\propto\mathcal{O}\left(s^{3}/ m_{i}^{4} \Lambda^2_\pi \right)$ due to the longitudinal polarizations of the external KK states. This growth is, however, unphysical, and when the full set of diagrams is taken into account (including intermediate GW states), the amplitude grows as $\mathcal{O}(s/\Lambda_{\pi}^{2})$~\cite{Chivukula:2023sua}. 
For example, at high energies, the leading order contribution to the annihilation amplitude reads, \looseness=-1 
\begin{equation}
        \mathcal{M}_{00}^{(1)} (SS \to h^{(i)} h^{(j)})= - \frac{i \kappa_i \kappa_j }{ 24 }  \left(  1 + 3 \cos 2 \theta \right)~,
\end{equation}
where we expand the matrix element $\mathcal{M}_{\lambda,\bar{\lambda}}$ in powers of $s$,
\begin{equation}
    \mathcal{M}_{\lambda\bar{\lambda}}(s,\theta) = \sum_{\sigma\in \mathbb{Z}} {\mathcal{M}}^{(\sigma/2)}_{\lambda\bar{\lambda}}(\theta) s^{\sigma/2}.
    \label{eqn:expantionDef}
\end{equation}
As another example, the leading order scattering amplitude for helicity-0 vector DM annihilation  is given by
\begin{equation}
\mathcal{M}_{00}^{(1)} (V_0 V_0 \to h^{(i)} h^{(j)}) =  \frac{1}{24} \left[\kappa_i \kappa_j \left( 1 + 3 \cos 2 \theta \right) \right].
\end{equation}
Note that these leading order contributions can be written in terms of the KK-DM couplings $\kappa_i$ directly, without reference to the rather more complicated KK mode self-coupling structures that appear in individual diagrams.
 The ability to express the leading order growth in the scattering amplitude entirely in terms of the $\kappa_i$ parameters is a result of the ``sum rules" which enforce cancellations found in~\cite{Chivukula:2023sua,Chivukula:2023qrt}.

\begin{figure}[t]
    \centering
    \includegraphics[width=0.49\linewidth]{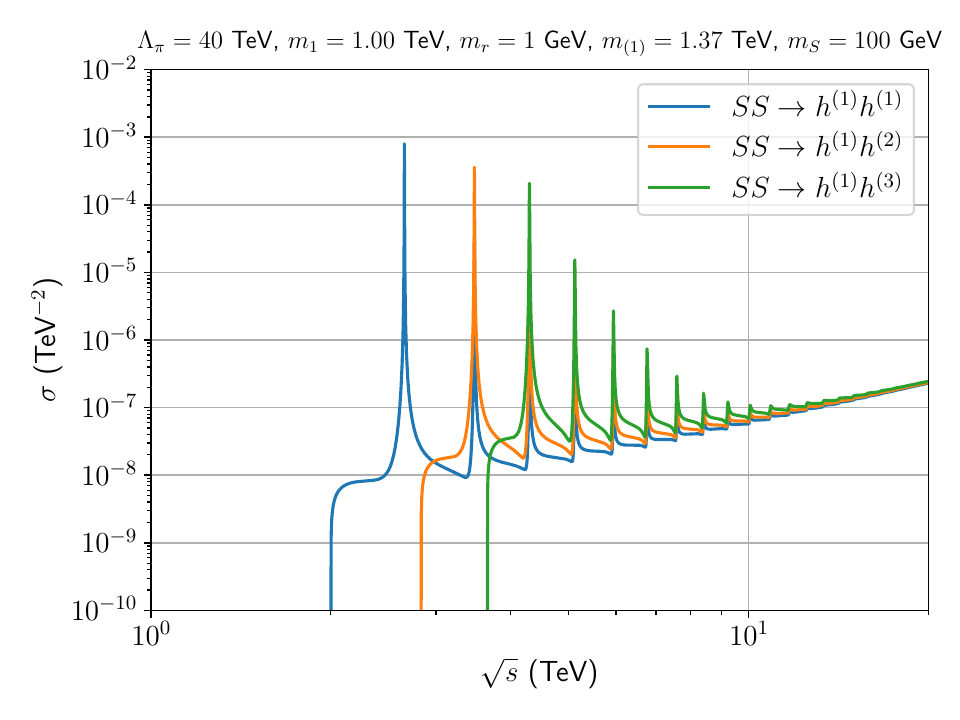}
    \includegraphics[width=0.49\linewidth]{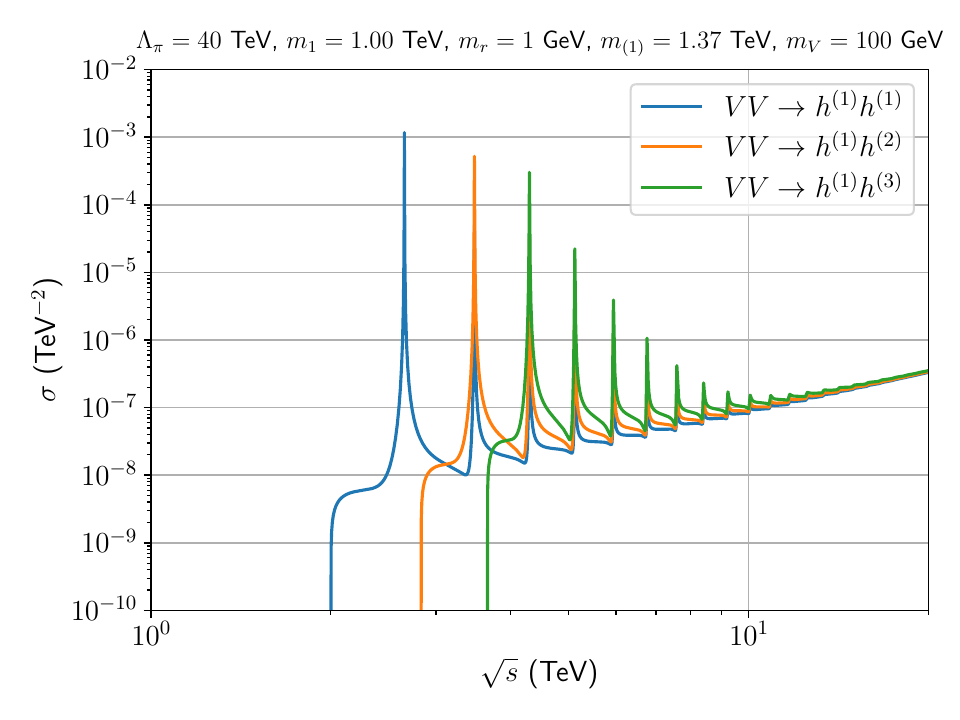}
    \caption{Cross-section for the scalar process $S S \rightarrow h^{(i)} h^{(j)}$ and vector process $V V \rightarrow h^{(i)} h^{(j)}$ (right panel) for a dark matter mass of 100 GeV. The cross-sections are computed by summing over a truncated KK tower of $25$ internal spin-$0$ and $25$ internal spin-$2$ KK modes.\looseness=-1
    \label{fig:CrossSectsPhiPhi} } 
\end{figure}

We numerically evaluate the DM to KK mode cross sections so that all wave functions and the masses of the spin-2 KK sector states and GW scalar sector states are obtained to the required precision. The numerical demonstration that resulting overall amplitudes grow linearly with $s$ (and the corresponding cross-sections like $s^2$) at high energy represents a check of the numerical accuracy of our results. In the left panel of Fig.~\ref{fig:CrossSectsPhiPhi}, we plot the illustrative behavior of various scalar DM to KK mode cross-sections for the model parameters shown. We observe the resonances when $\sqrt{s}\simeq 2 m_{i}$, and also see that the cancellations at high energy lead to small (relative to the resonant peaks) cross sections which grow only linearly with $s$ at high-energy. We also note that the resonances occurring due to the intermediate GW scalars are significantly suppressed compared to those due to spin-2 KK states; only resonances due to the exchange of spin-2 KK states are visible in the figure. Similar behavior can be observed for the initial state of brane-localized fermions and vectors. In the right panel of Fig.~\ref{fig:CrossSectsPhiPhi}, we present the annihilation to spin-2 KK modes via the diagrams in Fig.~\ref{fig:branescalar} for vector DM candidates. Again, we notice that the cross-sections for the vector-initiated states are larger than the corresponding scalar-initiated states due to the polarization states of the massive vector, and we find that the fermion case is intermediate. 

Finally, comparing Figs.~\ref{fig:SigmaSSKKPortal}~and~\ref{fig:CrossSectsPhiPhi},  we note that, once all of the proper cancellations have been taken into account, the cross-sections corresponding to the DM annihilation to KK mode processes are always substantially smaller (both on resonance and at high-energies) than the corresponding DM annihilation cross sections to SM particles.

\subsection{Annihilation to GW scalar final states}


The final set of processes includes the mixed spin-2 KK mode and the spin-0 (radion + GW scalar) modes $h^{(i)}r^{(j)}$ in the final state, corresponding to the Feynman diagrams in Fig.~\ref{fig:ProcessInQuestion2},  as well as a pair of spin-0 modes $r^{(i)}r^{(j)}$ in the final state, as illustrated in Fig.~\ref{fig:ProcessInQuestion3}.

These diagrams require new sets of interactions involving one spin-2 KK mode and two (massive) radion/GW states or three radion/GW states; none of these were required in previous work, so this is the first time they are being calculated and studied. We expand the Lagrangian to the appropriate order in coupling to extract these interactions and derive the corresponding Feynman rules. We document the corresponding Lagrangians, interactions and Feynman rules in appendix~\ref{sec:laghr}. As an example, we briefly summarize the calculation of $S~S~\to h^{(i)}r^{(j)}$. As before, there are two regions of interest: the funnel region where the internal spin-2 KK modes or the radion goes on-shell, and the high energy limit where an anomalous growth would naively seem to be present. 
In appendix~\ref{sec:scathr}, we present a sketch of the calculation for the process $S S\to h^{(i)}r^{(j)}$. We find that all anomalously growing terms are canceled out and the residual high-energy terms linear in COM energy-squared\footnote{In the numerical evaluation we evaluate terms to subleading orders  with sufficient numerical precision.} are
\begin{equation}
    \mathcal{M}_0^{\left( 1 \right)}(SS \to h^{(i)}r^{(j)}) = \frac{i \kappa_i \kappa_{\left( j \right)} }{24}  \left( 1 + 3 \cos 2 \theta \right).
\end{equation}
Again, the fact that the leading high-energy behavior of this scattering amplitude can be written purely in terms of the couplings of the modes to brane-localized particles, the $\kappa_i$ and $\kappa_{(j)}$ parameters, is a result of a (new) sum-rule as explained in appendix~\ref{sec:scathr}.

\begin{figure}[t]
    \centering
\includegraphics[width=0.49\linewidth]{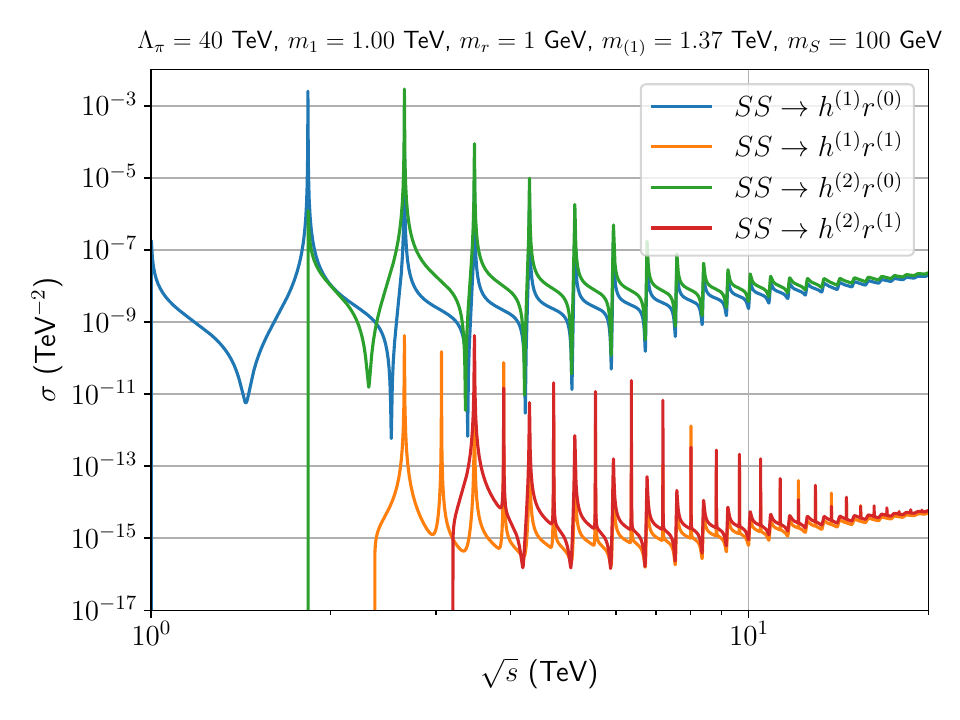}
\includegraphics[width=0.49\linewidth]{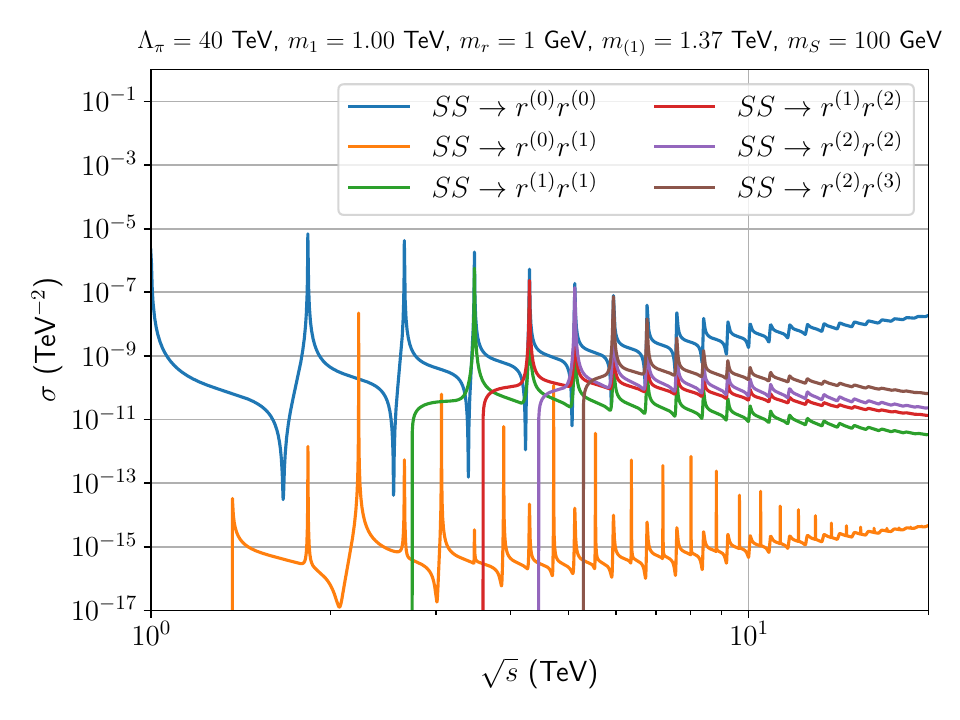}
    \caption{Cross-section for the process $S S \rightarrow h^{\left( i\right)} r^{\left( j \right)}$ (left panel), and $S S \rightarrow r^{\left( i \right)} r^{\left( j \right)}$ (right panel) for the mass of the brane scalar of $0.5$ TeV $m_{r}=1$ GeV. The cross-sections are computed by summing over a truncated KK tower of $25$ internal spin-$0$ and $25$ internal spin-$2$ KK modes.\label{fig:CrossSectsPhiPhiHR}} 
    
\end{figure}

The resulting cross-section for the process in question is obtained numerically by summing over the truncated spin-$0$ and spin-$2$ towers are presented in the left panel of Fig.~\ref{fig:CrossSectsPhiPhiHR}. From Fig.~\ref{fig:CrossSectsPhiPhiHR}, we note that since the coupling constants of higher spin-$0$ modes are several orders of magnitude smaller than those of the zeroth spin-$0$ mode,  the cross-section of processes with higher spin-$0$ modes in the final state is suppressed compared to the processes with the zeroth spin-$0$ mode in the final state. A sketch of the calculation for vector and fermion dark matter candidates is documented in appendix~\ref{sec:scathr}. \looseness=-1

Finally there are processes with two massive radions ($r^{(0)}r^{(0)}$), two GW scalars $r^{(i)}r^{(j)}$, or GW scalar massive radion final states $r^{(0)}r^{(i)}$, as depicted in Fig.~\ref{fig:ProcessInQuestion3}. In the stabilized RS1 model, the three scalar vertex is complicated, and the resulting amplitudes involve many coupling structures. These structures are summarized in appendix~\ref{sec:laghr}.  After a series of complex calculations, we are left with a leading order high-energy amplitude that can be simply written as, 
\begin{equation}
    \mathcal{M}_0^{\left( 1 \right)}(SS \to r^{(i)}r^{(j)}) = - \frac{i \kappa_{\left( i \right)} \kappa_{ \left( j \right)} }{24} \left( 1+ 3   \cos 2 \theta \right). 
    \label{eq:newlabel}
\end{equation}
Again, the high-energy behavior of this amplitude is determined entirely by the couplings $\kappa_{(i)}$ due to a sum rule examined in appendix~\ref{sec:laghr}.

The resulting cross-section for the process in question obtained numerically by summing over the truncated spin-$0$ and spin-$2$ towers are presented in the right panel of Fig.~\ref{fig:CrossSectsPhiPhiHR}. We have set the effective scale $\Lambda_{\pi}=40$ TeV and adjusted the parameters of the stabilizing potential to obtain $m_{1}=1$ TeV, $m_{(1)} \simeq 1.4$ TeV, $m_{r}=1$ GeV for a choice of scalar DM mass of 100 GeV.  We also note that the cross-sections $SS\to h^{\left( i \right)} r^{\left( 0 \right)} $ are enhanced compared to $SS\to h^{\left( i \right)} r^{\left( j \neq 0 \right)}$.  The most intuitive way of thinking about the qualitative difference between the cross-sections for the $h^{\left( i \right)} r^{\left( 0 \right)}$ and $h^{\left( i \right)} r^{\left( j \neq 0 \right)}$, is that any interactions terms between the RS and GW sectors come suppressed by the VEV of the bulk scalar and are suppressed by  $\mathcal{O}(\frac{u}{k})$  \cite{Csaki:2004ay,Chivukula:2023sua}.
As before, 
the calculation with vector and fermion DM initial states are documented in appendix~\ref{sec:laghr}.

Finally, by comparing Figs.~\ref{fig:SigmaSSKKPortal}~and~\ref{fig:CrossSectsPhiPhiHR}, we see that the mixed DM to KK/GW and GW/GW annihilation cross-sections (once all anomalous high-energy growth has been tamed) are significantly smaller than the DM to SM cross sections. Note here that the portion of the amplitude scaling like $s^1$, shown in Eq.~(\ref{eq:newlabel}), is suppressed by the product of small couplings $\kappa_{(i)} \kappa_{(j)}$, and instead contributions which are proportional to $s^0$ dominate until COM energies as large as $\sqrt{s} \simeq 10^{3}$ TeV.


\subsection{Cross Section Summary}

\begin{figure}[t]
\vspace{-0.2cm}
\centering
{\includegraphics[width=0.49\linewidth]{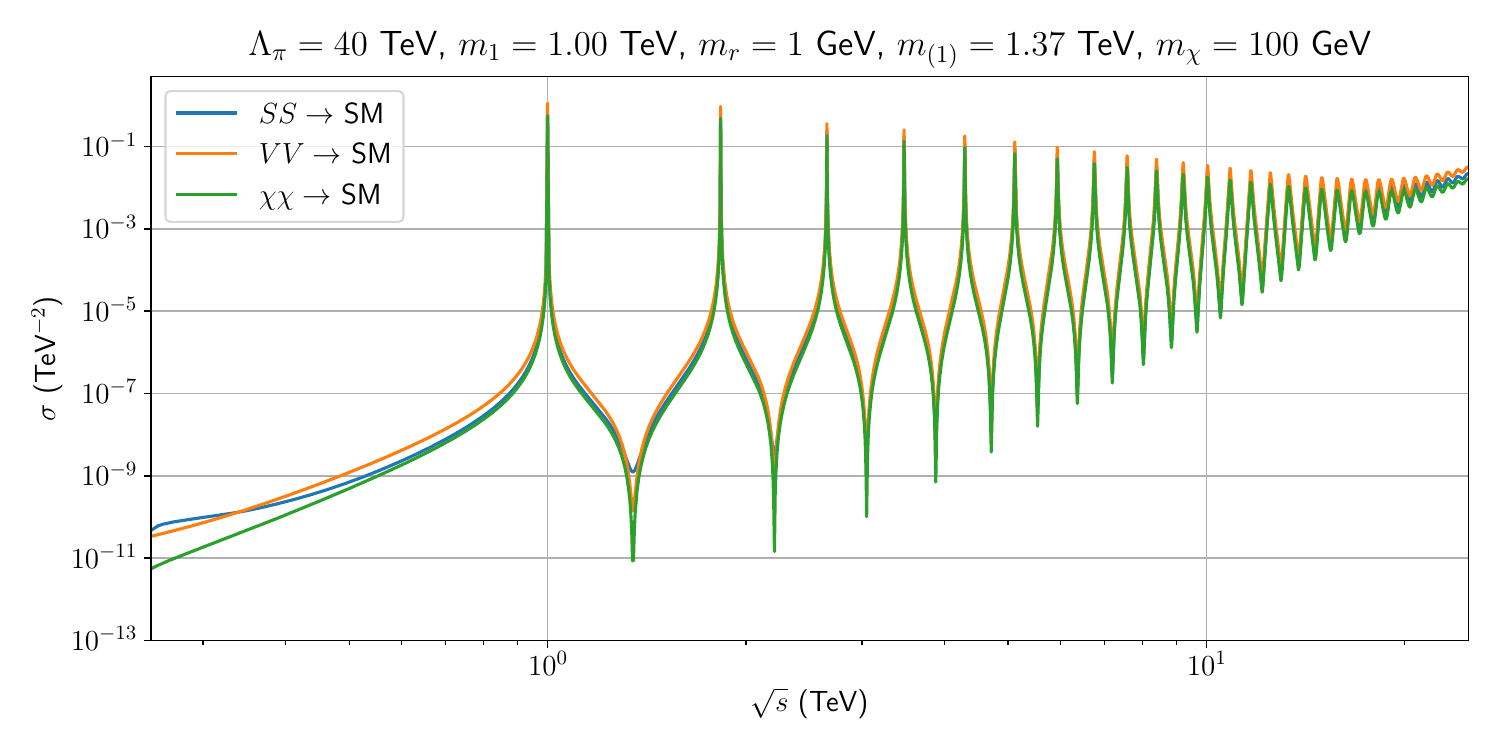}} 
{\includegraphics[width=0.49\linewidth]{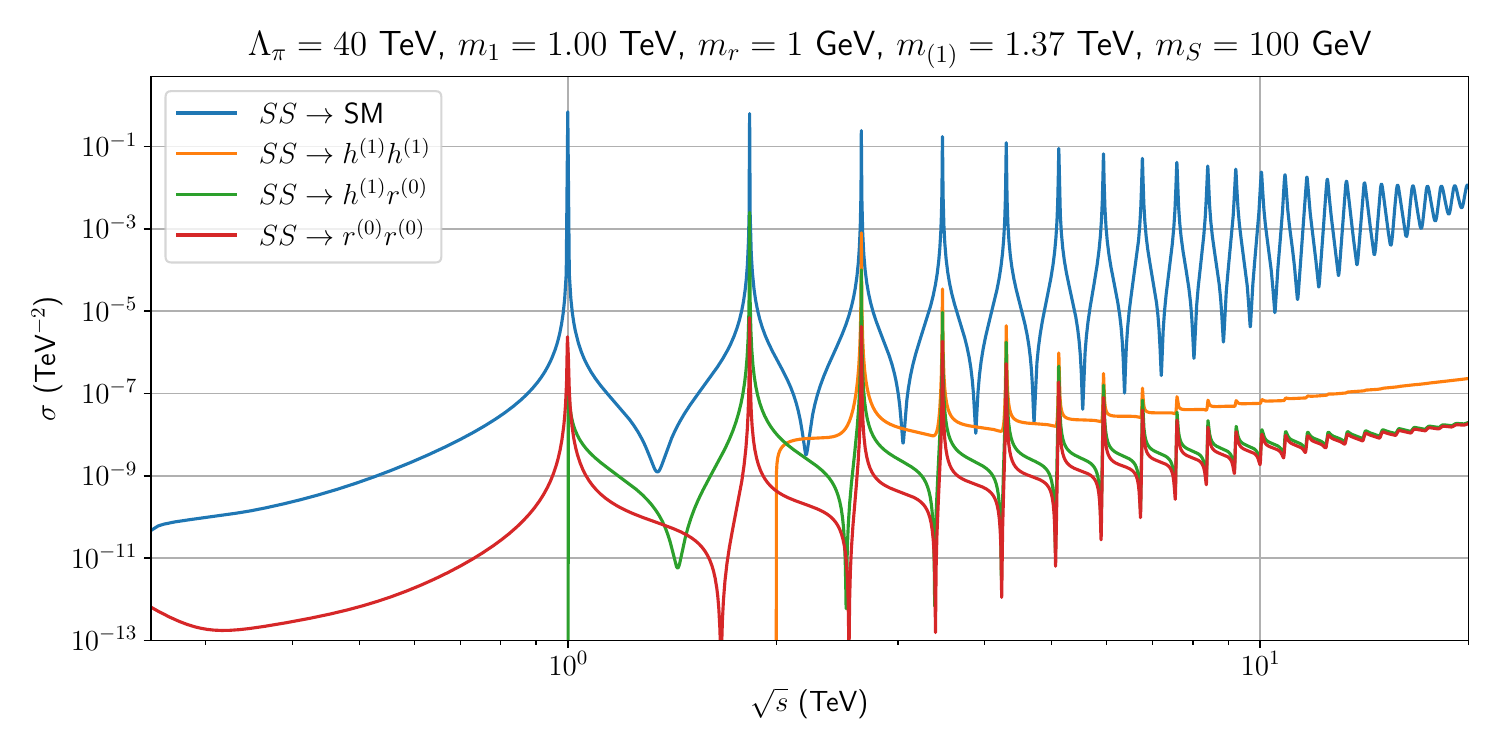}} 
\caption{Contributions to the  SM annihilation cross-section of brane localized matter for each species of dark matter, left panel.  For the case of scalar DM, we compare the different annihilation channels - and illustrate the dominance of the annihilation into SM final states.
\label{fig:TotalrawCrossSections}}
\end{figure}

In the left panel of Fig.~\ref{fig:TotalrawCrossSections}, we present the annihilation cross-sections of brane-localized scalars, vectors and fermions, respectively, to SM final states. While annihilation into SM particles from vector initial states is slightly larger than that from fermion or scalar initial states, the difference is at most a factor of 2-3, even at resonance; all of the cross-sections behave very similarly. On the right panel of Fig.~\ref{fig:TotalrawCrossSections}, we compare the cross-sections for brane-localized scalars annihilating into various final states and note that, as described above, the annihilation cross-section into KK final states is well-behaved and is smaller than that for SM final states. As mentioned previously, neither the light radion nor large $\tilde{k}r_c$ approximations are used in the results presented here, and all contributions (including the SM and spin-2 KK + GW sector) have been considered in the full numerical evaluation.

\section{Velocity Averaged cross-sections and relic density}

\label{sec:relic-density}

In order to compute the relic density of DM particles, we require the thermal velocity averaged cross-section, which can be expressed as~\cite{Dodelson:2003ft,Kolb:1990vq}\footnote{In the natural units where the Boltzmann constant $k_B$ is set to unity.},
\begin{equation} 
 \braket{ \sigma_{\Phi \Phi \rightarrow f f } v} = \frac{ 2 \pi^2 T \int_{4 m_\Phi^2}^\infty  ds \sqrt{s} \left( s -  4 m_\Phi^2 \right) K_1 \left( \frac{\sqrt{s}}{T} \right) \sigma_{\Phi \Phi \rightarrow f f} (s)}{\left(   4 \pi m_\Phi^2 T K_2 \left(\frac{m_\Phi}{T} \right) \right)^2}. \label{eqn:sigmaVfinaleqn}
\end{equation}
In the above expression, $K_{1}$, $K_{2}$ are, respectively, the first and second modified Bessel functions of the second kind, $T$ is the temperature of the thermal bath, and $f$ represents the particles that the DM particle $\Phi$ annihilates to, i.e., $S,V,\chi,h_{i},r_{j}$. Assuming standard thermal WIMP freeze-out relic abundance, with a typical freeze-out temperature of order $m_\Phi/20$, we estimate that a velocity-averaged cross-section of $\langle \sigma v_{\rm rel} \rangle\simeq 10^{-26}\, {\rm cm}^{3} / {\rm s} $  can provide for the observed Planck inferred relic density of the Universe~\cite{Planck:2018vyg}. 

\begin{figure}[htb]
\vspace{-0.9cm}
\centering
{\includegraphics[width=0.49\linewidth]{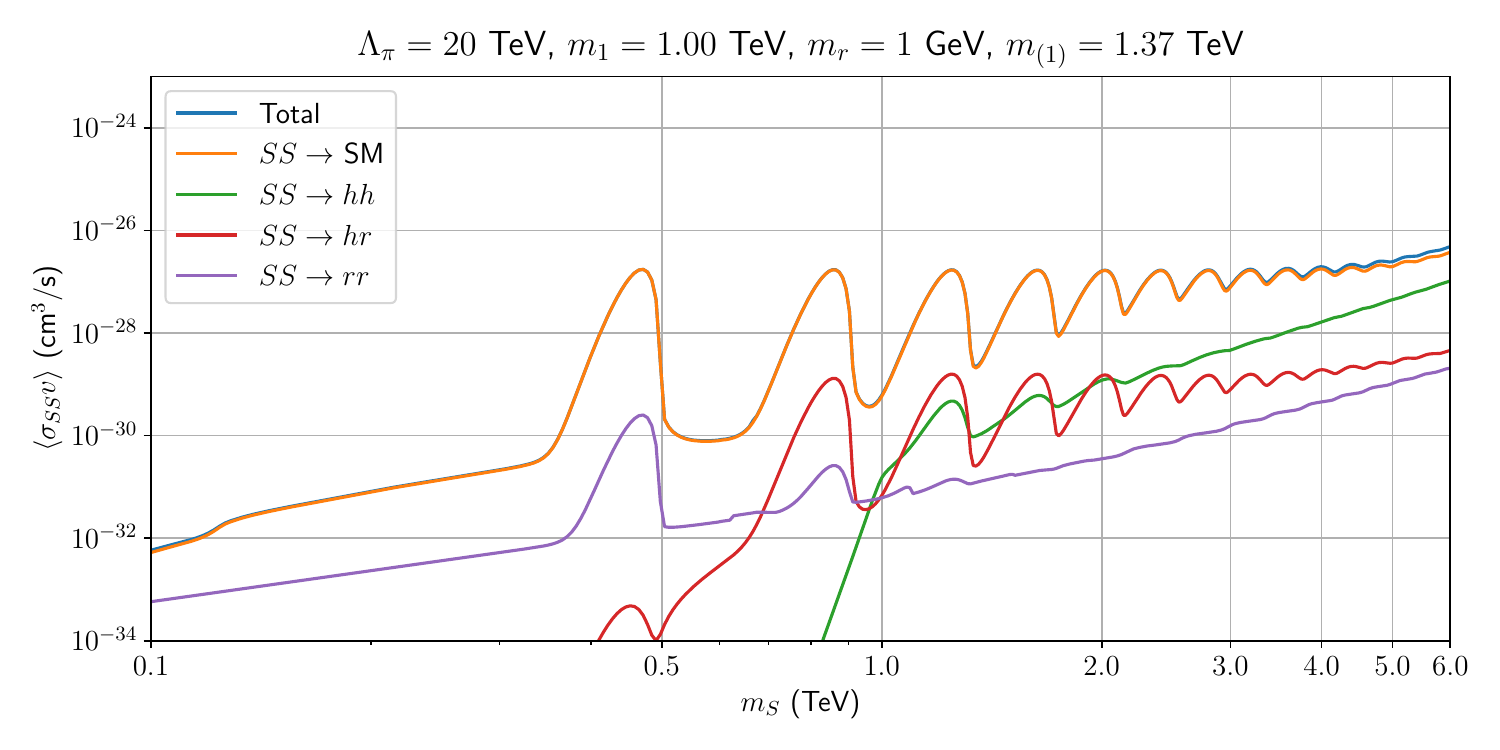}} 
{\includegraphics[width=0.49\linewidth]{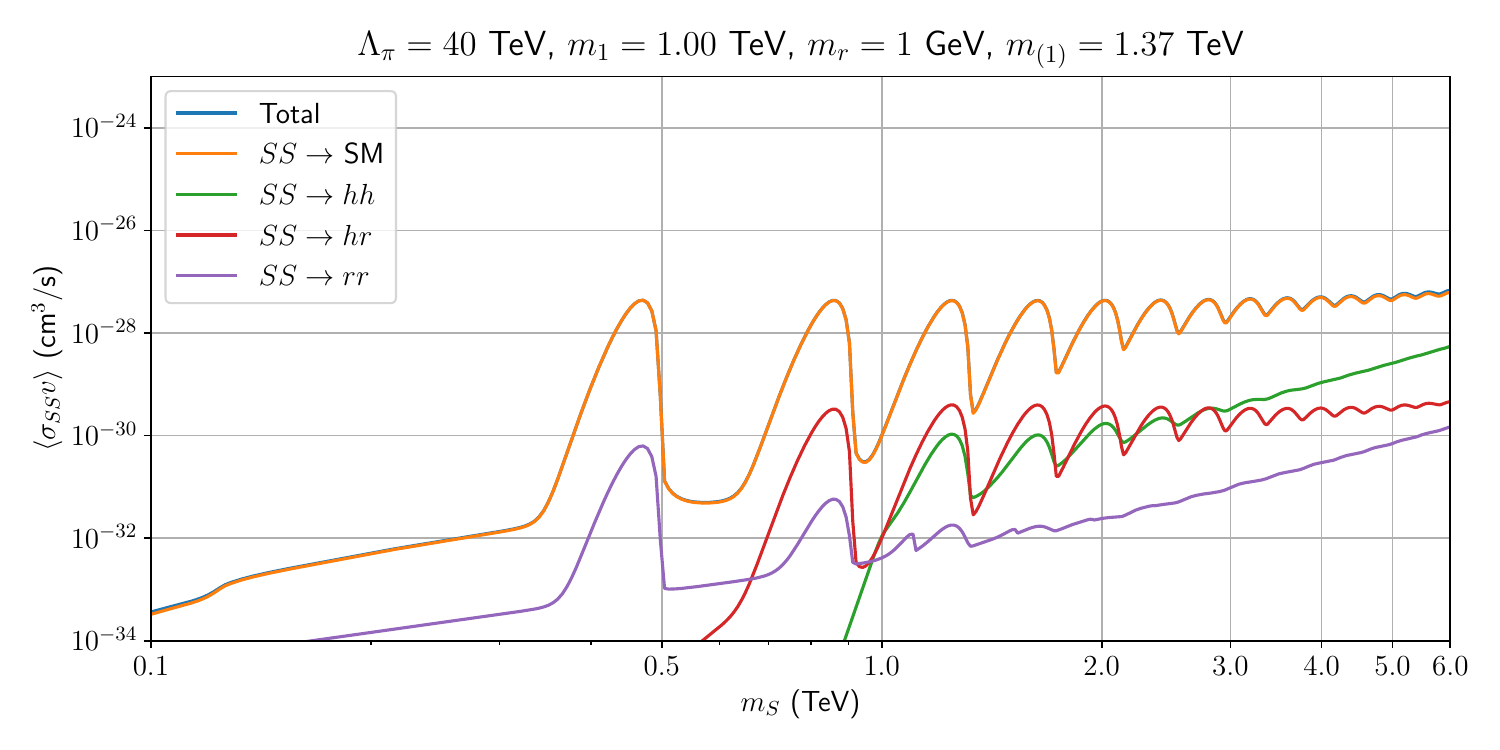}} \\
{\includegraphics[width=0.49\linewidth]{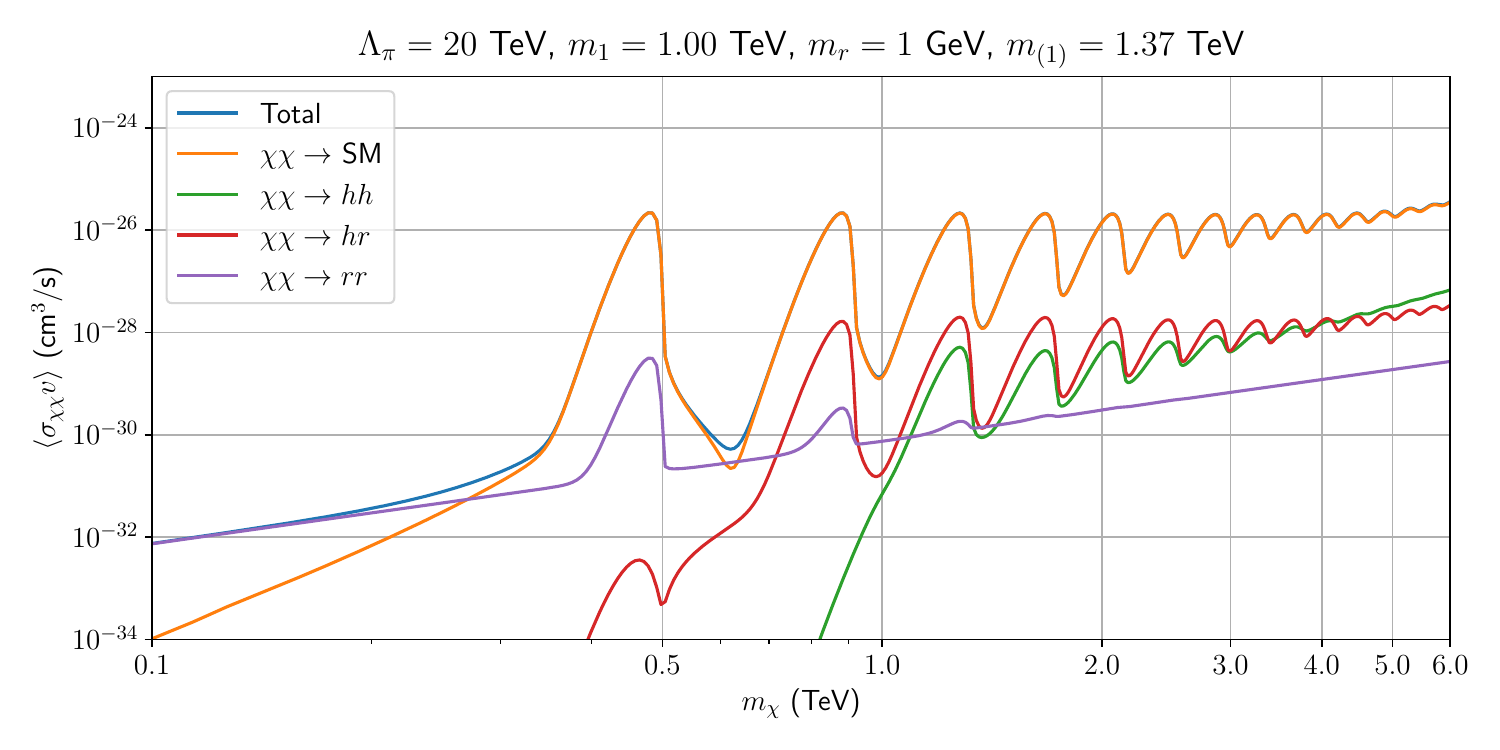}}
{\includegraphics[width=0.49\linewidth]{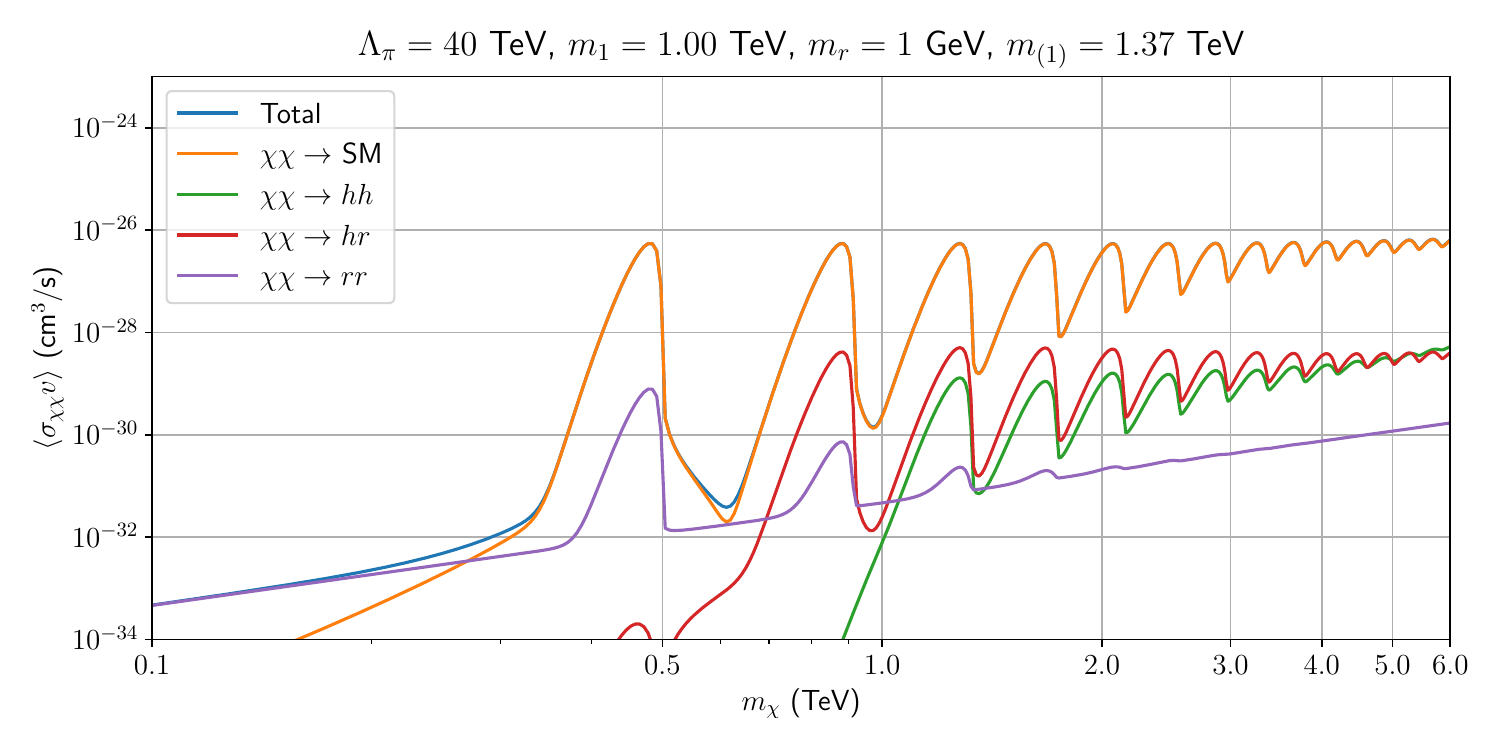}} \\
{\includegraphics[width=0.49\linewidth]{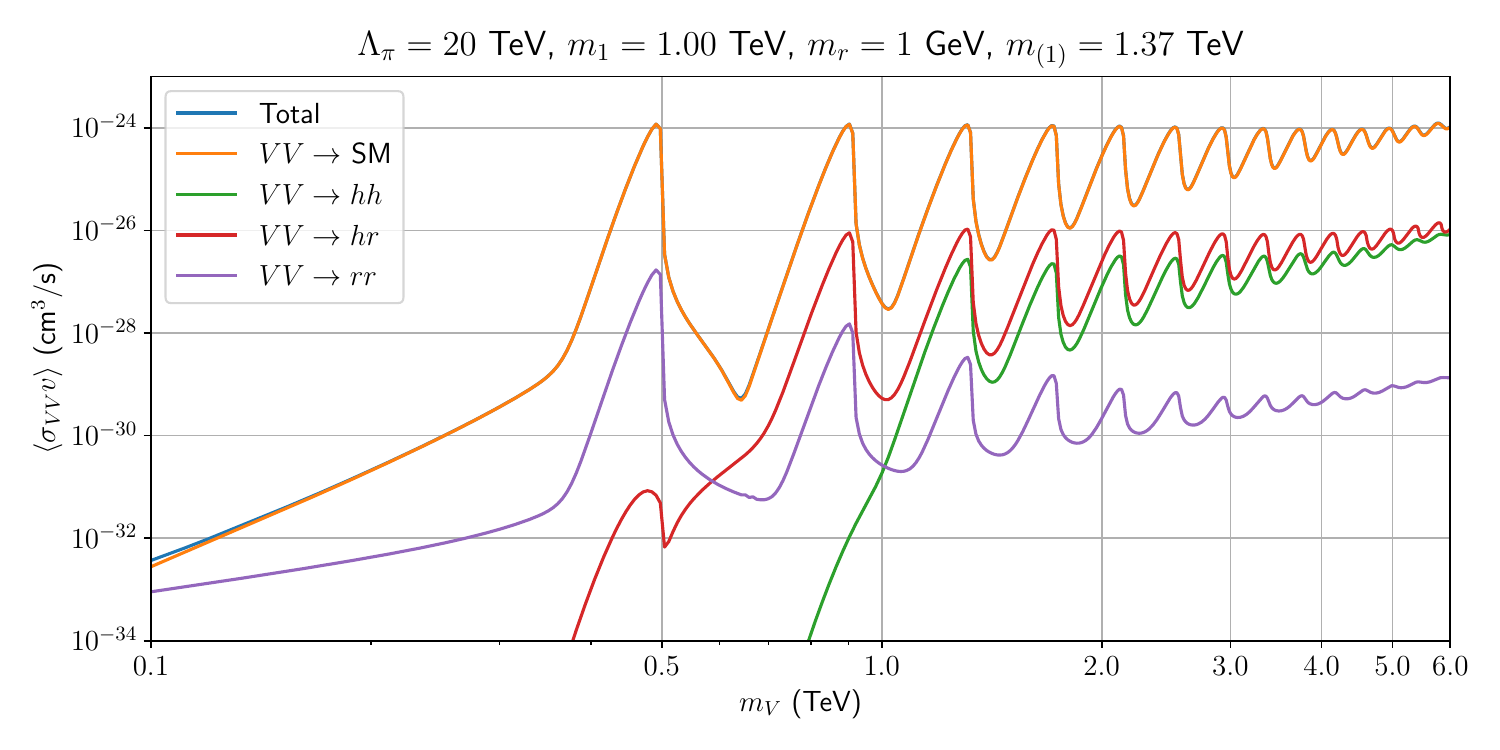}} 
{\includegraphics[width=0.49\linewidth]{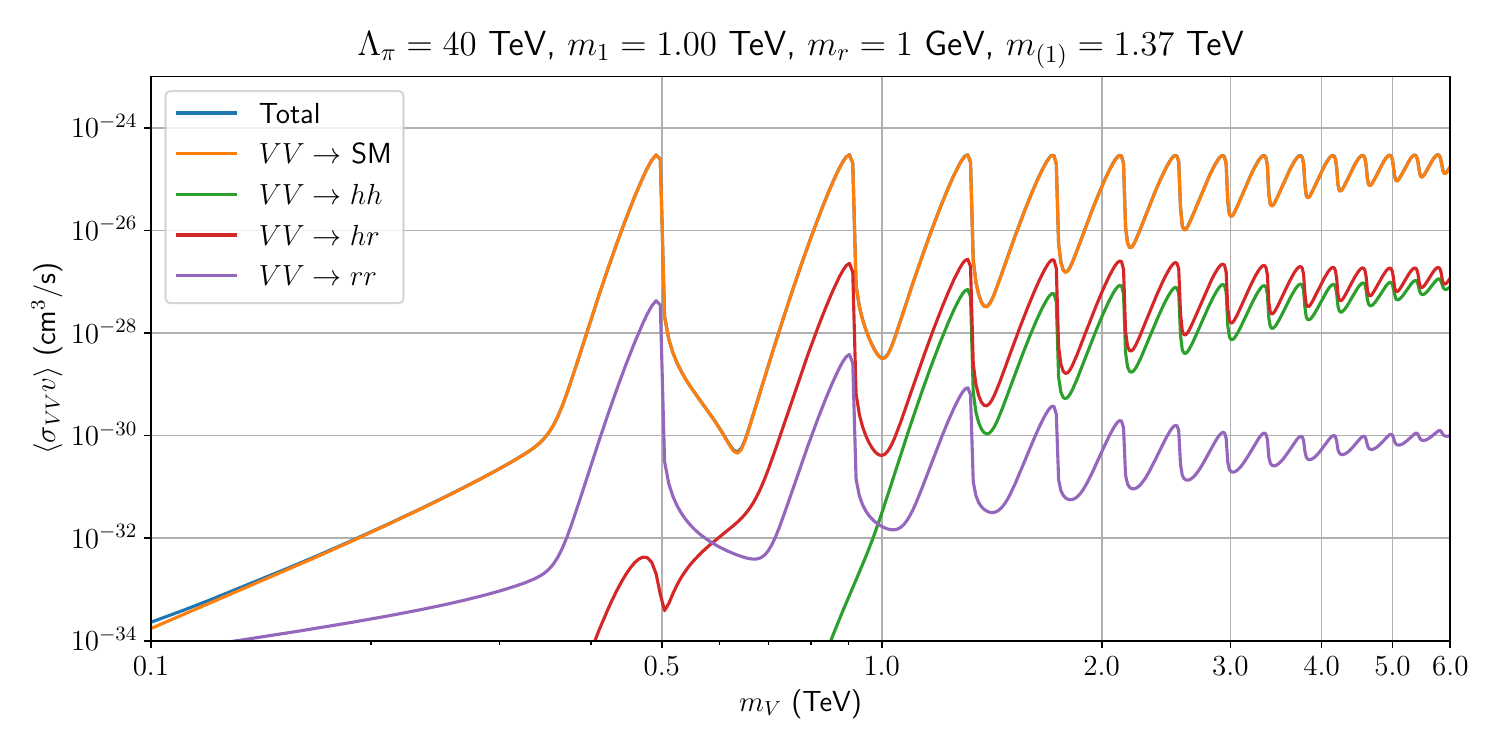}} \\
{\includegraphics[width=0.49\linewidth]{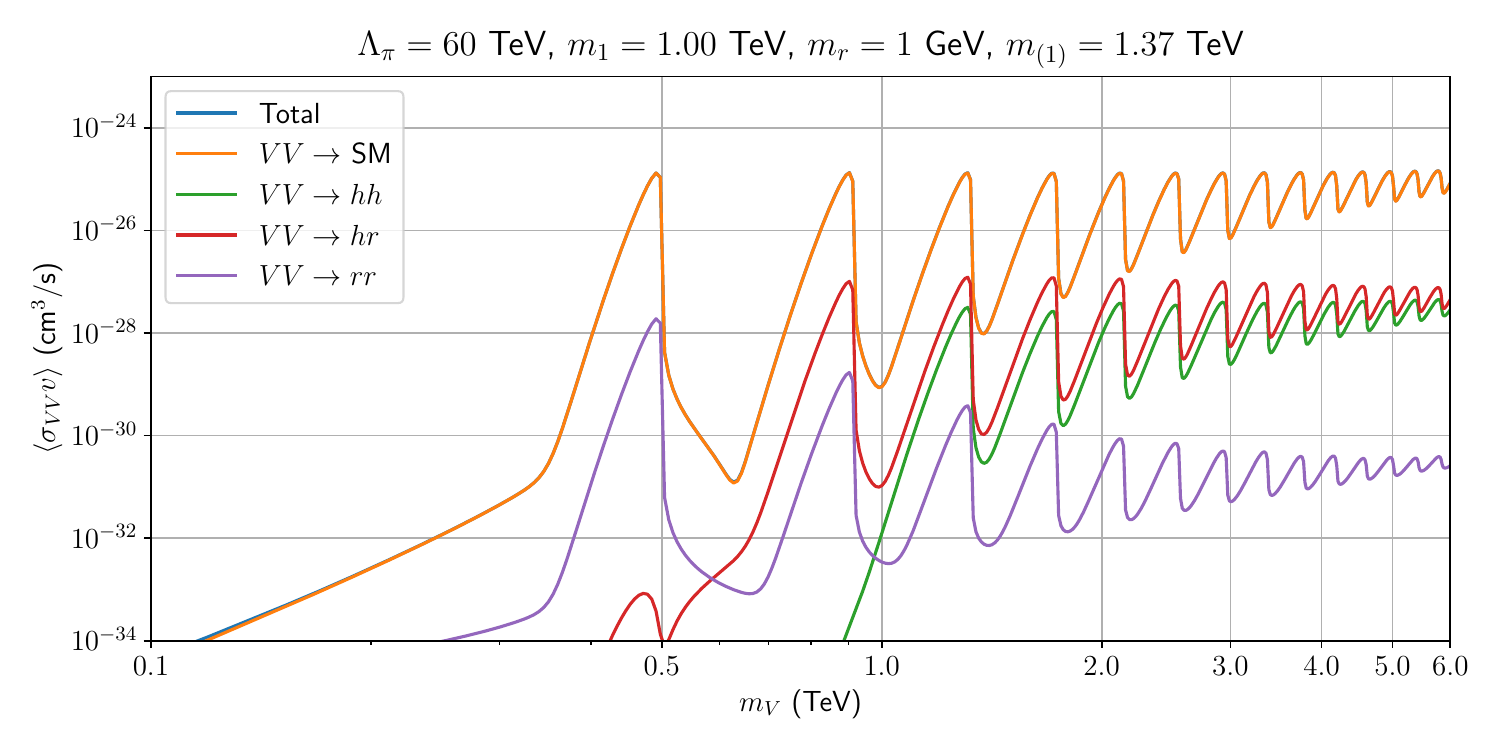}} 
{\includegraphics[width=0.49\linewidth]{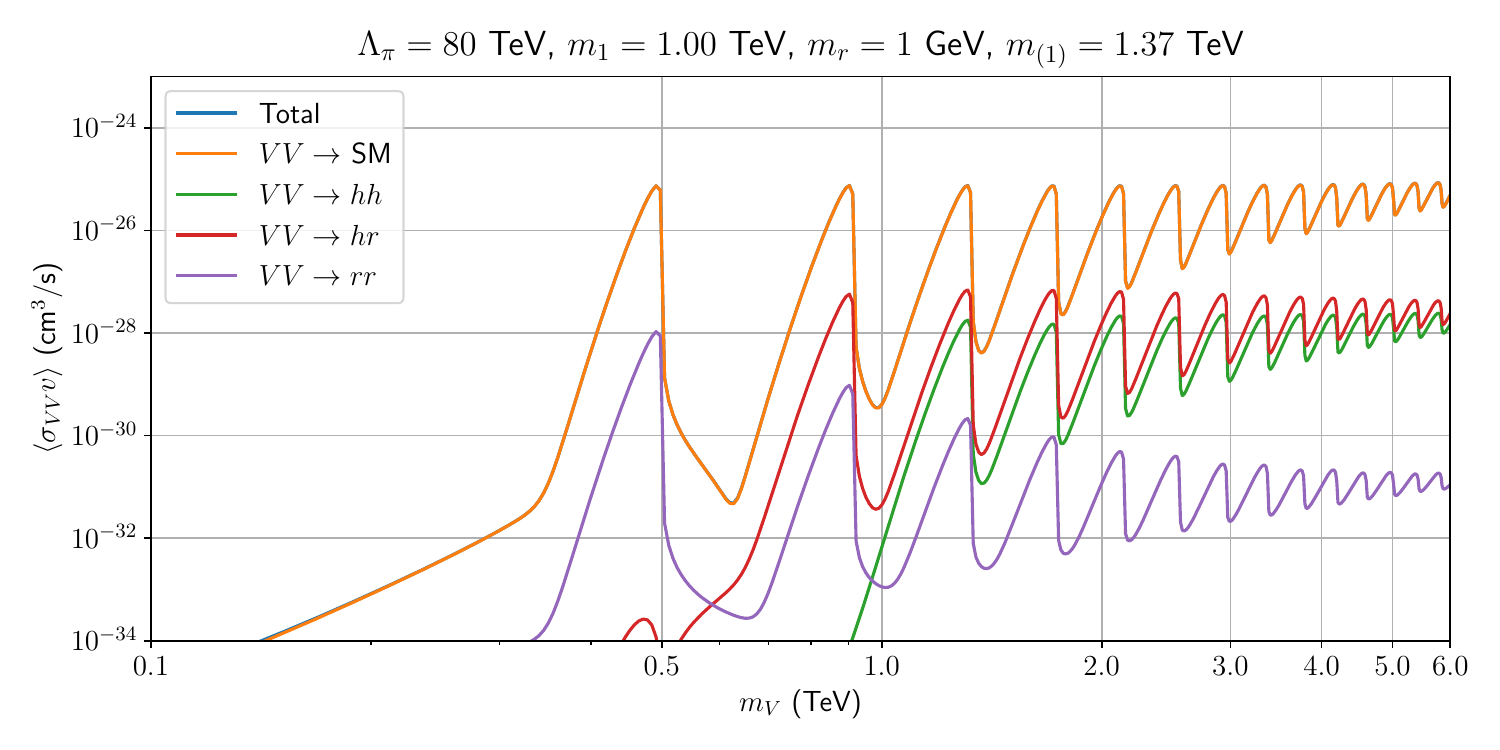}} 
\caption{ Contributions to the total velocity averaged annihilation cross-section of brane localized matter from each of the channels studied here; note that the total cross-section largely overlaps that for SM final states in most panels. 
The first row shows the velocity averaged cross-section for scalar dark matter for $\rm \Lambda_{\pi}=20$ and 40 TeV, respectively, while the second and third row of panels do the same for fermion and vector DM, respectively, and the fourth row for vector DM candidates and $\Lambda_{\pi}=60$ and 80 TeV. As explained in the text, a thermal annihilation cross-section $\langle\sigma v\rangle\approx  10^{-26}$ cm$^3$/s (corresponding to $x=\frac{m_{\Phi}}{T}\simeq 20$) is required to obtain the DM abundance observed in the universe today. This cross-section is never reached for scalar DM candidates, while it is for fermions and vectors in the case of (near) resonant DM annihilation.
\label{fig:TotalCrossSections}}
\end{figure}

In separate panels of Fig.~\ref{fig:TotalCrossSections}, we plot the total velocity averaged cross-section (blue curve) resulting from the annihilation of scalar, vector, and fermion dark matter candidates as a function of the mass of the dark matter species.  We also show the contributions to the total from various final states; note that the total (blue) curve tends to overlap the (orange) curve for SM final states in most panels.  
In the plots in Fig.~\ref{fig:TotalCrossSections}, we take representative values of the model parameters, specifically, we fix $m_{1}\simeq 1$ TeV, and $m_{r}=1$ GeV (for which we obtain $m_{(1)}\simeq 1.4$ TeV) for several choices of $\Lambda_{\pi}$: 20 TeV and 40 TeV for all dark matter species, and also 60 TeV and 80 TeV for vector dark matter.

One may observe that, in all cases shown, the resonant peak heights of the velocity-averaged cross-sections fall as $\Lambda_\pi$ increases. This dependence is phenomenologically important since it controls the region of $\Lambda_\pi$ that yields an appropriate dark matter relic abundance. This dependence on $\Lambda_\pi$, the interaction strength of the KK graviton,  is counterintuitive since resonant amplitudes saturate the unitarity bound on the peak. The corresponding particle cross-sections on the peak would be strictly independent of the total decay width (depending only on the mass of the resonance and the branching ratios for the incoming and outgoing states) in the zero-width approximation. However, this behavior is modified for the velocity-averaged cross-sections.

We illustrate the origin of this dependency of the annihilation cross-section on $\Lambda_\pi$ for the contribution to the annihilation cross-section of a brane localized vector boson to Higgs bosons via the first spin-$2$ mode with mass $m_{1}$. Near the peak, the corresponding matrix element is of the Breit-Wigner form,
\begin{equation}
    \mathcal{M} \simeq \frac{i}{24} \frac{\kappa_1^2 s \left( s + 4 m_V^2 \right)}{s - m_1^2 + i m_1 \Gamma_{h^{\left( 1 \right)}}} \left( 1 + 3 \cos 2 \theta \right),
\end{equation}
where the decay width of the $1$st KK mode can be approximated as $\Gamma_{h^{\left( 1 \right)}} \simeq \alpha m_1^3/\Lambda^2_\pi$ for some numerical prefactor $\alpha$. 
Next, we expand the velocity-averaged cross-section~\cite{Cannoni2016}
\begin{align}
    \braket{\sigma v} = a + b v^2 + c v^4 + \ldots~,
\end{align}
in terms of the relative thermal velocity of dark matter particles
\begin{align}
    v & = \sqrt{\dfrac{16 T}{\pi m_{V}}}~.
\end{align}
Doing so on annihilation resonance peak, evaluated at the dark matter mass fixed to half of the first KK mode ($m_V = m_1/2$), in the relative thermal velocity up to $4$th order, we get
\begin{equation}
    \braket{\sigma v} = \frac{\mathcal{A}{\left( v \right)}}{m_1^2} - \frac{\mathcal{B} v^4}{m_1^2} \frac{\Lambda_\pi^4}{m_1^4} + \ldots,
\end{equation}
where $\mathcal{A}{\left( v \right)}$ is a positive $4$th degree polynomial in $v$ corresponding to dominantly the {\it{s}} and {\it{p}} wave contribution,   and $\mathcal{B}$ is a positive numerical coefficient corresponding to the  {\it{d}} wave coefficient. Note that despite the $\Lambda_\pi$ dependent contribution being suppressed by $v^4$, it is still sizable as it is also enhanced by $\Lambda_\pi^4 / m_1^4$. For the typical freeze-out temperatures of $T = m_{\Phi}/ 20$, the DM species' average relative thermal velocity is $v \simeq 1/2$. However, for the set of model parameters considered in this work, $\Lambda_\pi / m_1 = \order{10}$, leading to an observable suppression in the annihilation cross-section for increasing $\Lambda_\pi$. We find the same pattern repeated for fermions and scalars. For fermions, where the leading term is {\it{p}} wave, the $\Lambda_\pi$  dependence in 
$ \braket{\sigma v}$ starts at the $v^{6}$ co-efficient, while for scalars it starts at $v^{8}$.\footnote{Note that, because of a discontinuity in the cross-section at the threshold, when $v\to 0$, this expansion breaks down for extremely narrow resonances at very large values of $\Lambda_\pi/m_1$ well beyond the values considered here.}
This analysis explains why $ \braket{\sigma v}$ decreases as a function of $\Lambda_{\pi}$. In our full numerical evaluation, we calculate every possible annihilation channel described in the paper. However, the dominant contribution to the cross-section originates from SM annihilation, which has this behavior.\looseness=-1

In Fig.~\ref{fig:TotalCrossSections}, 
we have already noted that the blue and orange curves almost overlap, indicating that $\langle \sigma v_{\rm rel} \rangle$ is dominated by the SM contribution. We next note that due to the {\it{`$s$-wave'}} annihilation\footnote{In the non-relativistic limit, we can expand $\langle \sigma v\rangle$ as $\langle \sigma v\rangle\simeq a + b~v^{2}~+~ c~v^{4} +\cdots$. The pieces with the coefficients $a,~b,~c$ are known as {\it{`$s$-wave'}},{\it{`$p$-wave'}},{\it{`$d$-wave'}} contribution respectively. } of vector dark matter to SM final states through spin-2 KK portal annihilation to SM particles near threshold, they easily satisfy the relic density with $\langle \sigma v_{\rm rel} \rangle \geq 10^{-26}\, {\rm cm}^{3} / {\rm s} $ at resonances of massive KK modes, i.e when $2m_{\Phi}\simeq m_{KK}$. In the bottom two rows of Fig.~\ref{fig:TotalCrossSections}, we present the plots for a vector DM candidate and show that the observed relic density can be satisfied for the vector DM candidate on resonance. \looseness=-1

The fermions and scalars are suppressed due to a {\it{`$p$-wave'}} and {\it{`$d$-wave'}} annihilation, respectively, for the dominant channel.  In general, we find that fermion DM can saturate DM relic abundance up to $\Lambda_{\pi}\leq 25$ TeV for resonant DM annihilation through KK gravitons.
For scalars, for the parameter choices in Fig.~\ref{fig:TotalCrossSections}, the cross-section is too small even on resonance. We can attempt to reduce the effective scale $\Lambda_{\pi}$ to ensure that the scalar DM candidate satisfies the relic abundance. However, this will be ruled out by collider constraints on the spin-2 KK modes. \looseness=-1

Finally, note that the contributions to the annihilation cross-section from the gravitational scalar modes (radion and GW scalars) are negligible in the light radion region investigated in this paper. The situation is quite different for a heavy radion, as shown in~\cite{Chivukula:2024}.

\section{Collider constraints}

\label{sec:ColliderConstraints}

\begin{figure}[t]
\vspace{-0.5cm}
\centering
{\includegraphics[width=0.65\linewidth]{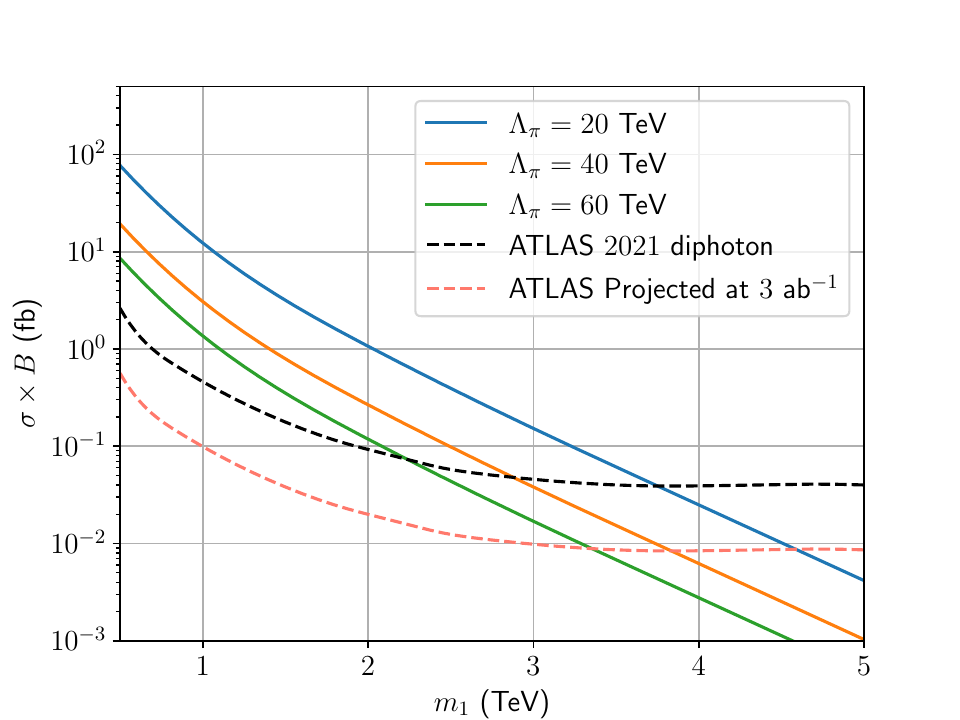}} 
\caption{Production cross section times branching ratio and corresponding experimental upper limit on the lightest KK graviton state from the process $pp\to h^{ ( 1 )}\to \gamma \gamma$ from~\cite{ATLAS:2021uiz}. The black dotted line represents the current constraints at $\rm 139~fb^{-1}$ luminosity, while the red dotted line represents the projected limits at 3000 $\rm fb^{-1}$ integrated luminosity.}
\label{fig:LHCconstraints}
\end{figure}

Next, we consider limits on the masses of KK-gravitons in the RS model at high-energy colliders for $\Lambda_\pi$ in the 20-60 TeV range.
KK gravitons can be produced at hadron colliders like LHC and detected via their decay to SM final states. Gluon-initiated states dominate the production, while the most sensitive channels for detection are high $p_{T}$ dijet and diphoton resonance searches. 
ATLAS and CMS experiments have searched 
for high mass resonances in both channels mentioned above. 
The strongest current constraints on KK gravitons come from diphoton final states using the full run-2 data of $139~\rm fb^{-1}$ at 13 TeV LHC energy~\cite{ATLAS:2021uiz}.  In this case, therefore, we simply re-interpret the existing $\sigma\times {\rm BR}$ limits in terms of the scale $\Lambda_{\pi}$ and the mass of the first (lightest) KK graviton. To this end, we implement the RS model in Feynrules~\cite{Alloul:2013bka}, followed by a cross-section computation using Madgraph 5~\cite{Alwall:2014hca} at 13 TeV centre of mass energy with CTEQ6L parton distribution functions~\cite{Dulat:2015mca}. The factorization and renormalization scales were chosen to be $\mu_{f}=\mu_{r}=\frac{H_{T}}{2}$, where $H_{T}$ is the sum of the transverse momentum of all final state particles.   

In Fig.~\ref{fig:LHCconstraints} we present the production cross-section ($\sigma$) multiplied by the 
branching ratio to diphoton final states as a function of the mass of the lightest spin-2 KK mode for various choices of the effective RS scale $\Lambda_{\pi}$. We also plot the 95\% exclusion for spin-2 particle production from the ATLAS diphoton search~\cite{ATLAS:2021uiz}. We observe that the lightest KK graviton masses of $m_{1}\simeq3.7,2.8, 2.3$ TeV  are already excluded for $\Lambda_{\pi}= 20, 40, 60$ TeV, respectively. We also provide a simple luminosity scaling to indicate the reach to constrain $m_{1}$ at  HL-LHC with $\rm 3000~fb^{-1}$ luminosity, for which we observe that the lightest KK graviton masses of $m_{1}\simeq 4.6,3.8, 3.3$ TeV could potentially be excluded by HL-LHC for $\Lambda_{\pi}= 20, 40, 60$ TeV, respectively.

Finally, we note that the LHC searches also constrain radion to diphoton final states. However, we find that for our parameter choices of $\Lambda_{\pi}\geq 20$ TeV, the $\rm \sigma\times BR$ limits for the light radion considered here are less constraining than the corresponding KK graviton bounds discussed above.

\section{Direct Detection via Radion Exchange}

\label{sec:DDconstraints}


\begin{figure}[t]
    \centering
    \includegraphics[width=.4\linewidth]{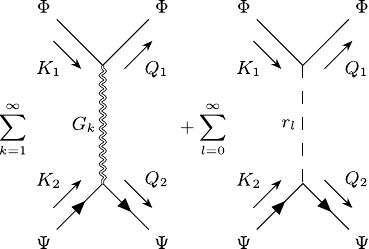}
\caption{Diagrams for the brane matter ($\Phi = \left( S, V, \chi \right)$) scattering off the brane fermion ($\Psi$) via the exchange of the KK mode. We ignore purely gravitational interactions as they are Planck-suppressed. \label{fig:DDdiags}}
\end{figure}


\begin{figure}[t]
\centering
{\includegraphics[width=0.49\linewidth]{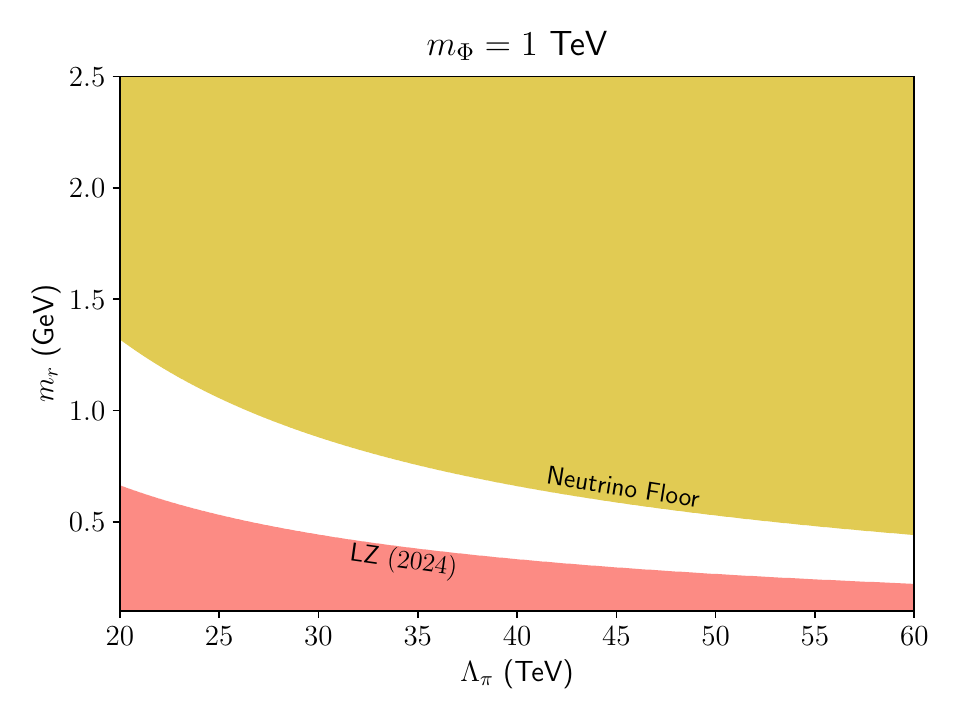}} 
{\includegraphics[width=0.49\linewidth]{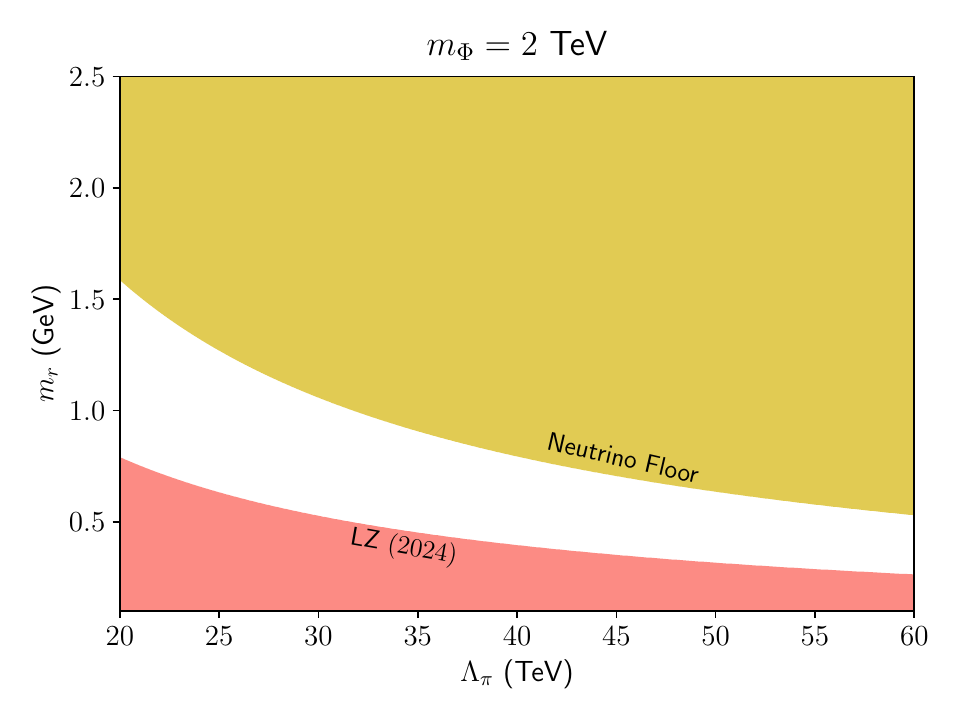}} \\
{\includegraphics[width=0.49\linewidth]{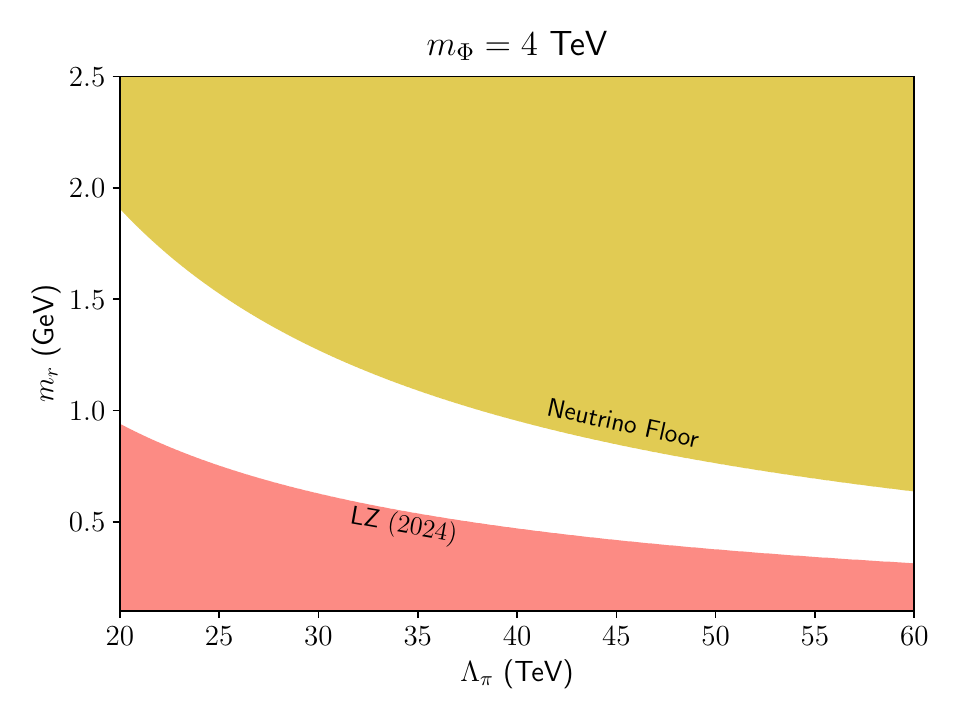}} 
{\includegraphics[width=0.49\linewidth]{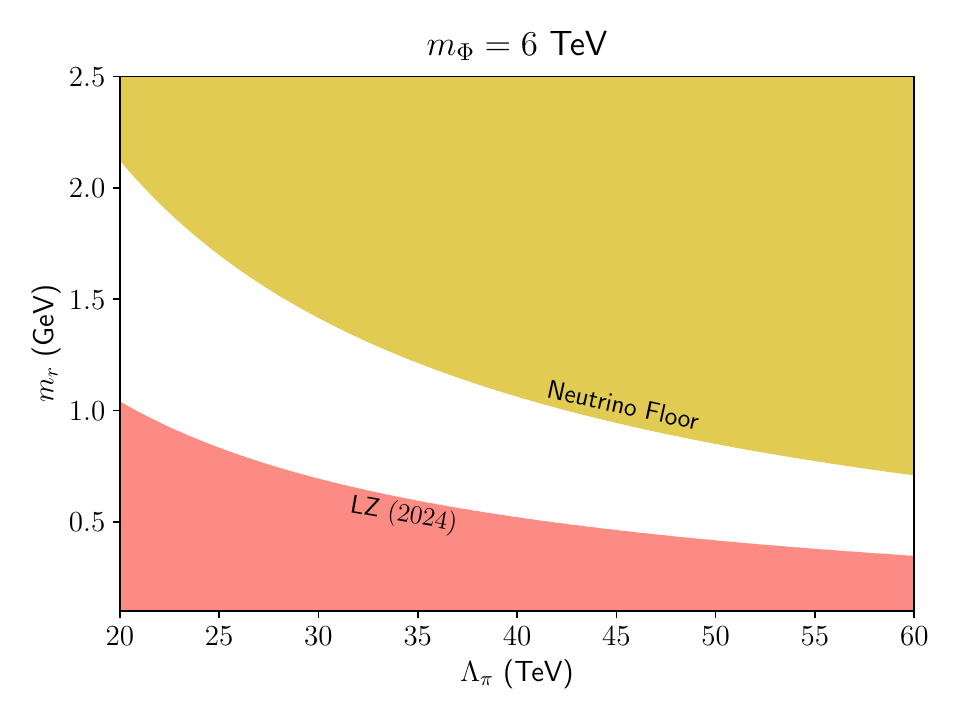}}
\caption{The direct detection constraints on the model corresponding to a variety of dark matter masses $m_{\Phi}=1-6$ TeV ($\Phi=S,V, \chi$) as a function of the radion mass which, in the light-radion limit, dominates the detection cross-section. The yellow band corresponds to the neutrino floor, where the direct detection cross-section for the model is unlikely to be tested by the current direct detection experiments. The red regions in the plots indicate the current constraints from the LZ experiment~\cite{LZCollaboration:2024lux}. }
\label{fig:dd}
\end{figure}

Having assessed the relic density and collider constraints, we next discuss the direct detection constraints on KK portal dark matter models, and show that they are dominated by radion-exchange for the model regions considered here. 

In the non-relativistic limit, the scattering of DM by a heavy nucleus via the gravity sector can be described by two components.  There is a spin-independent (SI) coupling to the energy-momentum tensor, which at low momentum transfer is proportional to the dark matter mass; this resolves the entire nucleus coherently, leading to an enhanced cross-section proportional to the square of the number of scattering centers
(nucleons). There is also a spin-dependent (SD) term, which couples to the nucleon spin, typically without any coherent enhancement. In our case, the SI term is significantly more constraining than the SD part - and the SI interaction is the same whether the DM is scalar, fermion, or vector.

We follow the standard prescription for direct detection as described in~\cite{Drees:1993bu,Hisano:2010yh,Hill:2011be,Hill:2014yxa}.
As SM fermions in Fig.~\ref{fig:DDdiags} correspond to quarks inside the nucleus, we need to consider their interaction vertices of the form  
\begin{align}
\vcenter{\hbox{\includegraphics[height=2.13cm]{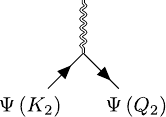}}} \hspace{0.07cm} &~\begin{gathered} \propto~V_2^{\mu \nu} \equiv - \frac{1}{4} \bar{u}_\Psi \left( Q_2 \right) \left[ \gamma^\mu \left( K_2^\nu + Q_2^\nu \right) + \gamma^\nu \left( K_2^\mu + Q_2^\mu \right)  \right.  \\
\hspace{0.4cm} \left. - 2 \eta^{\mu \nu} \left( \slashed{K}_2 + \slashed{Q}_2 - 2 m_\Psi \right) \right] u_\Psi \left( K_2 \right) ,  \end{gathered}\label{eqn:Spin2VertexNucleon}\\
\vcenter{\hbox{\includegraphics[height=2.13cm]{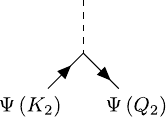}}} \hspace{0.07cm} &~\begin{gathered}\propto V_0 \equiv - \sqrt{\frac{2}{3}} \bar{u}_\Psi \left( Q_2 \right) \left[  \frac{3}{4} \left( \slashed{Q}_2 + \slashed{K}_2 \right) - 2 m_\Psi\right] u_\Psi \left( K_2 \right), \end{gathered} \label{eqn:Spin0VertexNucleon}
\end{align}
which we need to express in terms of the corresponding twist-$2$ and twist-$0$ nuclear operators. Starting with Eq.~(\ref{eqn:Spin2VertexNucleon}), we can separate it into the traceless and scalar parts~\cite{Rueter:2017nbk} as
\begin{equation}
    V_2^{\mu \nu} = \left( V_2^{\mu \nu} - \frac{1}{4} \eta^{\mu \nu} V_2 \right) + \frac{1}{2} \eta^{\mu \nu} V_2 = \tilde{T}_\Psi^{\mu \nu} + \frac{1}{4} \eta^{\mu \nu} T_\Psi, \label{eqn:Spin2VertexDecomp}
\end{equation}
where we have identifed the trace-less part of $V_2^{\mu \nu}$ with the twist-$2$ operator and the scalar part with the twist-$0$ operator. Similarly, for Eq.~(\ref{eqn:Spin0VertexNucleon}),
\begin{equation}
    V_0 = - \frac{1}{2} \sqrt{\frac{2}{3}} T_\Psi. \label{eqn:Spin0VertexDecomp} 
\end{equation}
As the ``dark-mater wind'' velocity is non-relativistic, we can work in the zero momentum transfer approximation where $Q_2 \approx K_2$. \looseness=-1

Using the above information, the resulting non-zero contribution to the $S$-matrix element can be approximated as
\begin{equation}
    \mathcal{M}_{\pm,\pm}  = - i \frac{ m_N^2 m_\Phi^2}{3} \left[ \zeta_\Psi^N \sum_{k=1}^\infty \frac{\kappa_k^2}{m_k^2}  - 2 f_{T_\Psi}^N  \sum_{n=0}^\infty \frac{ \kappa_{\left( n \right)}^2}{m_{\left( n \right)}^2} \right], \label{eqn:MFull}
\end{equation}
where subscripts of the matrix elements denote the spins of the quark $\Psi$, and $\zeta_\Psi^N$ and $f_{T_\Psi}^N$ are  $\order{1}$ and $\order{0.01}$ coefficients representing the valence quarks, and the mass fractions of the quark $\Psi$ in the nucleon $N$  of the nucleus. To simplify the above result further, we recall that, in the light radion regime $m_r \simeq \order{1~\text{GeV}}$, $m_1 \simeq \order{1~\text{TeV}}$, and $\kappa_{\left( 0 \right)} = \kappa_1 = 1 / \Lambda_\pi$. Thus, as $f_{T_\Psi}^N$ is a numerical quantity of $\order{0.01}$ and $\zeta_\Psi^N$ is a numerical quantity of $\order{1}$~\cite{Hisano:2010yh}, we can observe that $\zeta_\Psi^N / m_1^2 \simeq \order{1~\text{TeV}^{-2}}$ and $f_{T_\Psi}^N / m_r^2 \simeq \order{0.01~\text{GeV}^{-2}} = \order{10^4~\text{TeV}^{-2}}$. Therefore, we can approximate Eq.~(\ref{eqn:MFull}) as 
\begin{equation}
 \mathcal{M}_{\pm,\pm}  \simeq -  \frac{2 m_N^2 m_\Phi^2 f_{T_\Psi}^N  }{3}   \sum_{n=0}^\infty \frac{ \kappa_{\left( n \right)}^2}{m_{\left( n \right)}^2}.
\end{equation}
Furthermore, given that for all $n > 0$ we have $\kappa_{\left( n \right)} \ll \kappa_{\left( 0 \right)}$ and $m_{\left( n \right)} \gg m_r$, we conclude that the dominant term in the above sum arises from the first term. Hence, our approximation of Eq.~(\ref{eqn:MFull}) is simply
\begin{equation}
 \mathcal{M}_{\pm,\pm}  \simeq -  \frac{2  f_{T_\Psi}^N m_N^2 m_\Phi^2  }{3}   \frac{ \kappa_{\left( 0 \right)}^2}{m_{r}^2} = -  \frac{2  f_{T_\Psi}^N m_N^2 m_\Phi^2 f_{T_\Psi}^N  }{3 \Lambda_\pi^2 m_r^2 } .
\end{equation}

Defining the convenient shorthand
\begin{equation}
    \bar{\mathcal{M}}^N_\Psi \equiv - \frac{2 f_{T_\Psi}^N m_\Phi}{\Lambda_\pi^2 m_r^2}, \label{eqn:LightRadionMApprox}
\end{equation}
we can write down the invariant amplitude corresponding to the interaction of the dark-matter $\Phi$ wind with the nucleus as 
\begin{equation}
    \overline{\sum \abs{\mathcal{M}}^2 } =  \frac{m_\Phi^2}{9} \sum_{\Psi = \{ u, d, s, c, b\}} \left( Z m_p^2 \bar{\mathcal{M}}^p_\Psi + \left( A - Z \right) m_n^2 \bar{\mathcal{M}}^n_\Psi \right)^2, \label{eqn:averagedM}
\end{equation}
where $p$ denotes the proton, $n$ denotes the neutron, $Z$ corresponds to the nuclear charge, and $A$ corresponds to the nuclear mass of the nucleus in question. To relate Eq.~(\ref{eqn:averagedM}) to the spin-independent scattering cross-section, we note that the corresponding phase space factor in the low momentum transfer approximation can be written as 
\begin{equation}
    \frac{1}{16 \pi s} \approx \frac{1}{16 \pi \left( m_\Phi + m_A \right)^2} = \frac{\mu_A^2}{16 \pi} \frac{1}{m_\Phi^2 m_A^2},  
\end{equation}
where we introduced the mass of nucleus $m_A$ and the reduced nuclear mass $\mu_A = m_\Phi m_A / \left( m_A + m_\Phi \right)$. Hence, the spin-independent cross-section relevant to DM direct detection is
\begin{equation}
    \sigma^{\text{SI}} \left( \Phi A \rightarrow \Phi A \right) \approx  \frac{\mu_A^2}{144 \pi m_A^2}  \left[ \sum_{\Psi = \{ u, d, s, c, b \}} \left( Z m_p^2 \bar{\mathcal{M}}^p_\Psi + \left( A - Z \right) m_n^2 \bar{\mathcal{M}}^n_\Psi \right) \right]^2. \label{eqn:FinalCrossSecSI}
\end{equation}
With the help of Eq.~(\ref{eqn:LightRadionMApprox}), we can write down an approximation of Eq.~(\ref{eqn:FinalCrossSecSI}) in the light radion regime as 
\begin{equation}
    \sigma^{\text{SI}} \left( \Phi A \rightarrow \Phi A \right) \simeq \frac{\alpha^2}{36 \pi} \frac{\mu_A^2 m_\Phi^2}{\Lambda_\pi^4 m_r^4 m_A^2} \left( Z m_p^2  + \left( A - Z \right) m_n^2 \right)^2 \simeq \frac{\alpha^2}{36 \pi} \frac{\mu_A^2  m_\Phi^2 m_n^2}{\Lambda_\pi^4 m_r^4} ,
    \label{eq:ddlr}
\end{equation}
where we have assumed that $f_{T_\Psi}^n \simeq f_{T_\Psi}^p$, and using standard calculations,  these quantities (see for example \cite{Hill:2011be}) can be expressed in terms of a simple numerical co-efficient $\alpha \simeq 0.07$.
 Furthermore, we approximated the term in brackets as $m_A^2 m_n^2$ as $m_p \simeq m_n$ and $m_A^2 \simeq A^2 m_n^2$.

To extract constraints on our model parameters from Eq. \ref{eq:ddlr},  we need to normalize  $\sigma^{SI}$ in Eq. \ref{eq:ddlr},  to the proton scattering cross-section such that it can be directly compared against the exclusion plots provided by \cite{LZCollaboration:2024lux}. This can be easily done with the help of the following equation, 
\begin{equation}
    \sigma^{\text{SI}}_p \left( \Phi A \rightarrow \Phi A \right) = \frac{1}{A^2}  \left( \frac{\mu_p}{\mu_A} \right)^2 \sigma^{\text{SI}} \left( \Phi A \rightarrow \Phi A \right),
\end{equation}
where the reduced proton mass is given by $\mu_p = m_\Phi m_p / \left( m_\Phi + m_p \right)$. 

We use the latest results from the LZ experiment~\cite{LZCollaboration:2024lux} to place constraints on the model parameters relevant to this work.  
 In Fig.~\ref{fig:dd}, we present the Direct Detection constraints in the plane of $\Lambda_{\pi}-m_{r}$. From Eq.~(\ref{eq:ddlr}), we observe that direct detection cross sections are enhanced in the light radion regime ($\sigma^{SI}\propto (1/(m_{r}^{4}\Lambda_{\pi}^{4})$). For a DM mass of order 1 TeV, the LZ experiment rules out the model up to $m_{r}\leq 0.7$ GeV for $\Lambda_{\pi}=20$ TeV, while for $\Lambda_{\pi}= 60$ TeV, the model is ruled out for $m_{r}\leq 0.2$ GeV. Note that for $\Lambda_{\pi}=20 (60)$ TeV, 
$\sigma_{SI}$ is below the neutrino floor for $m_{r}\geq 1.3 (0.7)$ GeV, at which point direct detection limits become insensitive due to the solar neutrino background. The Direct Detection bound increases with increasing DM mass since $\sigma^{SI}\propto m_{\Phi}^{2}$, which can 
be observed from the plots. For $m_{\Phi}=6$ TeV,  the LZ experiment rules out the model up to $m_{r}\leq 1.0 (0.5)$ GeV for $\Lambda_{\pi}=20 (60)$ TeV.

\section{Constraints on KK portal dark matter}

\label{sec:results}

In this section, we collect all of the constraints 
to present a picture of the KK-portal dark matter scenario within a stabilized RS1 model. In this paper, we restrict our attention to the case in which the radion is light ($m_{r} \simeq 100$ GeV or lower), in which case the ``back-reaction" of the GW sector on the gravitational background is negligible. The case of a heavy radion with significant changes to the background geometry will be discussed in~\cite{Chivukula:2024}. 

We focus on the case of  $\Lambda_\pi \simeq 20$ - $40$ TeV and, as described above, the region of resonant enhancement $2m_\Phi \simeq m_{KK}$, for which the thermal relic abundance of the dark matter particles can provide the density observed in the universe today. For regions in which the DM relic density constraints are satisfied, we then apply collider-physics constraints, Sec. \ref{sec:ColliderConstraints}, on the KK graviton mass which - due to the resonant condition $2m_\Phi = m_{KK}$ - directly constrains the DM mass as well. Finally, we consider whether the regions of parameter space indicated by these considerations can be probed by direction detection experiments via the radion-exchange interaction, Sec. \ref{sec:DDconstraints}.

\begin{figure}[t]
\centering
{\includegraphics[height=6.15cm]{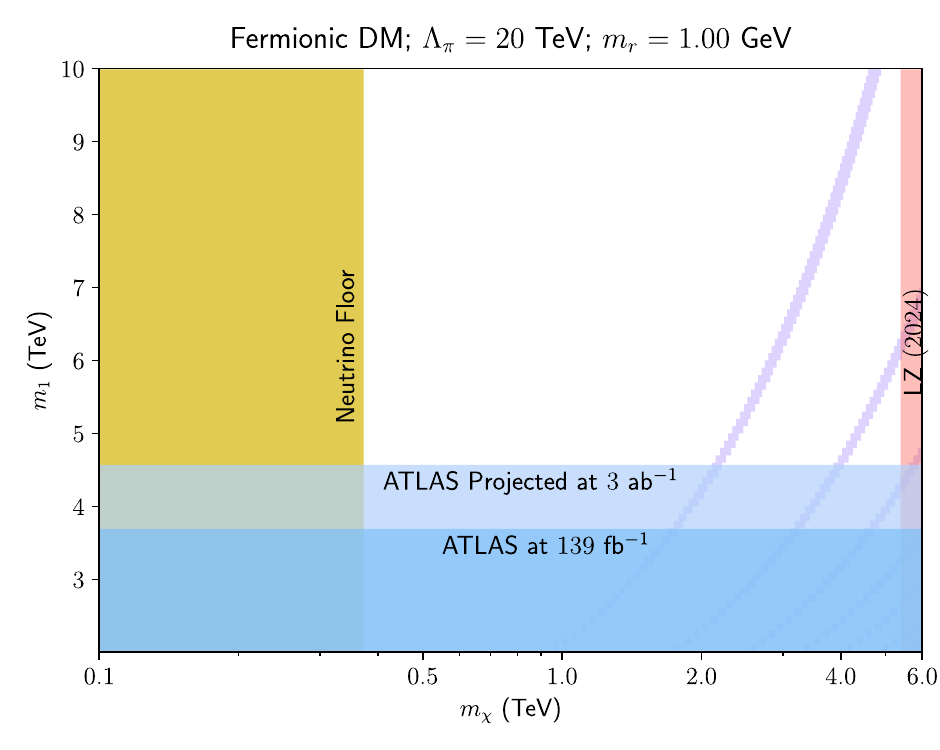}} {\includegraphics[height=6.15cm]{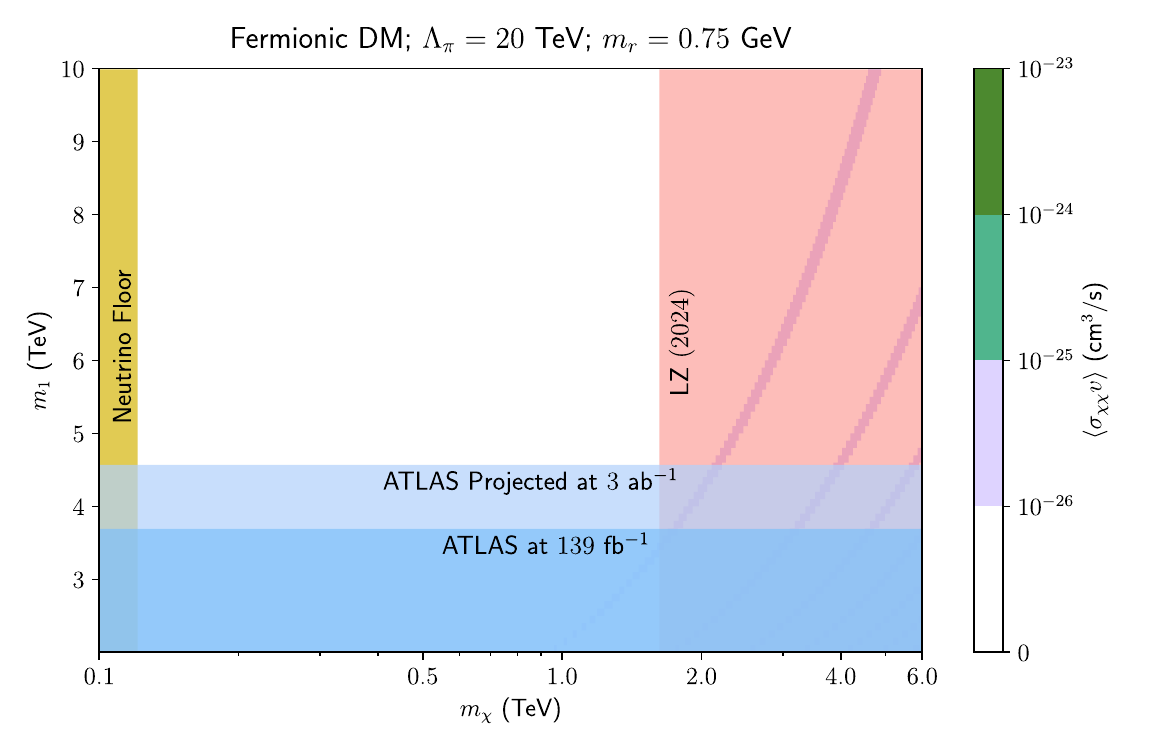}} 
\caption{Combined constraints for fermion dark matter candidates for $\Lambda_\pi = 20$ TeV and $m_r=1$ GeV (left) or 0.75 GeV (right). The lilac shaded curves represent parameter space combinations where the relic density constraint can be satisfied, corresponding to $\langle\sigma v\rangle\geq 10^{-26}\, {\rm cm}^{3}{\rm s}^{-1}$ (corresponding to $x=\frac{m_{\Phi}}{T}\simeq 20$). 
 In both plots the lower darker blue region represents the current ATLAS limit on the 1st spin-2 KK mode from diphoton searches at $\rm 139~ fb^{-1}$~\cite{ATLAS:2021uiz}, while the enlarged light blue region represents the projected reach (exclusion) of the high-luminosity LHC program.  Note that multiple resonances might be detectable at LHC. [Left pane] This pane summarizes limits when $m_{r}=1$ GeV. The region marked in salmon at the right of the plot is ruled out by the SI direct detection experimental results from the LZ experiment~\cite{LZCollaboration:2024lux}.  The region marked in yellow at the left of the plot is impacted by the neutrino fog. The white region in the middle represents the space that direct detection experiments can cover, i.e., in between the direct detection limits and the neutrino floor. [Right panel] This panel summarizes analogous limits when $m_{r}=0.75$ GeV, for which the direct detection constraints entirely exclude the regions which simultaneously satisfy the ATLAS collider limits and the relic density constraint. }
\label{fig:FFscan}
\end{figure}

\begin{figure}[t]
\centering
{\includegraphics[height=6.15cm]{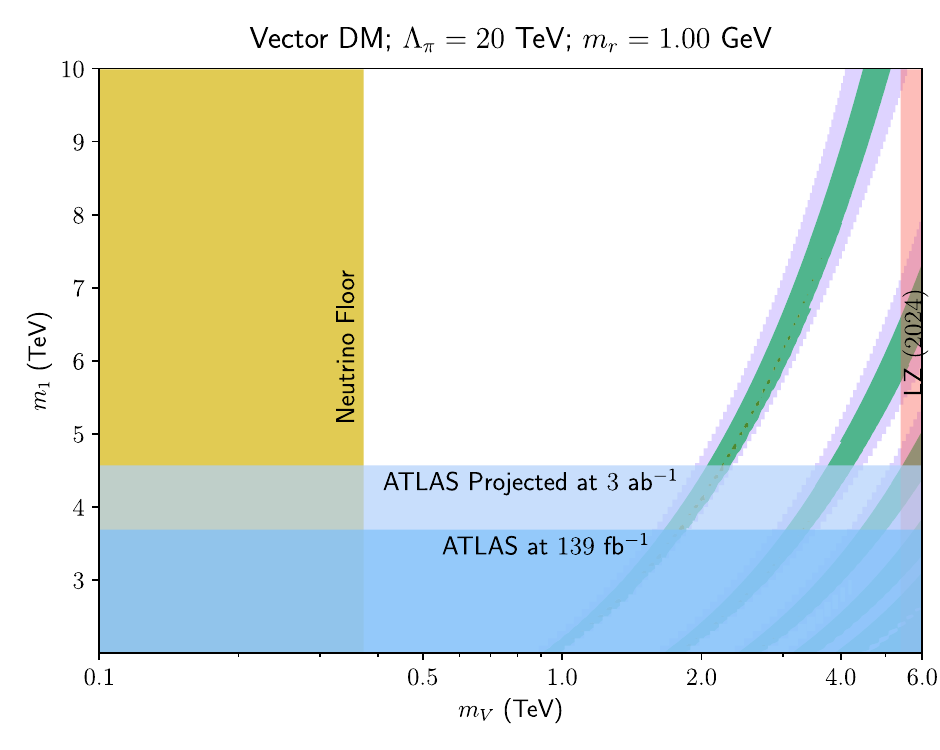}} {\includegraphics[height=6.15cm]{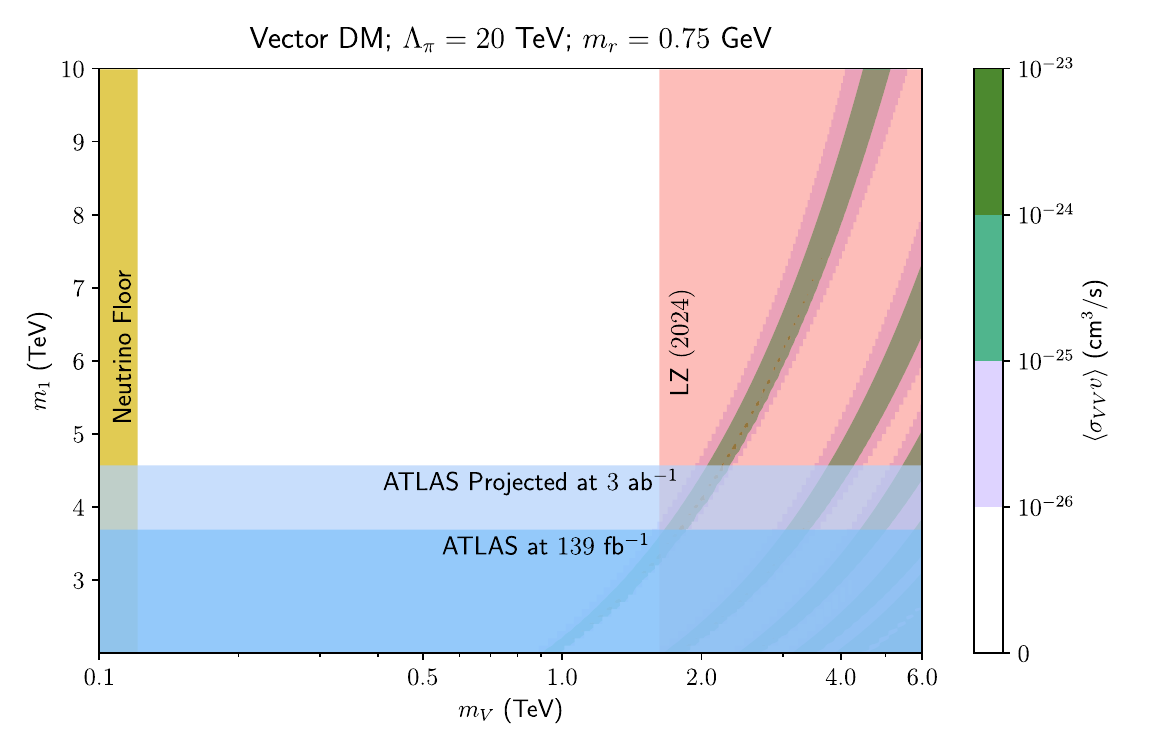}} 
\\
{\includegraphics[height=6.15cm]{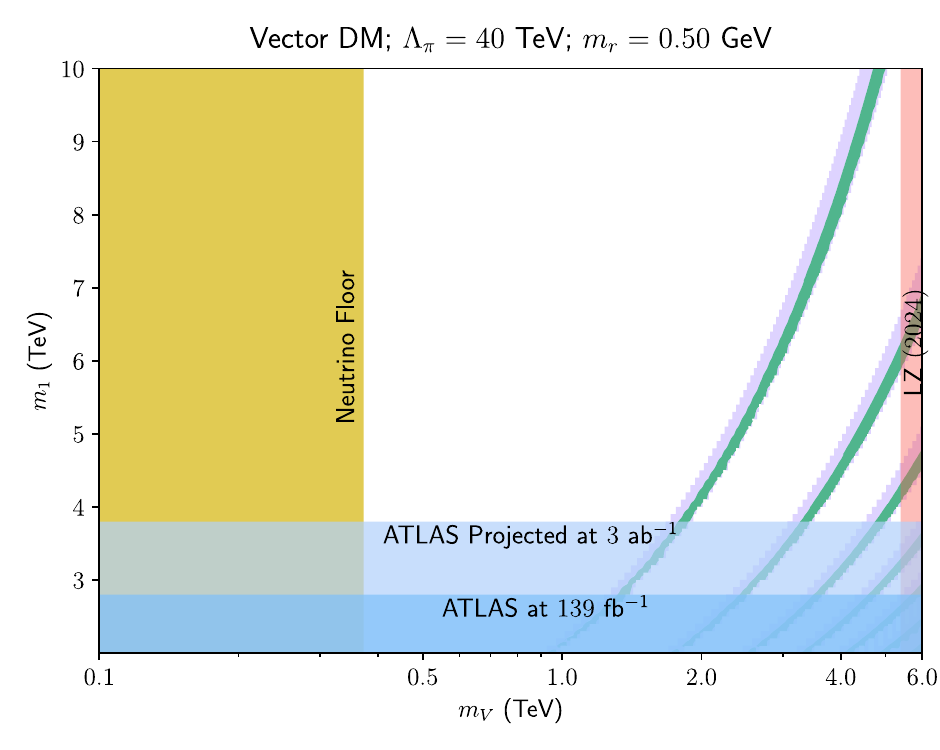}} 
{\includegraphics[height=6.15cm]{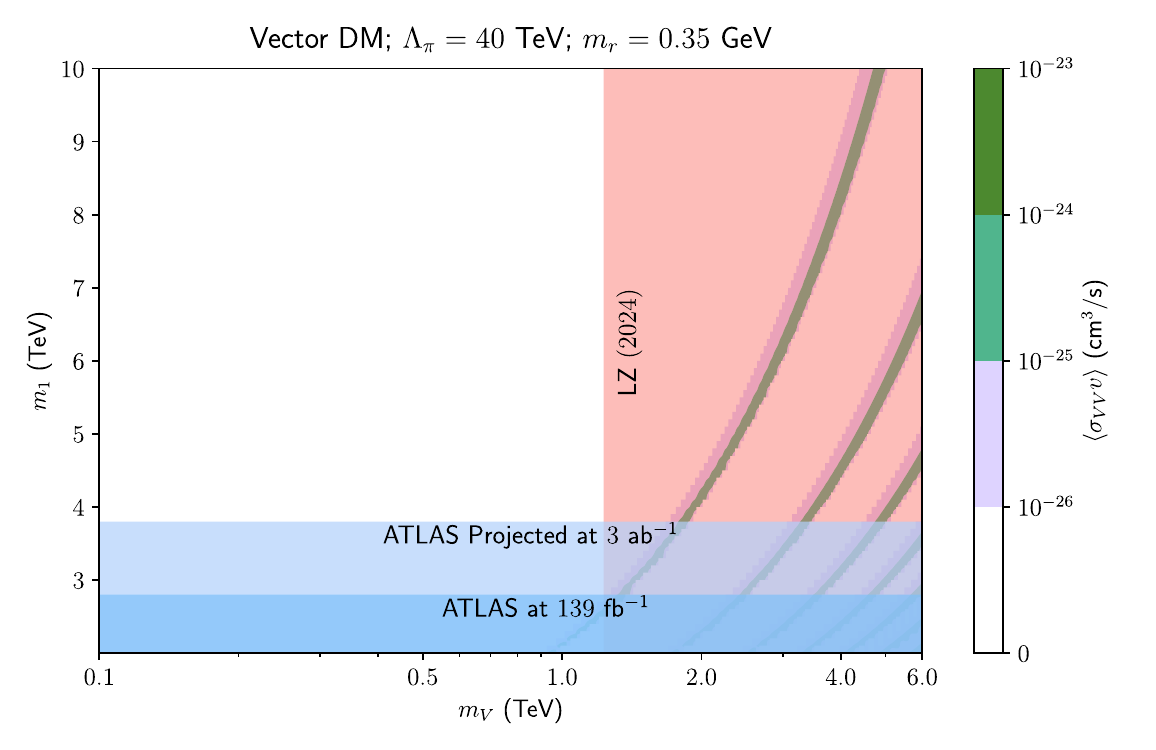}} 
\caption {Combined constraints for vector DM candidates, for $\Lambda_\pi=20$ TeV (upper row) and 40 TeV (lower row), using the same shading as in Fig. \ref{fig:FFscan}. The lilac and green regions indicate where the resonant DM annihilation cross-section equals or exceeds what is required to produce the correct DM relic density today. The lower darker blue regions represent the ATLAS limits on the 1st spin-2 KK mode from diphoton searches \cite{ATLAS:2021uiz}, while the enlarged region represents the projected reach (exclusion) of the high-luminosity LHC program. Note that multiple resonances might be detectable at the LHC.  
The region marked in salmon is ruled out by the SI direct detection experimental results from the LZ experiment \cite{LZCollaboration:2024lux}  while the yellow regions illustrate where the neutrino fog becomes important for the radion masses specified.
The right panels in both rows illustrate parameters for which the direct detection constraints entirely exclude the regions which simultaneously satisfy the ATLAS collider limits and the relic density constraint. }
\label{fig:VVscan}
\end{figure}

For the different possible dark matter scenarios considered here, we then find the following:

\begin{itemize}

\item  {\bf Scalar DM:} From the curves in the top panel of Fig.~\ref{fig:TotalCrossSections}, we see that for a scalar dark matter candidate  $\langle \sigma v_{\rm rel} \rangle\leq  10^{-26}\,{\rm cm}^3 / {\rm s}$ for $\Lambda_{\pi}=20$ TeV,  
indicating that for this scenario, the Universe is always overclosed.  As we discussed previously, the scalar DM annihilation cross section is ``d-wave" ($v^4$) suppressed, and hence would require a large coupling (even lower values of $\Lambda_{\pi}$) on resonance to saturate the relic density. However, lower values of $\Lambda_\pi$ are excluded by collider constraints.

\item {\bf Fermionic DM:} For fermions, the  annihilation cross section at threshold is ``p-wave", and thus $v^{2}$, suppressed. 
Therefore, from Fig.~\ref{fig:TotalCrossSections}, we require $\Lambda_{\pi}\simeq 20$ TeV to produce the correct relic density. 
In the two panels of Fig.~\ref{fig:FFscan}, we show a heat map of the thermal average resonance annihilation cross-section $\langle\sigma v\rangle$ as a function of the dark matter mass $m_{\chi}$ (where we set the lightest KK graviton mass to be $2m_\chi$ and $\Lambda_{\pi}=20$ TeV). (The two panels differ in the assumed radion mass and the corresponding direct detection signal, as discussed below.) The lilac regions in these panels correspond to the cross-section needed to explain the DM relic density. We see that DM masses up to order 5.5 TeV (and corresponding (resonant) KK graviton masses of twice this amount) are allowed. Even higher masses would naively appear to be allowed, however with $\Lambda_\pi=20$ TeV, the effective KK theory breaks down in this region because the KK gravitons become too heavy.\looseness=-1

Shown in dark blue in both panels of Fig.~\ref{fig:TotalCrossSections} are the current LHC bounds on the KK graviton mass for $\Lambda_\pi=20$ TeV. We observe in this case that DM masses below $m_{\chi}\leq 1.8$ TeV are ruled out since the corresponding LHC bound on the mass of the lightest KK graviton is around 3.6 TeV. We also display that the lightest KK graviton mass reach (exclusion) of HL-LHC should reach approximately 4.6 TeV, showing there are potential collider signatures of a model in which the first KK graviton is responsible for resonant annihilation in the early universe for $m_\chi$ up to approximately 2.3 TeV for this value of $\Lambda_\pi$.

In addition, there are also potentially viable parameters in which resonant DM annihilation occurs through the {\it second or third} KK level graviton resonances (corresponding to the two lilac curves at the right side of the panels in  Fig.~\ref{fig:FFscan}), corresponding to DM masses roughly between 3.0 - 4.0 TeV and 4.5 to 5.5 TeV respectively. In this case, there could be clear HL-LHC collider signals for the lighter KK graviton(s), though likely {\it not} for the heavy KK graviton responsible for resonant DM annihilation in the early universe. \looseness=-1

There are also potentially interesting direct detection signals/constraints for this model if the radion is sufficiently light. For $m_{r}\simeq$ 1 GeV, current LZ~\cite{LZCollaboration:2024lux} constraints rule out $m_{\chi}> 5.5$ TeV (region shaded in salmon in the left panel of Fig.~\ref{fig:FFscan}). The ``neutrino floor" in this case is at a much lower mass and, as illustrated in this figure, the region 1.8 TeV $ \le m_\chi \le$ 5.5 TeV is potentially observable in future direct detection experiments.
The direct detection signal is extremely sensitive to the radion mass, however. In the right panel of Fig.~\ref{fig:FFscan} we consider the case of $m_r\simeq 0.75$ GeV, and 
we see that current LZ constraints already {\it rule out} the fermion DM scenario with $\Lambda_\pi=20$ TeV. 
Conversely, for higher radion masses (not shown in these plots) the DM cross section becomes smaller and the direct detection signal will become obscured by the ``neutrino floor" (though the collider signatures for spin-2 states discussed above, if accessible, would still be present). \looseness=-1

Finally, increasing the scale $\Lambda_{\pi}$ we find that these parameter choices lead to an overclosure of the Universe ($\langle \sigma v\rangle \leq  10^{-26}~\text{cm}^3 / \text{s}$), as illustrated for $\Lambda_\pi = 40$ TeV in the right panel of the second row of  Fig.~\ref{fig:TotalCrossSections}.

\item {\bf Vector DM:} We summarize the constraints and prospects for vector DM in Fig.~\ref{fig:VVscan}  for $\Lambda_{\pi}=20$ TeV (upper row) and $\Lambda_{\pi}=40$ TeV (lower row), using the same notation and shading as used for fermion DM in Fig.~\ref{fig:FFscan}. (The plots in each row differ according to the radion masses considered, which impacts the direct detection window.) We have already observed in Fig.~\ref{fig:TotalCrossSections} that $\rm \langle \sigma v\rangle \geq   10^{-26}\rm cm^{3} / s$ on resonance for our parameter choices, thus satisfying the relic density requirement. This is illustrated in the heat maps shown, with the purple (and green) colors indicating the regions where the resonant thermal-average cross-section achieves or exceeds the values needed to explain the vector DM relic density. \looseness=-1

\begin{itemize}

\item We begin with the case of $\Lambda_\pi=20$ TeV, the upper row of Fig.~\ref{fig:VVscan}.
As discussed in the case of fermion dark matter above, the current ATLAS collider constraints on the lightest graviton KK mass can again, using the resonance condition $m_{KK}=2m_\Phi$, be shown to provide a lower bound $m_V \ge 1.9$ TeV and a prospective reach (at HL-LHC) of up to $m_V \simeq 2.3$ TeV if resonant with the lightest KK graviton. Also, as in the fermion case, there is the possibility of a heavier vector DM candidate resonant with a higher KK graviton mode: e.g., $m_V\simeq 3.0-4.0$ TeV resonant with the second KK graviton or $m_V \simeq 4.5-5.5$ TeV resonant with the third KK graviton. In this case, the lighter KK state(s) may be accessible at HL-LHC, though the heavier state would likely not be.

Direct detection signals/constraints are also plotted in Fig.~\ref{fig:VVscan}, and depend sensitively on the radion mass precisely as described in the fermion DM case above. Specifically, for a radion mass of 1 GeV (left panel, upper row), the region of allowed vector DM mass should be visible in future direct detection experiments. On the other hand, if the radion mass is 0.75 GeV, the vector DM scenario with $\Lambda_\pi=20$ TeV is excluded by current LZ \cite{LZCollaboration:2024lux} constraints (right panel, upper row).
Finally, for heavier radion masses, the direct detection would be obscured by the neutrino fog (not illustrated).

\item In the lower row of Fig.~\ref{fig:VVscan}, we present the same scan but for $\Lambda_{\pi}=40$ TeV. In this case, the KK gravitons are narrower due to their smaller couplings, and the parameter regions that satisfy the relic density constraint shrink. The collider signals/constraints (see Fig. \ref{fig:LHCconstraints}) are a bit weaker, and we find that masses below 1.5 TeV are ruled out (corresponding to a lightest KK graviton mass of order 3.0 TeV), while one as heavy as 1.9 TeV (corresponding to a lightest KK mass of 3.8 TeV) should be directly observable at the HL-LHC. As in the cases discussed above, it is also possible to accommodate the dark matter relic abundance via resonance annihilation through the second or third KK gravitons, starting at vector DM masses of order 2.8 TeV, in which case the lighter graviton(s) may be accessible at the HL-LHC.

The direct detection signals/constraints shift substantially for $\Lambda_\pi=40$ TeV. The salmon colored region in the left-hand plot is now for a radion mass of 0.5 GeV, in which case the direct detection signal can potentially cover all of the remaining parameter space. However, for a radion mass of order 0.35 GeV, the current LZ \cite{LZCollaboration:2024lux} constraints exclude this scenario. 
Finally, the neutrino floor obscures the direct detection signal for higher radion masses (not illustrated).

\end{itemize}
As we have seen previously (see Fig. \ref{fig:TotalCrossSections}), 
it is also possible to achieve the correct relic abundance for vector DM with substantially higher values of $\Lambda_\pi$. Existing LHC bounds on TeV-scale KK graviton masses are weaker for larger values of $\Lambda_\pi$ (see Fig. \ref{fig:LHCconstraints}), and the reach of HL-LHC extends only to lower masses.  Correspondingly, some narrow ranges of parameter space for higher values of $\Lambda_\pi$ and TeV-scale masses satisfying the DM relic density constraints will still be present. Depending on the radion mass, the direct detection signals in this parameter regime may also be observable. Alternatively, for larger values of $\Lambda_\pi$, there are regions KK masses and DM masses satisfying the relic abundance than those considered here. In this case, the HL-LHC is unlikely to be able to directly probe the KK gravitons - though, again, depending on the radion mass, direct detection signals may be possible.

\end{itemize}
 
Finally, we note that while we have illustrated our results by assuming a radion mass of order 1 GeV, the thermal relic abundance and collider limits results apply in any case in which the background geometry (and hence the spin-2 and spin-1 KK spectra and couplings) are close to AdS -- that is, so long as the ``back reaction" due to the GW sector is small (in the GW model considered here, massless of order 100 GeV or lower). When the radion is heavier, the KK spectrum will shift substantially, modifying both the relic abundance calculations and collider constraints.  In a forthcoming publication \cite{Chivukula:2024}, we will investigate models where the radion becomes heavy (of order several hundred GeV), and the geometry deviates significantly from AdS.

\section{Conclusion}

\label{sec:conclusion}

In this paper, we have revisited dark-matter scenarios within radius stabilized Randall-Sundrum models where the dark matter candidates are Standard Model (SM) singlets confined to the TeV brane and interact with the SM via spin-2 and spin-0 gravitational Kaluza-Klein (KK) modes. Applying our previous work, which has shown that scattering amplitudes of massive spin-2 KK states involve an intricate cancellation between various diagrams, we compute the thermal relic density of DM particles in these models, including all contributions to the annihilation cross-sections. 

Considering the resulting DM abundance, collider searches, and the absence of a signal in direct DM detection experiments, we show that spin-2 KK portal DM models are highly constrained. In particular, we confirm that within the usual thermal freeze-out scenario, scalar dark matter models are essentially ruled out and show that fermion dark matter is constrained to a narrow region of the parameter space, while vector dark matter models are still viable for a certain region of parameter space. Specifically, we find that vector DM models with masses ranging from 1.1 TeV to 5.5 TeV are phenomenologically viable for theories in which the scale of couplings of the KK modes is of order 40 TeV or lower, while fermion DM models are viable for a similar mass range if the KK coupling scale is of order 20 TeV.

We have focused here on the case where all SM particles are confined to the IR TeV-brane. We briefly comment on models in which the light fermions live in the bulk, potentially closer to the Planck-brane instead. A model of this sort, in which the right-handed top-quark is localized close to the TeV-brane, can potentially address the fermion mass hierarchy \cite{Gherghetta:2000qt,Huber:2000ie}. In this case, the SM gauge fields must live in bulk, and the graviton KK modes, which are localized near the TeV-brane, will have (parametrically) the same interactions with the massless gluon and photon (and potentially comparable interactions with the weak-scale massive $W$ and $Z$ bosons) as discussed here. Overall, the annihilation cross-sections in the early universe will be dominated by the decay of the KK-gravitons to gauge-bosons, resulting in a somewhat lower cross-section similar to those found above.\footnote{The collider bounds for KK gravitons in the context of bulk SM matter has been considered in \cite{Lee:2013bua}, where the authors found that the bounds are weaker due to suppressed couplings of the KK modes to photons and gluons.} Finally, the direct detection bounds will also change somewhat in the absence of direct quark couplings, but since the bulk of these couplings arises from the coupling of the radion to the gluon field-strength squared, which results in a coupling proportional to the nucleon mass, these results too will not change greatly. Therefore, while we defer a detailed study of these models to future work, we do not anticipate that the phenomenologically interesting regions will change significantly from those described here for SM particles confined to the TeV-brane.

Finally, we reiterate that the phenomenology of a model with a heavy radion and the corresponding consideration of the effects of the radion stabilization dynamics on the DM abundance will be covered in a forthcoming paper~\cite{Chivukula:2024}.

\section{Acknowledgements}
RSC, EHS, and XW were supported, in part, by the US National Science Foundation under
Grant No. PHY-2210177. XW is also partially supported by the European Union - Next Generation EU through the MUR PRIN2022 Grant n.202289JEW4. The work of KM was supported in part by the National Science Foundation under Grant No. PHY-2310497. DS is partially supported by the University of New
South Wales, Sydney startup grant PS-71474. 
JAG and GS acknowledge the support
they have received for their research through the provision of
an Australian Government Research Training Program
Scholarship. GS acknowledges the support provided by
the University of Adelaide and the Australian Research
Council through the Centre of Excellence for Dark Matter Particle Physics (CE200100008). GS is thankful to Prof. Anthony Williams for helpful and fruitful discussions.\\

{\it We dedicate this work to the memory of Rohini Godbole (1952-2024) -- role model, mentor, and friend.}
\appendix

\section{Decay Widths of KK gravitons and radions to SM particles.}
\label{sec:smdecaywidth}

\subsection{Spin-$2$ modes}
\label{sec:decayspin2}
For the decay of the massive spin-$2$ mode of mass $m_i$ into standard model species localized on the TeV brane, we have the following decay widths
\begin{align}
    \Gamma_{h^{\left( i \right)} \rightarrow H H} &= \frac{\kappa_i^2}{960 \pi m_i^2} \left( m_i^2 - 4 m_H^2 \right)^{\frac{5}{2}}, \label{eqn:decaywidthhHH} \\
    \Gamma_{h^{\left( i \right)} \rightarrow W W} &= \frac{\kappa_i^2}{480 \pi m_i} \sqrt{1 - \frac{4 m_W^2}{m_i^2}} \left( 56 m_i^2 m_W^2 + 13 m_i^4 + 48 m_W^4 \right),\label{eqn:decaywidthhWW} \\
    \Gamma_{h^{\left( i \right)} \rightarrow Z Z} &= \frac{\kappa_i^2}{960 \pi m_i} \sqrt{1 - \frac{4 m_Z^2}{m_i^2}} \left( 56 m_i^2 m_Z^2 + 13 m_i^4 + 48 m_Z^4 \right),\label{eqn:decaywidthhZZ}  \\
    \Gamma_{h^{\left( i \right)} \rightarrow \gamma \gamma} &=  \frac{\kappa_i^2 m_i^3}{80 \pi}, \label{eqn:decaywidthhPP} \\
    \Gamma_{h^{\left( i \right)} \rightarrow g g} &= \frac{\kappa_i^2 m_i^3}{10 \pi}, \label{eqn:decaywidthhGG}  \\
    \Gamma_{h^{\left( i \right)} \rightarrow \psi \psi} &= \frac{N_c \kappa_i^2 m_i^3}{160 \pi} \left( 1 - \frac{4 m_\psi^2}{m_i^2} \right)^{\frac{3}{2}} \left(1 + \frac{8 m_\psi^2}{3 m_i^2} \right), \label{eqn:decaywidthhpsipsi} 
\end{align}
where $\psi$ is a placeholder for standard model fermions, $N_c$ is the counting factor appearing due to the color charge ($N_c = 3$ for quarks, and $N_c = 1$ for leptons, note that the counting factor is responsible for a difference by a factor of $8$ between Eqs.~(\ref{eqn:decaywidthhPP})~and~(\ref{eqn:decaywidthhGG}) as there are $8$ gluons in QCD), and $m_H$, $m_W$, $m_Z$ are the masses of Higgs, $W$, and $Z$ bosons respectively.

\section{Spin-$0$ modes}
\label{sec:decaywidthradion}
We consider the decay of the spin-$0$ KK mode of mass $m_{\left( n \right)}$. We note that the interaction vertex with the brane-localized vector boson is proportional to the mass of the brane-localized vector boson; hence, at the tree level, there will be no contribution to the decay width from decays into $\gamma$ or $g$. We have the following decays into the standard model species localized on the TeV brane
\begin{align}
    \Gamma_{r^{\left( n \right)} \rightarrow H H} &= \frac{\kappa_{\left( n \right)}^2}{192 \pi m_{\left( n \right)}} \sqrt{ 1 - \frac{4 m_H^2}{m_{\left( n \right)}^2} } \left( 2 m_H^2 + m_{\left( n \right)}^2 \right)^2 \label{eqn:GammarHH} , \\
    \Gamma_{r^{\left( n \right)} \rightarrow W W} &= \frac{\kappa_{\left( n \right)}^2}{96 \pi m_{\left( n \right)}} \sqrt{ 1 -  \frac{4 m_W^2}{m_{\left( n \right)}^2}} \left( 12 m_W^4 - 4 m_{\left( n \right)}^2 m_W^2 + m_{\left( n \right)}^4 \right), \label{eqn:GammarWW} \\ 
    \Gamma_{r^{\left( n \right)} \rightarrow Z Z} &= \frac{\kappa_{\left( n \right)}^2}{192 \pi m_{\left( n \right)}} \sqrt{ 1 -  \frac{4 m_Z^2}{m_{\left( n \right)}^2}} \left( 12 m_Z^4 - 4 m_{\left( n \right)}^2 m_Z^2 + m_{\left( n \right)}^4 \right), \label{eqn:GammarZZ} \\
    \Gamma_{r^{\left( n \right)} \rightarrow \psi \psi} &= \frac{N_c \kappa_{\left( n \right)}^2}{48 \pi} m_\psi^2 m_{\left( n \right)} \left[ 1 -  \frac{4 m_\psi^2}{m_{\left( n \right)}^2} \right]^{\frac{3}{2}}, \label{eqn:GammarFF} 
\end{align}
where, once again, $\psi$ is a placeholder for standard model fermions, $N_c$ is the counting factor appearing due to the color charge ($N_c = 3$ for quarks and $N_c = 1$ for leptons), and $m_H$, $m_W$, $m_Z$ are the masses of Higgs, $W$, and $Z$ bosons respectively.

\section{Decay Widths of KK gravitons and radions to RS particles.}
\label{sec:RSdecaywidth}
\subsection{Spin-$2$ modes}
For the decay of a spin-$2$ KK mode into lighter KK modes, we have three different channels: decay into a pure spin-$2$ state, decay into a mixed state, and decay into a pure spin-$0$ state. The corresponding decay widths can be written as 
\begin{align}
    \Gamma_{{h}^{\left( i \right)} \rightarrow {h}^{\left( j \right)} {h}^{\left( k \right)} } =~& \frac{a_{ijk}^2}{17280 \pi \Lambda_\pi^2} \left[ \left( m_k - m_j - m_i \right) \left( m_k + m_j - m_i \right) \left( m_k - m_j + m_i \right) \right. \nonumber \\ 
    &\times \left. \left( m_k + m_j + m_i \right) \right]^{\frac{5}{2}} \left[ 26 m_k^6 \left( m_i^2 + m_j^2 \right)  + 14 m_k^4 \left( 26 m_i^2 m_j^2 + 9 m_i^4 + 9 m_j^4 \right) \right. \nonumber \\
    &+ 26 m_k^2 \left( 14 m_j^4 m_i^2 + 14 m_j^2 m_i^4 + m_i^6 + m_j^6 \right) + m_k^8 + 26 m_j^2 m_i^6  \nonumber \\
    &\left. + 126 m_i^4 m_j^4 + 26 m_j^6 m_i^2 + m_i^8 + m_j^8 \right] / \left( m_k^4 m_j^4 m_i^7 \right), \label{eqn:Gammahhh} \\
    \Gamma_{h^{\left( i \right)} \rightarrow h^{\left( j \right)} r^{\left( n \right)}} =~& \frac{a_{i'j'\left( n \right)}^2}{480 \pi \Lambda_\pi^2 m_i^3} \left( 76 + \frac{22 \left( m_i^2 + m_j^2 - m_{\left( n \right)}^2 \right)^2}{m_i^2 m_j^2} + \frac{\left( m_i^2 + m_j^2 - m_{\left( n \right)}^2 \right)^4}{m_i^4 m_j^4} \right) \nonumber \\
    &\times \sqrt{ \left(m_i^2 - m_j^2 - m_{\left( n \right)}^2 \right)^2 - 4 m_j^2 m_{\left( n \right)}^2}, \label{eqn:Gammahhr} \\
    \Gamma_{h^{\left( i \right)} \rightarrow r^{\left( n \right)} r^{\left( m \right)}} =~& \frac{\left( d_{i \left( n \right) \left( m \right)} + 2 \tilde{d}_{ i \left( n \right) \left( m \right)} \right)^2}{480 \pi \Lambda_\pi^2 m_i^7}  \left[ \left( m_i^2 - m_{\left( n \right)}^2 - m_{\left( m \right)}^2 \right)^2 - 4 m_{\left( n \right)}^2 m_{\left( m \right)}^2 \right]^{\frac{5}{2}}, \label{eqn:Gammahrr}
\end{align}
where the ``a-type'' overlap integrals are defined as
\begin{align}
    a_{ijk} &= \frac{1}{\psi_1{\left( \pi \right)}} \int_{-\pi}^{\pi} d \phi e^{-2 A{\left( \phi \right)}} \psi_i{\left( \phi \right)} \psi_j{\left( \phi \right)} \psi_k{\left( \phi \right)}, \label{eqn:ProofAppendixAtypeOverlapDef} \\
    a_{i' j' \left( n \right)} &= \frac{1}{r_c^2 \psi_1{\left( \pi \right)}} \int_{-\pi}^\pi d \phi e^{-2 A{\left( \phi \right)}} \partial_\phi \psi_i{\left( \phi \right)} \partial_\phi \psi_j{\left( \phi \right)} \gamma_n{\left( \phi \right)}, \label{eqn:ProofAppendixAtypeOverlapDefSpin0} \\
\end{align}
and the ``d-type'' overlap integrals are defined in Eqs.~(\ref{eqn:OverlapDtype})~and~(\ref{eqn:OverlaptildeDtype}). 

\subsection{Spin-$0$ modes}
Analogously to the decay of a massive spin-$2$ mode, there are three different channels for the decay of a massive spin-$0$ mode: spin-$2$ channel, mixed state channel, and spin-$0$ channel. The corresponding decay widths are 
\begin{align}
    \Gamma_{r^{\left( n \right)} \rightarrow h^{\left( i \right)} h^{\left( j \right)}} =~& \frac{ a_{i' j' {\left( n \right)}}^2}{96 \pi \Lambda_\pi^2 m_{ \left( n \right)}^3} \sqrt{  \left( m_{\left( n \right)}^2 - m_i^2 - m_j^2 \right)^2 - 4 m_i^2 m_j^2 }  \nonumber \\
    &\times \left( 76 + \frac{22 \left(  m_i^2 + m_j^2 - m_{\left( n\right)}^2 \right)^2}{m_i^2 m_j^2} + \frac{\left( m_i^2 + m_j^2 - m_{\left( n\right)}^2 \right)^4}{m_i^4 m_j^4} \right), \label{eqn:Gammarhh} \\
    \Gamma_{r^{\left( n \right)} \rightarrow h^{\left( i \right)} r^{\left( m \right)}} =~&  \frac{ \left( d_{i \left( n \right) \left( m \right)} + 2 \tilde{d}_{i \left( n \right) \left( m \right)} \right)^2}{96 \pi \Lambda_\pi^2 m_{\left( n\right) }^3 m_i^4} \left[ \left( m_{\left( n \right)}^2 - m_{\left( m \right)}^2 - m_i^2 \right)^2 - 4 m_{\left( m \right)}^2 m_i^2 \right]^{\frac{5}{2}}, \label{eqn:Gammarrh} \\
    \Gamma_{r^{\left( n \right)} \rightarrow r^{\left( m \right)} r^{\left( l \right)}} =~& \frac{\left( \sqrt{6} \left( m_{\left( n \right)}^2 + m_{\left( m \right)}^2 + m_{\left(  l \right)}^2 \right) e_{\left( n \right) \left( m \right) \left( l \right)} - 3 \tilde{e}_{\left( n \right) \left( m \right) \left( l \right)} \right)^2}{144 \pi \Lambda_1^2 m_{\left( n \right)}^3}, \label{eqn:Gammarrr}
\end{align}
where the ``a-type'' overlap integral is defined in Eq.~(\ref{eqn:ProofAppendixAtypeOverlapDefSpin0}), ``d-type'' overlap integral defined in Eqs.~(\ref{eqn:OverlapDtype})~and~(\ref{eqn:OverlaptildeDtype}), and ``e-type'' overlap integrals defined in Eqs.~(\ref{eqn:OverlapCtype})~and~(\ref{eqn:OverlaptildeCtype}).

\subsection{Wave functions and Decay Widths}

\begin{figure}[t]
    \centering
\includegraphics[width=0.49\linewidth]{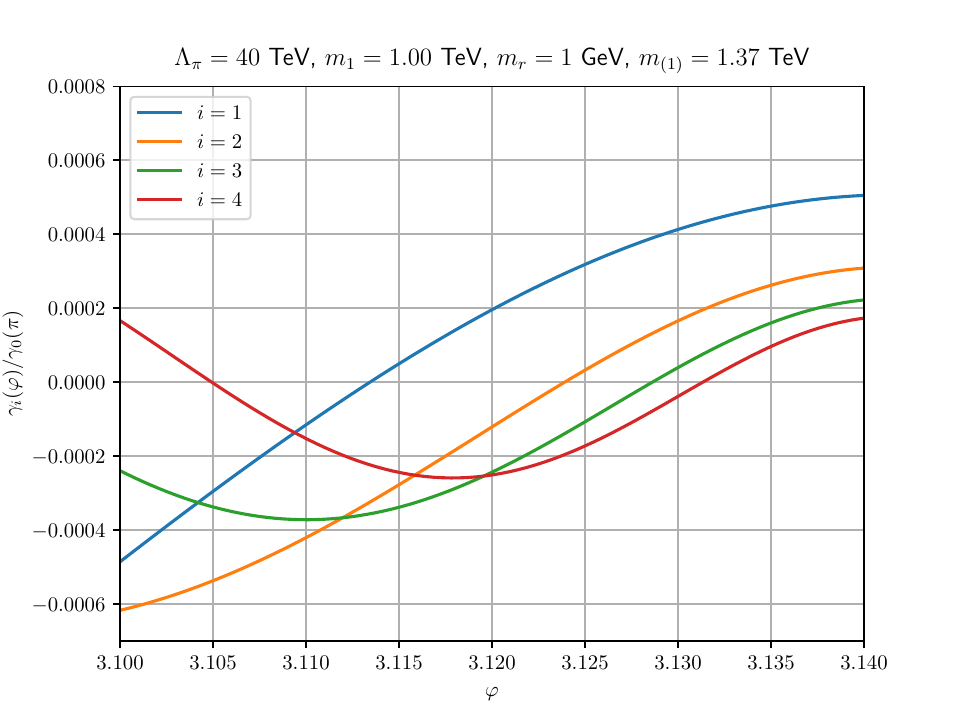}
\includegraphics[width=0.49\linewidth]{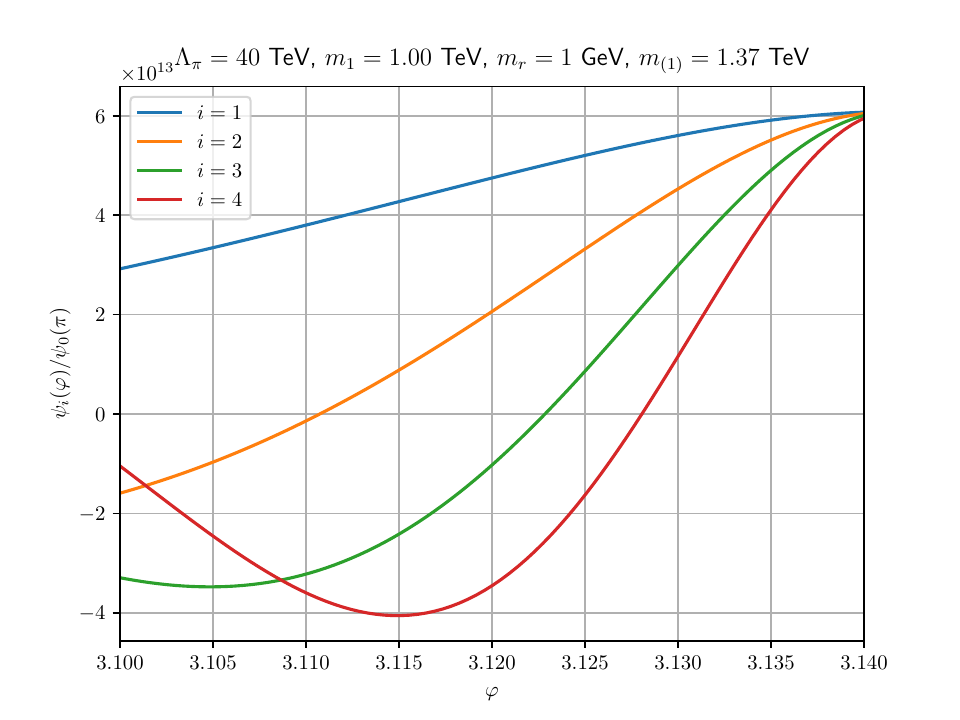} \\
    \caption{Plots of the spin-$0$ [left panel] and spin-$2$ [right panel] wave-functions scaled by the corresponding $0$th mode on the TeV brane in the vicinity of the TeV brane.} 
    \label{fig:WFplotsaroundTeV}
\end{figure}

\begin{figure}[H]
    \centering
    \includegraphics[width=0.7\linewidth]{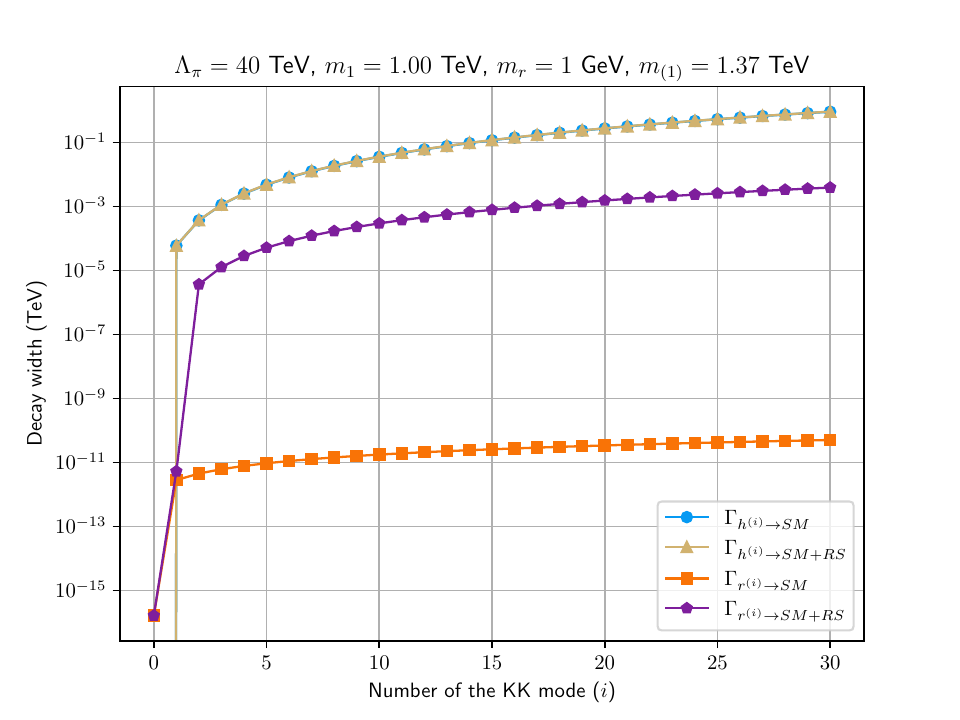}
    \caption{Plots of the spin-$0$ (labelled $r^{\left( i \right)}$) and spin-$2$ (labelled $h^{\left( i \right)}$) decay widths into brane SM species (labelled "SM") and both brane SM species and other KK modes of the lower mass (labelled ``SM+RS''). \label{fig:GammaPlots}}
\end{figure}

This section briefly discusses the properties of the wave functions and decay widths of the spin-2 KK and GW scalar sectors. The wave functions and the mass-spectrum of the spin-2 KK modes and the spin-0 GW scalars have been discussed in detail in~\cite{Chivukula:2021xod,Chivukula:2023sua}. Here, we elucidate some of their features by numerically solving the St\"urm-Liouville equations for spin-2 KK and spin-0 GW modes. 
A plot of the first $4$ modes scaled by the $0$th mode at the TeV brane for both spin-$0$ and spin-$2$ is depicted in Fig.~\ref{fig:WFplotsaroundTeV},  where we plot the ratio of wave-functions $\gamma_{i}(\varphi)/\gamma_{0}(\pi)$ for spin-0 and $\psi_{i}(\varphi)/\psi_{0}(\pi)$ for spin-2 as a function of the extra-dimensional co-ordinate $\varphi$.
As the coupling to the SM is directly
proportional to the wave function on the TeV brane,  we observe that massive spin-$2$ modes couple to the brane matter more strongly than the massless mode (graviton). However, the coupling strength of spin-$0$ modes decreases with an increase in their mass, with an initial jump from $0$th to $1$st mode. Furthermore, we plot the decay widths for each massive mode (plotted in Fig.~\ref{fig:GammaPlots}).


\section{Interactions and coupling structures}
\label{sec:newcouplings}

This paper required a number of interactions for the spin-2 KK and the spin-0 GW sector not documented previously in the literature, including the relevant works of this collaboration ~\cite{SekharChivukula:2019yul,SekharChivukula:2019qih,Chivukula:2020hvi,Chivukula:2022kju,Chivukula:2021xod,Chivukula:2022tla,Chivukula:2023sua}. These include the interactions corresponding to Fig.~\ref{fig:ProcessInQuestion2}, and Fig.~\ref{fig:ProcessInQuestion3}, i.e, processes corresponding to $h^{ \left( i \right)}r^{ \left( j \right)}$, and $r^{ \left( i \right)}r^{ \left( j \right) }$ final states. A set of sum-rules in unitary gauge ensures that amplitudes with these final states do not grow anomalously with the center of mass energy. We document the interactions relevant for the fully-stabilized model, described in~\cite{Chivukula:2022tla,Chivukula:2021xod}. 

To calculate the matrix elements required for the cross-section evaluation, we need the cubic and quartic self-interactions of the 5D tensor field $\hat{h}_{\mu\nu}$, as well as the $\hat{h} \hat{h} \hat{r}$, $\hat{h} \hat{r} \hat{r}$, and the $\hat{r} \hat{r} \hat{r}$   cubic interactions. As far as $\hat{h} \hat{h} \hat{r}$ interactions are concerned, the tensor structure remains the same; the only change from our previous works ~\cite{SekharChivukula:2019yul,SekharChivukula:2019qih,Chivukula:2020hvi,Foren:2020egq} is due to the change in $A(y)$ via the stabilizing potential~\cite{Chivukula:2020hvi}. 
Thus, the primary difference between the stabilized and unstabilized cases as far as $\hat{h}\hat{h}\hat{r}$ is concerned relates to the KK decomposition of the 5D field $\hat{r} \left( x,y \right)$. In the unstabilized case, $\hat{r}$ generates only the single massless scalar state $\hat{r}{ \left( x \right)}$, which is referred to as the radion. 
In the stabilized case, $\hat{r}$ mixes with the GW scalar sector proportional to the VEV of the stabilizing field and develops a nontrivial $y$-dependence. The combined scalar system generates an infinite tower of massive scalars $\{\hat{r}^{\left( i \right)}{ \left( x \right) }$, of which the lightest state is identified as the radion. As such, to extract the $\hat{h} \hat{r} \hat{r}$, and the $\hat{r} \hat{r} \hat{r}$ interactions, we need to expand the Lagrangian in the canonical basis of the spin-2 and spin-0 system. This process has been described in great detail in~\cite{Chivukula:2021xod}. Here, we follow the same procedure to extract the $\hat{h} \hat{r} \hat{r}$ and the $\hat{r} \hat{r} \hat{r}$ interactions.
\subsection{Coupling definitions}
In~\cite{Chivukula:2022tla}
we defined generalized ``couplings" as overlap integrals of spin-2 and spin-0 wavefunctions of the form
\begin{align}
    x^{(p)}_{(k^{\prime}\cdots l) \cdots m^{\prime} \cdots n} &\equiv  \int_{-\pi}^{+\pi} d\varphi\hspace{10 pt}\varepsilon^{p}\, (\partial_{\varphi}\gamma_{k})\cdots \gamma_{l}\cdots (\partial_{\varphi}\psi_{m})\cdots \psi_{n} \label{generalizedx}
\end{align}
with $A(\varphi)$ being a generalized warp-factor and $\varepsilon\equiv\exp(-A)$. Any index in the above definition with a prime denotes a derivative with respect to the extra-dimensional coordinate. 
We also defined the so-called {\it{a}} type and {\it{b}} type couplings with the exponential pre-factors  $\varepsilon^{-2}$ and $\varepsilon^{-4}$ as, 
\begin{align}
    a_{(k^{\prime}\cdots l) \cdots m^{\prime} \cdots n} \equiv x^{(-2)}_{(k^{\prime}\cdots l) \cdots m^{\prime} \cdots n} \hspace{35 pt} b_{(k^{\prime}\cdots l) \cdots m^{\prime} \cdots n} = x^{(-4)}_{(k^{\prime}\cdots l) \cdots m^{\prime} \cdots n}
\end{align}
Additionally, we defined an overlap with the label ``$c$", associated with $p=-6$, with the generic definition,
\begin{align}
c_{k^{\prime}l^{\prime}m^{\prime}n^{\prime}} & \equiv x^{(-6)}_{k^{\prime}l^{\prime}m^{\prime}n^{\prime}} =  \int_{-\pi}^\pi d\varphi\hspace{5 pt} \varepsilon^{-6}(\partial_{\varphi}\psi_{k})(\partial_{\varphi}\psi_{l})(\partial_{\varphi}\psi_{m})(\partial_{\varphi}\psi_{n}) \label{sup-cklmnDEF}
\end{align}
In this work, we will introduce further symbols to describe the interactions $\hat{h} \hat{r} \hat{r}$, and the $\hat{r} \hat{r} \hat{r}$ after KK decomposition, called {\it{``d''}} and {\it{``e''}}
type couplings.
\subsection{Interactions and Feynman Rules}
\label{sec:laghr}
In this section, we will outline the interactions with $\hat{h} \hat{r} \hat{r}$ and the $\hat{r} \hat{r} \hat{r}$ terms. 
Focusing on the term containing two field $\hat{r}$ and one field $\hat{h}_{\mu \nu}$ the situation is a little bit more complicated, as upon the application of the background equations of motion  and the equation of motion for the field $\hat{r}$,  with the corresponding boundary conditions, we are left with 
\begin{align}
    S_{\hat{h} \hat{r} \hat{r}} =~& \frac{1}{M_5^{\frac{3}{2}}} \int d^4 x \int_{-\pi}^\pi d\phi~r_c \left[ \vphantom{\sum_{i \in \{ 1,2 \}} \frac{2 \left( -1 \right)^i \hat{r}' \hat{h}}{r_c^2 \varphi_0'^3}} \frac{e^{2 A }}{6} \partial^\alpha \hat{r} \left( 4 \hat{r} \partial_\alpha \hat{h} + 3 \hat{h} \partial_\alpha \hat{r} - 4 \hat{r} \partial_\beta \hat{h}^\beta_\alpha - 6 \hat{h}_{\alpha \beta} \partial^\beta \hat{r} \right) \right.  \nonumber\\
    ~&\left. + \frac{3 e^{2 A }}{ \varphi_0'^2} \partial^\alpha \hat{r}' \left( \hat{h} \partial_\alpha \hat{r}' - 2 \hat{h}_{\alpha \beta} \partial^\beta \hat{r}' \right) + \frac{e^{2 A} \hat{z} \hat{h}}{r_c^2 \varphi_0'^2} \left( \frac{24 \dot{V} r_c^2 \hat{r}'}{\varphi_0'} + 24 A' \hat{r}' + \varphi_0'^2 \hat{r} - 3 e^{2 A} \hat{z} \right)  \right. \nonumber\\
    ~&\left. - \frac{\hat{r} \hat{h}}{6 r_c^2} \left( \frac{72 \dot{V} r_c^2 \hat{r}'}{\varphi_0'} + 60 A' \hat{r}' + \varphi_0'^2 \hat{r} \right) - \frac{2 \hat{r}'^2 \hat{h}}{r_c^2 \varphi_0'^4} \left( 24 \dot{V}^2 r_c^4 + 48 \dot{V} r_c^2 A' \varphi_0'  + \varphi_0' \left( 6 \dot{V}' r_c^2 + 24 A'^2 \varphi_0' + \varphi_0'^3 \right) \right) \right. \nonumber\\
    ~& +\left. \sum_{i = 1,2} \frac{2 \left( -1 \right)^i \hat{r}' \hat{h}}{r_c^2 \varphi_0'^3} \left( 12 \dot{V} r_c^2 \hat{r}' + 12 A' \hat{r}' \varphi_0'  + \varphi_0'^3 \hat{r} - 3 e^{2 A} \varphi_0' \hat{z} \right) \delta_i  \right]. \label{eqn:ShrrFullKKexpand1}
\end{align}
The object $\hat{z}=\dfrac{1}{\sqrt{r_{c}}} \sum_{i = 0}^{+\infty} \mu_{(i)}^{2}\, \hat{r}^{(i)}(x)\,\gamma_{i}(\varphi) ~$ was introduced an auxiliary field to facilitate manipulations at the 5D level.
 At face value, it is worth pointing out that the interaction term Eq.~(\ref{eqn:ShrrFullKKexpand1}) is different from the interaction terms presented in Ref.~\cite{Chivukula:2020hvi} for the unstabilized RS model. This is not surprising, as by introducing brane stabilization though the bulk field $\hat{\Phi}$, we effectively introduced another coupling between the field $\hat{h}_{\mu \nu}$ and the stress-energy tensor of the bulk field $\hat{\Phi}$\footnote{We envisage that suitable integration by parts and use of equations of motion followed by additional total derivative terms should simplify Eq.~(\ref{eqn:ShrrFullKKexpand1}), to resemble the unstabilized RS1 model just like the $\hat{h} \hat{h} \hat{r}$ terms. We have not pursued this in this paper but have chosen to verify some additional rules numerically by plugging in solutions to the DFGK model.}. Similarly to previous interaction terms, the next stage is to KK expand Eq.~(\ref{eqn:ShrrFullKKexpand1}) in the {\it{``stiff-wall'' }} limit.  KK decomposing the various 5D fields and discarding the boundary term we get,
\begin{align}
    S_{\hat{h} \hat{r} \hat{r}} =~& \frac{1}{\sqrt{r_c} M_5^{\frac{3}{2}}} \sum_{i=0}^\infty \sum_{n =0}^\infty \sum_{m=0}^\infty \int d^4 x  \nonumber\\
    ~&\times \left[ \frac{ \partial^\alpha \hat{r}^{\left( n \right)}}{6} \left( 4 \hat{r}^{\left( m \right)} \partial_\alpha \hat{h}^{\left( i \right)} + 3 \hat{h}^{\left( i \right)} \partial_\alpha \hat{r}^{\left( m \right)} - 4 \hat{r}^{\left( m \right)} \partial_\beta \left( \hat{h}^{\left( i\right)} \right)^\beta_\alpha - 6 \hat{h}^{\left( i \right)}_{\alpha \beta} \partial^\beta \hat{r}^{\left( m \right)} \right) \int_{-\pi}^\pi d \phi e^{2A} \psi_i \gamma_n \gamma_m \right.  \nonumber\\
    ~&\left. + \partial^\alpha \hat{r}^{\left( n \right)} \left( \hat{h}^{\left( i \right)} \partial_\alpha \hat{r}^{\left( m \right)} - 2 \hat{h}^{\left( i \right)}_{\alpha \beta} \partial^\beta \hat{r}^{\left( m \right)} \right) \int_{-\pi}^\pi d \phi \frac{3 e^{2 A }}{ \varphi_0'^2} \psi_i \gamma'_n \gamma'_m\right. \nonumber\\
    ~&+\left. \hat{h}^{\left( i \right)} \hat{r}^{\left( n \right)} \hat{r}^{\left( m \right)} \int_{-\pi}^\pi d \phi \frac{e^{2 A} m_{\left( n \right)}^2 \psi_i \gamma_{n} }{\varphi_0'^2}  \left( \frac{24 \dot{V} r_c^2 \gamma'_m}{\varphi_0'} + 24 A' \gamma'_m + \varphi_0'^2 \gamma_m - 3 e^{2 A} m_{\left( m \right)}^2 \gamma_m \right)  \right. \nonumber\\
    ~&\left. - \hat{h}^{\left( i \right)} \hat{r}^{\left( n \right)} \hat{r}^{\left( m \right)} \int_{-\pi}^\pi d\phi  \frac{\psi_i \gamma_n}{6 r_c^2} \left( \frac{72 \dot{V} r_c^2 \gamma'_{m} }{\varphi_0'} + 60 A' \gamma'_{m} + \varphi_0'^2 \gamma_{m} \right) \right.  \nonumber\\
    ~&- \left. \hat{h}^{\left( i \right)} \hat{r}^{\left( n \right)} \hat{r}^{\left( m \right)} \int_{-\pi}^\pi d \phi \frac{2 \psi_i \gamma'_n \gamma'_m }{r_c^2 \varphi_0'^4} \left( 24 \dot{V}^2 r_c^4 + 48 \dot{V} r_c^2 A' \varphi_0'  + \varphi_0' \left( 6 \dot{V}' r_c^2 + 24 A'^2 \varphi_0' + \varphi_0'^3 \right) \right) \right], \label{eqn:ShrrFullKKexpand2}
\end{align}
where we can associate the first term in square brackets with the only term appearing in the same interaction term for the unstabilised RS model presented in Ref.~\cite{Chivukula:2020hvi}. We proceed to define the {\it{``d-type''}} overlap integrals (integrals with a factor of $e^{2 A}$) as 
\begin{equation}
    d_{i \left( n \right) \left( m \right)} = \frac{1}{\psi_1{\left( \pi \right)}} \int_{-\pi}^\pi d \phi e^{2 A{\left( \phi \right)}} \psi_i{\left( \phi \right)} \gamma_n{\left( \phi \right)} \gamma_m{\left( \phi \right)}. \label{eqn:OverlapDtype}
\end{equation}
However, clearly, Eq.~(\ref{eqn:OverlapDtype}) is not the only overlap integral appearing in Eq.~(\ref{eqn:ShrrFullKKexpand2}). Hence, we proceeded by defining a brane stabilization correction to the overlap integral $d_{i \left( n \right) \left( m \right)}$ as 
\begin{equation}
    \tilde{d}_{i \left( n \right) \left( m \right)} = \frac{1}{\psi_1{\left( \pi \right)}} \int_{-\pi}^\pi d \phi \frac{3 e^{2 A }}{ \varphi_0'^2} \psi_i{\left( \phi \right)}  \partial_\phi \gamma_n{\left( \phi \right)} \partial_\phi \gamma_m{\left( \phi \right)}, \label{eqn:OverlaptildeDtype}
\end{equation}
we further define auxiliary overlap integrals as
\begin{align}
\bar{d}_{i \left( n \right) \left( m \right):A} &= \frac{1}{\psi_1{\left( \pi \right)}} \int_{-\pi}^\pi d \phi \frac{e^{2 A} m_{\left( n \right)}^2  \psi_i \gamma_{n}}{\varphi_0'^2} \left( \frac{24 \dot{V} r_c^2 \gamma'_m}{\varphi_0'} + 24 A' \gamma'_m + \varphi_0'^2 \gamma_m - 3 e^{2 A} m_{\left( m \right)}^2 \gamma_m \right),\label{eqn:OverlapbarDtypeA} \\
\bar{d}_{i \left( n \right) \left( m \right):B} &= \frac{1}{\psi_1{\left( \pi \right)}}  \int_{-\pi}^\pi d\phi  \frac{\psi_i \gamma_n}{6 r_c^2} \left( \frac{72 \dot{V} r_c^2 \gamma'_{m} }{\varphi_0'} + 60 A' \gamma'_{m} + \varphi_0'^2 \gamma_{m} \right), \label{eqn:OverlapbarDtypeB} \\
\bar{d}_{i \left( n \right) \left( m \right):C} &=\frac{1}{\psi_1{\left( \pi \right)}}  \int_{-\pi}^\pi d \phi \frac{2 \psi_i \gamma'_n \gamma'_m }{r_c^2 \varphi_0'^4} \left( 24 \dot{V}^2 r_c^4 + 48 \dot{V} r_c^2 A' \varphi_0'  + \varphi_0' \left( 6 \dot{V}' r_c^2 + 24 A'^2 \varphi_0' + \varphi_0'^3 \right) \right), \label{eqn:OverlapbarDtypeC}
\end{align}
where we have used shorthand for the derivatives on the basis functions. With the help of Eqs.~(\ref{eqn:OverlapDtype})-(\ref{eqn:OverlapbarDtypeC}),  we can write down Eq.~(\ref{eqn:ShrrFullKKexpand2}) in a concise form
\begin{align}
    S_{\hat{h} \hat{r} \hat{r}} =~& \frac{1}{\Lambda_\pi} \sum_{i=0}^\infty \sum_{n =0}^\infty \sum_{m=0}^\infty \int d^4 x  \nonumber\\
    ~&\times \left[  \partial^\alpha \hat{r}^{\left( n \right)} \left( 4 \hat{r}^{\left( m \right)} \partial_\alpha \hat{h}^{\left( i \right)} + 3 \hat{h}^{\left( i \right)} \partial_\alpha \hat{r}^{\left( m \right)} - 4 \hat{r}^{\left( m \right)} \partial_\beta \left( \hat{h}^{\left( i\right)} \right)^\beta_\alpha - 6 \hat{h}^{\left( i \right)}_{\alpha \beta} \partial^\beta \hat{r}^{\left( m \right)} \right) \frac{d_{i \left( n \right) \left( m \right)}}{6}  \right.  \nonumber\\
    ~&\left. + \partial^\alpha \hat{r}^{\left( n \right)} \left( \hat{h}^{\left( i \right)} \partial_\alpha \hat{r}^{\left( m \right)} - 2 \hat{h}^{\left( i \right)}_{\alpha \beta} \partial^\beta \hat{r}^{\left( m \right)} \right) \tilde{d}_{i \left( n \right) \left( m \right)} + \hat{h}^{\left( i \right)} \hat{r}^{\left( n \right)} \hat{r}^{\left( m \right)} \left( \bar{d}_{i \left( n \right) \left( m \right):A} - \bar{d}_{i \left( n \right) \left( m \right):B} - \bar{d}_{i \left( n \right) \left( m \right):C} \right) \vphantom{\frac{d_{i \left( n \right) \left( m \right)}}{6}}\right], \label{eqn:ShrrFullKKexpand3}
\end{align}
where the unstabilized RS model limit is obtained by setting $\tilde{d}_{i \left( n \right) \left( m \right)} = 0$ and $\bar{d}_{i \left( n \right) \left( m \right):A} - \bar{d}_{i \left( n \right) \left( m \right):B} - \bar{d}_{i \left( n \right) \left( m \right):C} = 0$.  \par
Lastly, we focus on the interaction terms between three $\hat{r}$ fields. Analogously to Eq.~(\ref{eqn:ShrrFullKKexpand1}) upon the application of the background equations of motion and the equation of motion for the field $\hat{r}$  with the corresponding boundary conditions, we are left with an expression of the form
\begin{align}
    S_{\hat{r} \hat{r} \hat{r}} =~& \frac{1}{M_5^{\frac{3}{2}}} \int d^4 x \int_{-\pi}^\pi d\phi r_c \left[\vphantom{\sum_{i=1,2} \left( - 1 \right)^i \frac{2 \sqrt{6} e^{2 A}}{r_c^2 \varphi_0'^3}} - \frac{2 e^{4 A}}{\sqrt{6}} \hat{r} \partial_\alpha \hat{r} \partial^\alpha \hat{r} + \frac{e^{2 A}}{\sqrt{6} r_c^2 \varphi_0'^2} \hat{r} \left( \varphi_0'^4 \hat{r}^2 - 6 e^{2 A} \varphi_0'^2 \hat{r} \hat{z} + 18 e^{2 A } \hat{z}^2 \right) \right. \nonumber\\
    ~&\left. - \frac{24 \sqrt{6} e^{4 A}}{r_c^2 \varphi_0'^3} \left( \dot{V} r_c^2 + A' \varphi_0' \right) \hat{r}'  \hat{r} \hat{z} + \sqrt{\frac{2}{3}} \frac{32 e^{2 A}}{r_c^2 \varphi_0'} \left( \dot{V} r_c^2 + A' \varphi_0' \right) \hat{r}' \hat{r}^2 \right.\nonumber\\
    ~&\left. + \frac{2 e^{2A}}{\sqrt{6} r_c^2 \varphi_0'^4} \left( 144 \dot{V}^2 r_c^4 + 288 \dot{V} r_c^2 A' \varphi_0' + \varphi_0' \left( 12 \dot{V}' r_c^2 + 144A'^2 \varphi_0' + 7 \varphi_0'^3 \right) \right) \hat{r}'^2 \hat{r} \right.  \nonumber\\
    ~&\left. - \sum_{i=1,2} \left( - 1 \right)^i \frac{2 \sqrt{6} e^{2 A}}{r_c^2 \varphi_0'^3} \left( 8 \dot{V} r_c^2 \hat{r}' + 8 A' \hat{r}' \varphi_0' + \varphi_0'^3 \hat{r} -2 e^{2A} \varphi_0' \hat{z} \right) \hat{r}' \hat{r} \delta_i \right]. \label{eqn:SrrrFullKKexpand1}
\end{align}
Performing KK decomposition of Eq.~(\ref{eqn:SrrrFullKKexpand1}) and utilizing the ``stiff-wall'' limit to remove the boundary term we are left with
\begin{equation}
    S_{\hat{r} \hat{r} \hat{r}} = \frac{1}{\Lambda_\pi} \sum_{n=0}^\infty \sum_{m=0}^\infty \sum_{l=0}^\infty  \int d^4 x  \left[ - \frac{2}{\sqrt{6}} \hat{r}^{\left( n \right)} \partial_\alpha \hat{r}^{\left( m \right)} \partial^\alpha \hat{r}^{\left( l \right)} e_{\left(n \right) \left( m \right) \left( l \right)} + \hat{r}^{\left( n \right)} \hat{r}^{\left( m \right)} \hat{r}^{\left( l \right)} \tilde{e}_{\left(n \right) \left( m \right) \left( l \right)} \right],\label{eqn:SrrrFullKKexpand2}
\end{equation}
where we have introduced a ``e-type'' overlap integral (overlap integral with a factor of $e^{4 A}$) defined as 
\begin{equation}
 e_{\left(n \right) \left( m \right) \left( l \right)} = \frac{1}{\psi_1{\left( \pi \right)}} \int_{-\pi}^\pi d \phi e^{4 A{\left( \phi \right)}} \gamma_n{\left( \phi \right)} \gamma_m{\left( \phi \right)} \gamma_l{\left( \phi \right)}, \label{eqn:OverlapCtype}
\end{equation}
and the corresponding brane stabilization correction
\begin{align}
     \tilde{e}_{\left(n \right) \left( m \right) \left( l \right):A} =~& \frac{1}{r_c^2 \psi_1{\left( \pi \right)} } \int_{-\pi}^\pi d \phi \times \left[ \frac{e^{2A} \left( 144 \dot{V}^2 r_c^4 + 288 \dot{V} r_c^2 A' \varphi_0' + \varphi_0' \left( 12 \dot{V}' r_c^2 + 144 A'^2 \varphi_0' + 7 \varphi_0'^3 \right) \right)}{\sqrt{6} \varphi_0'^4}  \gamma'_n  \gamma'_m \gamma_l \right.  \nonumber\\
     ~&- \sqrt{\frac{2}{3}} \frac{ 4 e^{2 A} \left( \dot{V} r_c^2 + A' \varphi_0' \right) \left( 9 e^{2A} m_{\left( l \right)}^2 r_c^2 - 4 \varphi_0'^2 \right)}{ \varphi_0'^3} \gamma'_n \gamma_m \gamma_l  \nonumber\\
    ~& \left.  + \frac{e^{2A} \left( 18 e^{4A} m_{\left( j \right)}^2 m_{\left( l \right)}^2 r_c^4 - 6 e^{2A} m_{\left( l \right)}^2 r_c^2 \varphi_0'^2 + \varphi_0'^4 \right)}{2 \sqrt{6} \varphi_0'^2} \gamma_n \gamma_m \gamma_l \right]. \label{eqn:OverlaptildeCAtype}
\end{align}
The unstabilized limit of interaction term Eq.~(\ref{eqn:SrrrFullKKexpand2}) is obtained by setting $\tilde{e}_{\left(n \right) \left( m \right) \left( l \right)} = 0$.

The corresponding Feynman rules can be extracted as, 
\begin{align}
\vcenter{\hbox{\includegraphics[height=2.13cm]{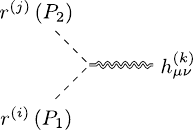}}\vspace{1.4cm}} \hspace{0.07cm} &~\begin{aligned}=~& \frac{i}{\Lambda_\pi} \left[
\vphantom{\left( \tilde{d}_{k \left(i \right) \left( j \right) } \left( P_1^\mu P_2^\nu + P_1^\nu P_2^\mu \right) + \eta^{\mu \nu} \left( \bar{d}_{k \left(i \right) \left( j \right) } - \tilde{d}_{k \left(i \right) \left( j \right) }P_1 \cdot P_2 \right) \right)}
d_{k \left(i \right) \left( j \right)} \left( P_1^\mu \left( P_2^\nu - 2 P_1^\nu \right) + P_2^\mu \left( P_1^\nu - 2 P_2^\nu \right)  \right. \right.  \\
&\left. \left. +  2 \eta^{\mu \nu} \left( P_1^2 + P_2^2 \right) + \eta^{\mu \nu} P_1 \cdot P_2 \right)  \right. \\
&\left.  + 2 \left( \tilde{d}_{k \left(i \right) \left( j \right) } \left( P_1^\mu P_2^\nu + P_1^\nu P_2^\mu \right) + \eta^{\mu \nu} \left( \bar{d}_{k \left(i \right) \left( j \right) } - \tilde{d}_{k \left(i \right) \left( j \right) }P_1 \cdot P_2 \right) \right) \right],  \label{eqn:Spin0KKint3} \end{aligned}\\
\vcenter{\hbox{\includegraphics[height=2.13cm]{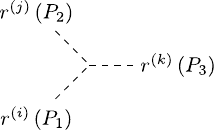}}} \hspace{0.07cm} & ~=~ -\frac{i}{\Lambda_\pi} \left[ \sqrt{\frac{2}{3}} e_{\left( i \right) \left( j \right) \left( k \right)}  \left(  P_1^2 + P_2^2 + P_3^2 \right) - \tilde{e}_{\left( i \right) \left( j \right) \left( k \right)} \right], \label{eqn:Spin0KKint1}
\end{align}
where
\begin{align}
    \tilde{e}_{\left( i \right) \left( j \right) \left( k \right)} &=  \tilde{e}_{\left(i \right) \left( j \right) \left( k \right):A} + \tilde{e}_{\left(i \right) \left( k \right) \left( j \right):A} + \tilde{e}_{\left(j \right) \left( i \right) \left( k \right):A} + \tilde{e}_{\left(j \right) \left( k \right) \left( i \right):A}  +\tilde{e}_{\left(k \right) \left( i \right) \left( j \right):A} + \tilde{e}_{\left(k \right) \left( j \right) \left( i \right):A}, \label{eqn:OverlaptildeCtype} \\
    \bar{d}_{k \left( i \right) \left( j \right)  } &= \frac{1}{2} \left( \bar{d}_{k \left( i \right) \left( j \right):A} - \bar{d}_{k \left( i \right) \left( j \right):B} - \bar{d}_{k \left( i \right) \left( j \right):C} + \bar{d}_{k \left( j \right) \left( i \right):A} - \bar{d}_{k \left( j \right) \left( i \right):B} - \bar{d}_{k \left( j \right) \left( i \right):C} \right).
\end{align}


\subsection{Relevant Sum-Rules}

List of relevant sum-rules:
\begin{align}
    \sum_{k=0}^\infty \kappa_k a_{i' k' \left( j \right)} &= \frac{\kappa_i \kappa_{\left( j \right)}}{\kappa_1} m_i^2, \label{eqn:SumRule5} \\
    \sum_{n=0}^\infty m_{\left( n \right)}^2 \kappa_{\left( n \right)} \left( d_{i \left( n \right) \left( j \right) } + 2 \tilde{d}_{i \left( n \right) \left( j \right)} \right) &= \frac{\kappa_i \kappa_{\left( j \right)}}{\kappa_1} m_{\left( j \right)}^2, \label{eqn:SumRule6} \\
    \sum_{k=0}^\infty \kappa_k d_{k \left( i \right) \left( j \right)} &= \frac{\kappa_{ \left( i \right)} \kappa_{\left( j \right)}}{\kappa_1},
    \label{eqn:SumRule7} \\
    \sum_{k=0}^\infty \kappa_k \tilde{d}_{k \left( i \right) \left( j \right)} &= 0. \label{eqn:SumRule8} 
\end{align}
Eqs.~(\ref{eqn:SumRule5}),~(\ref{eqn:SumRule7}),~and~(\ref{eqn:SumRule8}) can be easily proven through the application of integration by parts and the completeness relation of the wave-function of spin-$2$ modes. The proof of Eq.~(\ref{eqn:SumRule6}) is, however, a little more involved as it requires an application of integration by parts, completeness relation for the wave-functions of the spin-$0$ sector, and subsequent insertion of the completeness relation of the wave-functions of the spin-$2$ sector.

\subsection{Scattering amplitudes for $\Phi \Phi \to h^{\left( i \right) }r^{\left( j \right)}$ }
\label{sec:scathr}
We briefly summarize the calculation
for the scattering amplitude of $SS\to h^{ \left( i \right) }r^{ \left( j \right) }$, as $\sqrt{s}\to \infty$.
To study the high-energy behavior of the amplitude, we expand the $S$-matrix element for the process in question as a power series in the COM energy of the form
\begin{equation}
    \mathcal{M}_{\lambda_i} = \sum_{p \in \mathbb{Z}} \mathcal{M}^{\left( p / 2 \right)}_{\lambda_i} s^{p/2}.
\end{equation}
where $\lambda_i$ is the polarization state of the outgoing spin-$2$ mode $h^{\left( i \right)}$. Considering leading order terms, we get
\begin{align}
    \mathcal{M}_{\pm 2}^{\left( 0 \right)} &= - \frac{i \kappa_1}{4 \sqrt{6}} \left[ 4 \left( m_{\left( j \right)}^2 + 2 m_S^2 \right) \frac{\kappa_i \kappa_{\left( j \right)}}{\kappa_1} - 3 \sum_{k =0}^\infty \kappa_k a_{i'k'(j)} \left( 1 - \cos 2 \theta \right) \right] \nonumber 
    \\ & = - \frac{i \kappa_i \kappa_{\left( j \right)} }{\sqrt{6}} \left[  m_{\left( j \right)}^2 + 2 m_S^2   - \frac{3}{4} m_i^2 \left( 1 - \cos 2 \theta \right) \right], \\
    \mathcal{M}_{\pm 1}^{\left( 1/2 \right)} &= \pm \frac{i \kappa_1}{4 m_i}  \sqrt{\frac{3}{2}} \sin 2 \theta \sum_{k=0}^\infty \kappa_k a_{i' k' (j)} = \pm \frac{i \kappa_i \kappa_{\left( j \right)} m_i}{4}  \sqrt{\frac{3}{2}} \sin 2 \theta , \\
    \mathcal{M}_{0}^{\left( 2 \right)} &= \frac{i \kappa_1}{6 m_i^2} \left(   \sum_{k=1}^{\infty} \frac{\kappa_k a_{i'k'(j)}}{m_k^2} + \sum_{n = 0}^{\infty} \kappa_{ \left( n \right)} \left( d_{i \left( n \right) \left( j \right) } + 2 \tilde{d}_{i \left( n \right) \left( j \right)} \right)  - \frac{\kappa_i \kappa_{\left( j \right)} }{\kappa_1} \right),
\end{align}
where we have used Eq.~(\ref{eqn:SumRule5})
whenever possible. Hence, to ensure a well-defined limit as $M_i$ approaches zero, we require 
\begin{equation}
\sum_{k=1}^{\infty} \frac{\kappa_k a_{i'k'\left(j \right)}}{m_k^2} + \sum_{n = 0}^{\infty} \kappa_{\left( n \right)} \left( d_{i \left( n \right) \left( j \right)} + 2 \tilde{d}_{i \left( n \right) \left( j \right)} \right)  - \frac{\kappa_i \kappa_{\left( j \right)}}{\kappa_1} = 0. \label{eqn:SumRuleRRcontrib}
\end{equation}

\begin{figure}[H]
    \centering
    \includegraphics[width=.7\linewidth]{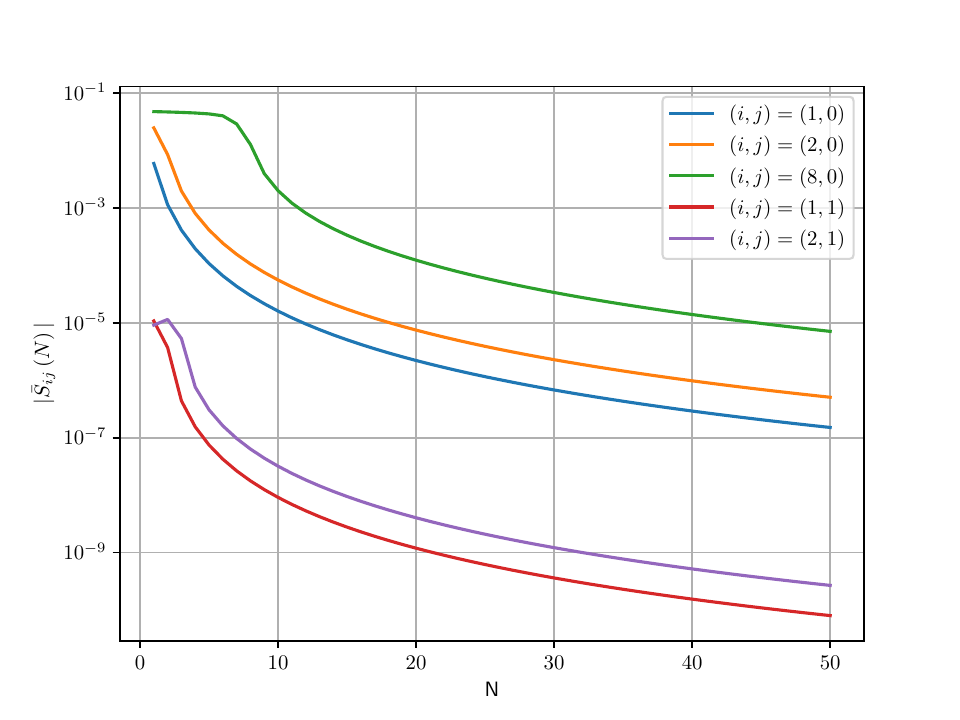}
    \caption{Numerical results for Eq.~(\ref{eqn:SumRuleRRcontribNumer}) given truncated KK towers for internal spin-$0$ and spin-$2$ modes as a function of the index of the highest KK mode in the internal towers $N$. \label{fig:SumRuleRRcontribNumer}}
\end{figure}

We resort to numerical verification to demonstrate that Eq.~(\ref{eqn:SumRuleRRcontrib}) holds. Let us start by defining a new function $\bar{S} \left( N \right)$ that would quantify the deviation of the sum rule Eq.~(\ref{eqn:SumRuleRRcontrib}) for truncated KK tower from zero as 
\begin{equation}
    \bar{S}_{ij} \left( N \right) = \sum_{k=1}^{N} \frac{\kappa_k a_{i'k'\left( j \right)}}{m_k^2} + \sum_{n = 0}^{N} \kappa_{\left( n \right)} \left( d_{i\left(n \right) \left(j \right)} + 2 \tilde{d}_{i \left( n \right) \left( j \right)} \right)  - \frac{\kappa_i \kappa_{\left( j \right)}}{\kappa_1} . \label{eqn:SumRuleRRcontribNumer}
\end{equation}
Evaluating Eq.~(\ref{eqn:SumRuleRRcontribNumer}) as a function of the number of states in the truncated KK tower is depicted on Fig.~\ref{fig:SumRuleRRcontribNumer}, verifying to the numerical precision shown that the sum rule is satisfied.

To proceed, we focus our attention on the contribution linear in the COM energy to $\mathcal{M}_0$. Applying the relevant sum-rules and coupling identities, 
whenever possible, we are left with 
\begin{equation}
    \mathcal{M}_0^{\left( 1 \right)} = \frac{i \kappa_1 }{24} \left[ \left( 1 + 3 \cos 2 \theta \right) \frac{\kappa_i \kappa_{ \left( j \right)}}{\kappa_1} - \frac{4 }{m_i^2} \left( m_{ \left( j \right)}^2 \frac{\kappa_i \kappa_{\left( j \right)}}{\kappa_1}  - \sum_{n=0}^\infty m_{ \left( n \right)}^2 \kappa_{\left( n \right)} \left( d_{i \left( n \right)  \left( j \right)} + 2 \tilde{d}_{i \left( n \right)\left( j \right)} \right) \right) \right].
\end{equation}

However, the terms inversely proportional to the mass of the outgoing spin-$2$ mode squared are zero as per Eq.~(\ref{eqn:SumRule6}). Hence, the residual terms linear in COM energy are
\begin{equation}
    \mathcal{M}_0^{\left( 1 \right)} = \frac{i \kappa_i \kappa_{ \left( j \right)} }{24}  \left( 1 + 3 \cos 2 \theta \right).
\end{equation}

The resulting cross-section for the process in question obtained numerically by summing over the truncated spin-$0$ and spin-$2$ towers are presented in Fig.~\ref{fig:CrossSectsPhiPhiHR}. 

Analogously to the previous case of the $S S \rightarrow h^{\left( i \right)} r^{\left( j \right)}$ process, we consider the high energy behavior for $V V \rightarrow h^{\left( i \right)} r^{\left( j \right)}$ by expanding the $S$-matrix elements in a power series in COM energy of the form
\begin{equation}
    \mathcal{M}_{\lambda_1, \lambda_2; \lambda_i} = \sum_{p \in \mathbb{Z}} \mathcal{M}^{\left( p / 2 \right)}_{\lambda_1, \lambda_2; \lambda_i} s^{p/2}, 
\end{equation}
where $\lambda_{\{ 1, 2 \}}$ are the polarizations of the incoming brane vector bosons, and $\lambda_i$ is the polarization of the $i$th outgoing spin-$2$ mode. Focusing our attention on the contributions of the highest order, we only one matrix element with the contribution of quadratic order
\begin{equation}
    \mathcal{M}_{0,0;0}^{\left( 2 \right)} = \frac{i \kappa_1}{6 m_i^2} \left[ \frac{\kappa_i \kappa_{\left( j \right)}}{\kappa_1} - \sum_{n=0}^\infty \kappa_{\left( n \right)} \left( d_{i \left( n \right) \left( j \right)} + 2 \tilde{d}_{i \left( n \right) \left( j \right)} \right) - \sum_{k=1}^{\infty} \frac{\kappa_k a_{i' k' \left( j \right)}}{m_k^2} \right].
\end{equation}
However, as verified earlier, the above contribution vanishes by Eq.~(\ref{eqn:SumRuleRRcontrib}). Considering the contributions of linear order, we have
\begin{align}
    \mathcal{M}_{\pm,\pm; 0}^{\left( 1 \right)} &= \frac{i \kappa_1 m_V^2}{3 m_i^2} \left[ \frac{\kappa_i \kappa_{\left( j \right)}}{\kappa_1} - \sum_{n=0}^\infty \kappa_{\left( n \right)} \left( d_{i \left( n \right) \left( j \right)} + 2 \tilde{d}_{i \left( n \right) \left( j \right)} \right)  - \sum_{k=1}^\infty \frac{\kappa_k a_{i' k' \left( j \right)}}{m_k^2} \right]= 0 , \\
    \mathcal{M}_{\pm,\mp;0}^{\left( 1 \right)} &= \frac{i \kappa_1}{8 m_i^2} \left( 1 - \cos 2 \theta \right) \sum_{k=0}^\infty \kappa_k a_{i' k' \left( j \right)} = \frac{i \kappa_i \kappa_{\left( j \right)} }{8} \left( 1 - \cos 2 \theta \right), \\
    \mathcal{M}_{0,0;0}^{\left( 1 \right)} &= \frac{i \kappa_1 }{24} \left[ \left( 1 + 3 \cos 2 \theta \right) \frac{\kappa_i \kappa_{\left( j \right)}}{\kappa_1} - \frac{4 }{m_i^2} \left( m_{ \left( j \right)}^2 \frac{\kappa_i \kappa_{\left( j \right)}}{\kappa_1}  - \sum_{n=0}^\infty m_{ \left( n \right)}^2 \kappa_{\left( n \right)} \left( d_{i \left( n \right) \left( j \right)} + 2 \tilde{d}_{i \left( n \right) \left( j \right)} \right) \right) \right]  \nonumber \\ &= \frac{i \kappa_i \kappa_{\left( j \right)} }{24}  \left( 1 + 3 \cos 2 \theta \right),
\end{align}
where we've applied Eqs.~(\ref{eqn:SumRuleRRcontrib}),~(\ref{eqn:SumRule5}),~(\ref{eqn:SumRule7}),~and~(\ref{eqn:SumRule8}). Similarly at the leading $\sqrt{s}$ order 
\begin{align}
    \mathcal{M}_{\pm,0;0}^{\left( 1/2 \right)} &= \mathcal{M}_{0,\pm;0}^{\left( 1/2 \right)} = \pm \frac{i \kappa_1 m_V}{\sqrt{2} m_i^2} \cos \theta \sin \theta \sum_{k=0}^\infty \kappa_k a_{i' k' \left( j \right)} =  \pm \frac{i \kappa_i \kappa_{\left( j \right)}}{2 \sqrt{2}} m_V \sin 2 \theta.
\end{align}

\begin{figure}[t]
    \centering
    \includegraphics[width=0.49\linewidth]{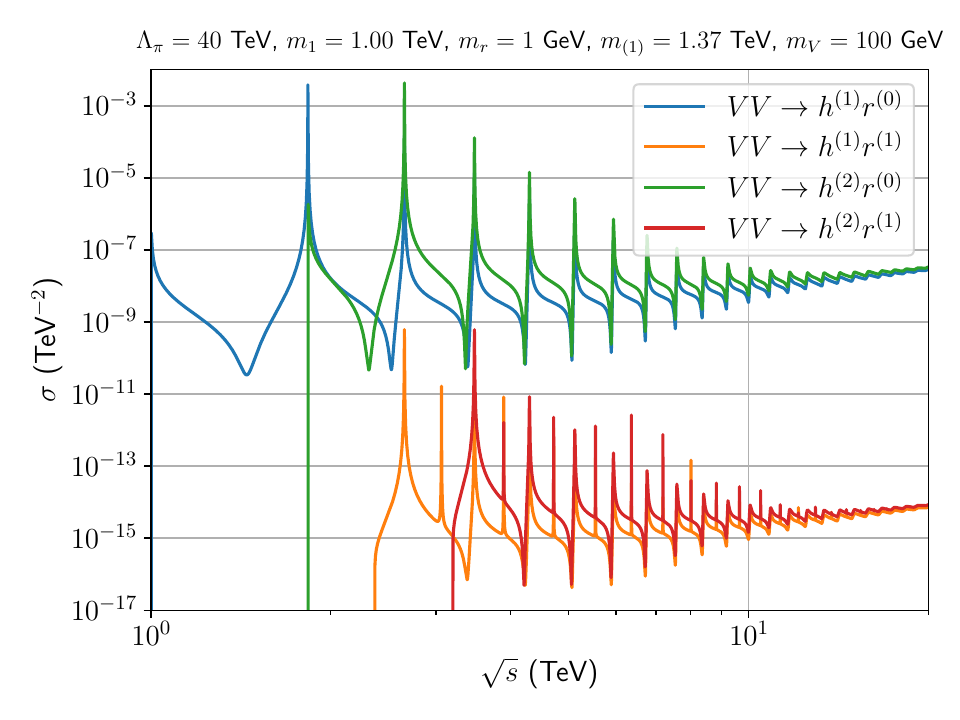}
\includegraphics[width=0.49\linewidth]{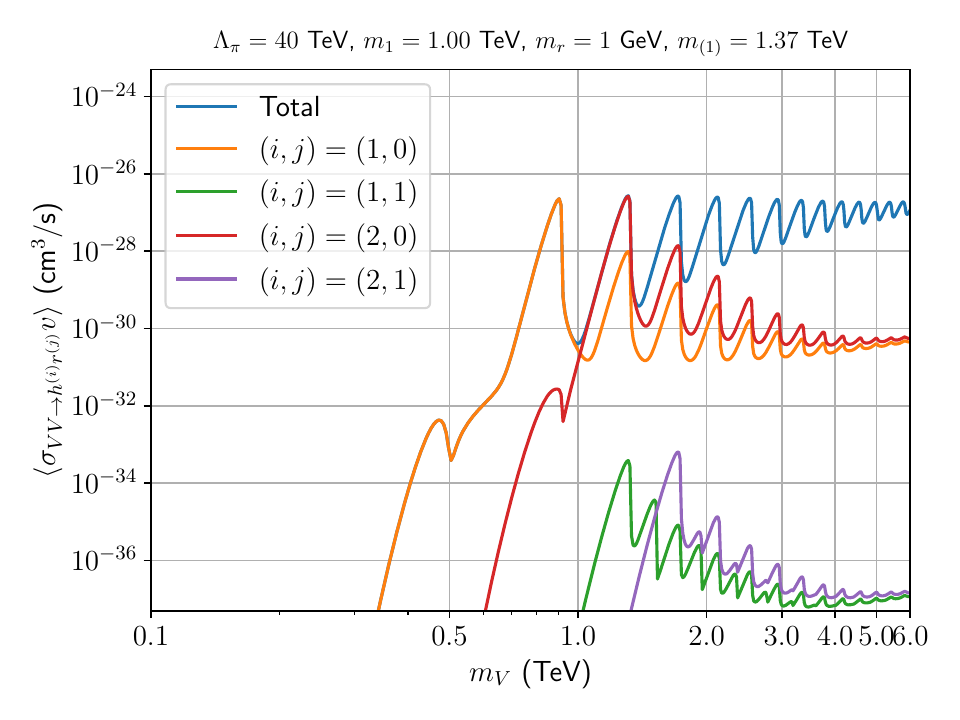}
    \caption{Cross-section (left panel) and velocity averaged cross-section (right panel) for the process $V V \rightarrow h^{\left( i \right)} r^{\left( j \right)}$ for the mass of the brane vector of $0.5$ TeV. The cross-sections are computed by summing over the truncated KK tower of $50$ internal spin-$0$ and $50$ internal spin-$2$ KK modes. \label{fig:CrossSectsAAHR} } 
\end{figure}

The resulting cross-sections for the processes in question, obtained numerically by summing over the truncated spin-$0$ and spin-$2$ towers, are presented in Fig.~\ref{fig:CrossSectsAAHR}, with the corresponding slice of the velocity averaged cross-section at the typical freeze-out temperatures depicted on the right panel of Fig.~\ref{fig:CrossSectsAAHR}.

\begin{figure}[t]
    \centering
    \includegraphics[width=0.49\linewidth]{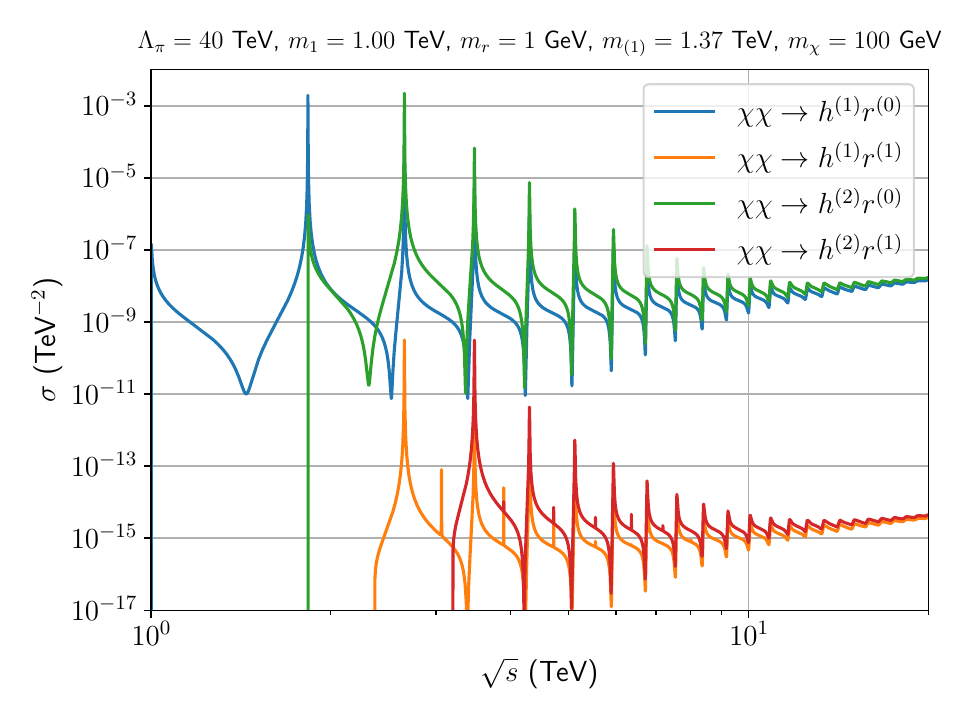}
    \includegraphics[width=0.49\linewidth]{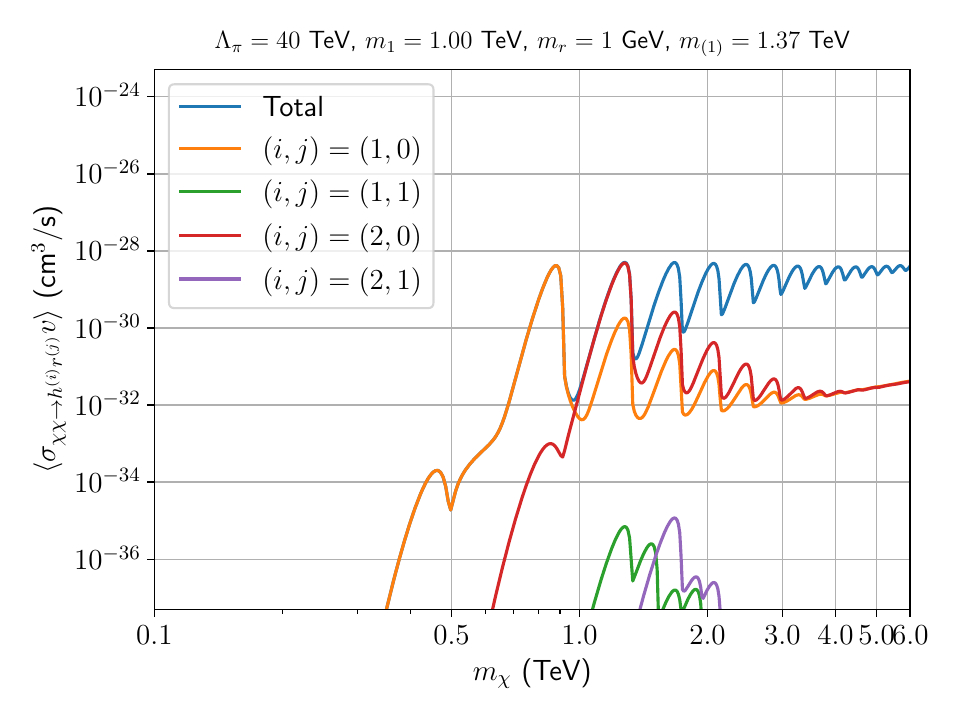}
    \caption{Cross-section (left panel) and velocity averaged cross section (right panel) for the process $\mathcal{\chi} \mathcal{\chi} \rightarrow h^{\left( i \right)} r^{\left( j \right)}$ for the mass of the brane fermion of $0.5$ TeV. The cross-sections are computed by summing over the truncated KK tower of $50$ internal spin-$0$ and $50$ internal spin-$2$ KK modes. \label{fig:CrossSectsFFHR} } 
\end{figure}

Finally we repeat our high energy analysis for $\chi\chi\to h^{ \left( i \right)}r^{ \left( j \right)}$ in the center of momentum energy $s$ as 
\begin{equation}
        \mathcal{M}_{s_1, s_2; \lambda_i} = \sum_{p \in \mathbb{Z}} \mathcal{M}^{\left( p / 2 \right)}_{s_1, s_2; \lambda_i} s^{p/2}, 
\end{equation}
where $s_{1,2}$ are the spins of the incoming brane fermions, and $\lambda_i$ is the polarization of the outgoing spin-$2$ KK mode. Focusing on the final polarization state of $\lambda_i = 0$, the leading order contributions are
\begin{align}
    \mathcal{M}_{\pm, \pm; 0}^{\left( 3/2\right)} &= \pm \frac{i \kappa_1 m_\mathcal{\chi}}{6 m_i^2} \left[ \frac{\kappa_i \kappa_{\left( j \right)}}{\kappa_1} - \sum_{n=0}^\infty \kappa_{\left( n \right)} \left( d_{i \left( n \right) \left( j \right)} + 2 \tilde{d}_{i \left( n \right) \left( j \right)} \right) - \sum_{k=1}^\infty \frac{\kappa_k a_{i' k' \left( j \right)} }{m_k^2}  \right] = 0, \\
    \mathcal{M}_{\pm,\mp; 0}^{\left( 1 \right)} &= \frac{i \kappa_1 \sin 2 \theta}{8 m_i^2} \sum_{k=0}^\infty \kappa_k a_{i' k' \left( j \right)}  = \frac{i \kappa_i \kappa_{\left( j \right)} \sin 2 \theta}{8},
\end{align}
where we applied Eqs.~(\ref{eqn:SumRuleRRcontrib})~and~(\ref{eqn:SumRule6}). The matrix element is slightly more complicated at the leading $\sqrt{s}$ order. However, it can be greatly simplified by applying Eqs.~(\ref{eqn:SumRuleRRcontrib}),~(\ref{eqn:SumRule6}),~and~(\ref{eqn:SumRule5}), leaving us with
\begin{equation}
    \mathcal{M}_{\pm,\pm;0}^{\left( 1/2 \right)} = \pm \frac{i m_\mathcal{\chi} \kappa_1 \kappa_0 a_{i'0' \left( j \right)} }{6 m_i^2} \pm \frac{m_\mathcal{\chi} \kappa_i \kappa_{\left( j \right)}}{12} \left( 1 + 3 \cos 2 \theta \right) = \pm \frac{m_\mathcal{\chi} \kappa_i \kappa_{\left( j \right)}}{12} \left( 1 + 3 \cos 2 \theta \right),
\end{equation}
where we've recognized that $a_{i'0' \left( j \right)}=0$ for any values of $i$ and $j$ as $\partial_\phi \psi_0 \left( \phi \right)= 0$. Note that, similarly to the two spin-$2$ mode final state, we have a different angular dependency of the leading term for fermions. The resulting cross-sections for the processes in question, obtained numerically by summing over the truncated spin-$0$ and spin-$2$ towers, are presented in Fig.~\ref{fig:CrossSectsFFHR}, with the corresponding slice of the velocity averaged cross-section at the typical freeze-out temperatures depicted on Fig.~\ref{fig:CrossSectsFFHR}.

\subsection{Scattering 
Amplitudes for $\Phi\Phi\to r^{\left( i \right)}r^{\left( j \right)}$}
\begin{figure}[t]
    \centering
    \includegraphics[width=.7\linewidth]{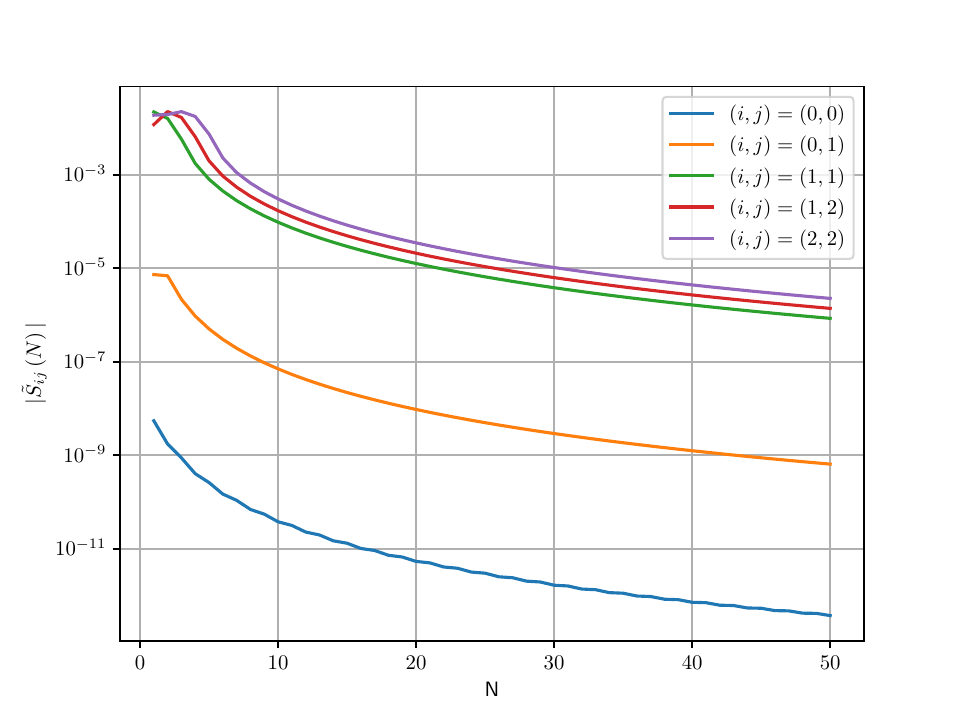}
    \caption{Numerical results for function $\tilde{S}_{ij}(N)$,  as in Eq.~(\ref{eqn:SumRuleFuncRR}), given truncated KK towers for internal spin-$0$ and spin-$2$ modes as a function of the index of the highest KK mode in the internal towers $N$. \label{fig:SumRuleFuncRR}}
\end{figure}

In this section, we document the results for the high energy amplitudes of $\Phi\Phi\to r^{\left( i \right)}r^{ \left( j \right)}$, starting with $SS\to r^{\left( i \right)}r^{\left( j \right)}$. For scalars in both initial and final states, we only have a single $S$-matrix element, which we expand as a power series in the COM energy analogously to previous cases
\begin{equation}
        \mathcal{M} = \sum_{p \in \mathbb{Z}} \mathcal{M}^{\left( p / 2 \right)} s^{p/2}, 
\end{equation}
In this case, the leading order contribution is of order $s$ and can be written as 
\begin{align}
    \mathcal{M}^{\left( 1 \right)} =~& \frac{i \kappa_1 }{24} \left[8 \sum_{n=0}^\infty \kappa_{\left( n \right)} e_{\left( i \right) \left( j \right) \left( n \right) } -4 \left( \frac{ \kappa_{\left( i \right)} \kappa_{\left( j \right)} }{\kappa_1} +  \kappa_0 \left( d_{0 \left( i \right) \left( j \right)} - 2 \tilde{d}_{0 \left( i \right) \left( j \right)} \right) \right) - \sum_{k=0}^\infty \kappa_k \left( d_{k \left( i \right) \left( j \right)} + 2 \tilde{d}_{k \left( i \right) \left( j \right)} \right) \right. \nonumber\\
    ~& + \left. 4 \left( \left( m_{\left( i \right)}^2 + m_{\left( j \right)}^2 \right) \sum_{k=1}^\infty \frac{ \kappa_k \left( d_{k \left( i \right) \left( j \right)} + 2 \tilde{d}_{k \left( i \right) \left( j \right)} \right)}{m_k^2} + 4 \sum_{k=1}^\infty \frac{\kappa_k \bar{d}_{k \left( i \right) \left( j \right)} }{m_k^2} \right)  \right. \nonumber\\
    ~& \left.  - 3 \cos 2 \theta \sum_{k=0}^\infty \kappa_k \left( d_{k \left( i \right) \left( j \right)} + 2 \tilde{d}_{k \left( i \right) \left( j \right) } \right) \right].
\end{align}
Inserting Eqs.~(\ref{eqn:SumRule7})~and~(\ref{eqn:SumRule8}) into the above result, we are left with 
\begin{align}
    \mathcal{M}^{\left( 1 \right)} =~& \frac{i \kappa_1 }{6} \left[2 \sum_{n=0}^\infty \kappa_{\left( n \right)} e_{\left( i \right) \left( j \right) \left( n \right)} - \frac{ \kappa_{\left( i \right)} \kappa_{\left( j \right)} }{\kappa_1} - \kappa_0 \left( d_{0 \left( i \right) \left( j \right) } - 2 \tilde{d}_{0 \left( i \right) \left( j \right)} \right) \right. \nonumber\\
    ~& + \left.  \left( m_{ \left( i \right)}^2 + m_{ \left( j \right)}^2 \right) \sum_{k=1}^\infty \frac{ \kappa_k \left( d_{k \left( i \right) \left( j \right)} + 2 \tilde{d}_{k \left( i \right) \left( j \right)} \right)}{m_k^2} + 4 \sum_{k=1}^\infty \frac{\kappa_k \bar{d}_{k \left( i \right) \left( j \right) } }{m_k^2} \right] \nonumber\\
    ~& - \frac{i \kappa_{\left( i \right)} \kappa_{\left( j \right)} }{24} \left( 1+ 3   \cos 2 \theta \right) .
    \label{eqn:SMatrixRR01}
\end{align}
Focusing our attention on the term in the square brackets, we define a new function $\tilde{S}_{ij} \left( N \right)$ as
\begin{align}
    \tilde{S}_{ij} \left( N \right) =~& 2 \sum_{n=0}^N \kappa_{\left( n \right)} e_{\left( i \right) \left( j \right) \left( n \right) } - \frac{ \kappa_{\left( i \right)} \kappa_{\left( j \right)} }{\kappa_1} - \kappa_0 \left( d_{0 \left( i \right) \left( j \right)} - 2 \tilde{d}_{0 \left( i \right) \left( j \right)} \right) \nonumber\\
    ~&+ \left( m_{ \left( i \right)}^2 + m_{ \left( j \right)}^2 \right) \sum_{k=1}^N \frac{ \kappa_k \left( d_{k \left( i \right) \left( j \right)} + 2 \tilde{d}_{k \left( i \right) \left( j \right)} \right)}{m_k^2} + 4 \sum_{k=1}^N \frac{\kappa_k \bar{d}_{k \left( i \right) \left( j \right)} }{m_k^2}.
    \label{eqn:SumRuleFuncRR}
\end{align}
The values of the above function for several values of $N$ are plotted in Fig.~\ref{fig:SumRuleFuncRR}.

From Fig.~\ref{fig:SumRuleFuncRR}, we observe that as we increase the ``height'' of the truncated KK tower summed over in the internal propagators, the value of Eq.~(\ref{eqn:SumRuleFuncRR}) approaches zero. Hence, we can conclude that we have a new, rather complicated, sum rule
\begin{align}
    0 ~=~& 2 \sum_{n=0}^\infty \kappa_{\left( n \right)} e_{\left( i \right) \left( j \right) \left( n \right) } - \frac{ \kappa_{\left( i \right)} \kappa_{\left( j \right)} }{\kappa_1} - \kappa_0 \left( d_{0 \left( i \right) \left( j \right)} - 2 \tilde{d}_{0 \left( i \right) \left( j \right)} \right) \\
    ~& + \left( m_{\left( i \right)}^2 + m_{ \left( j \right)}^2 \right) \sum_{k=1}^\infty \frac{ \kappa_k \left( d_{k \left( i \right) \left( j \right)} + 2 \tilde{d}_{k \left( i \right) \left( j \right)} \right)}{m_k^2} + 4 \sum_{k=1}^\infty \frac{\kappa_k \bar{d}_{k \left( i \right) \left( j \right)} }{m_k^2} .
    \label{eqn:SumRuleRR}
\end{align}
Inserting Eq.~(\ref{eqn:SumRuleRR}) into Eq.~(\ref{eqn:SMatrixRR01}) we are left with 
\begin{equation}
    \mathcal{M}^{\left( 1 \right)} = - \frac{i \kappa_{\left( i \right)} \kappa_{\left( j \right)} }{24} \left( 1+ 3   \cos 2 \theta \right). \label{eqn:SMatrixRR01Leading}
\end{equation}
As noted previously, the portion of the amplitude scaling like $s^1$, shown in Eq.~(\ref{eqn:SMatrixRR01Leading}), is suppressed by the product of small couplings $\kappa_{(i)} \kappa_{(j)}$, and instead contributions which are proportional to $s^0$ dominate until COM energies as large as $\sqrt{s} \simeq 10^{3}$ TeV.

\begin{figure}[t]
    \centering
\includegraphics[width=0.49\linewidth]{img/CrossSectionSSRR.pdf}
\includegraphics[width=0.49\linewidth]{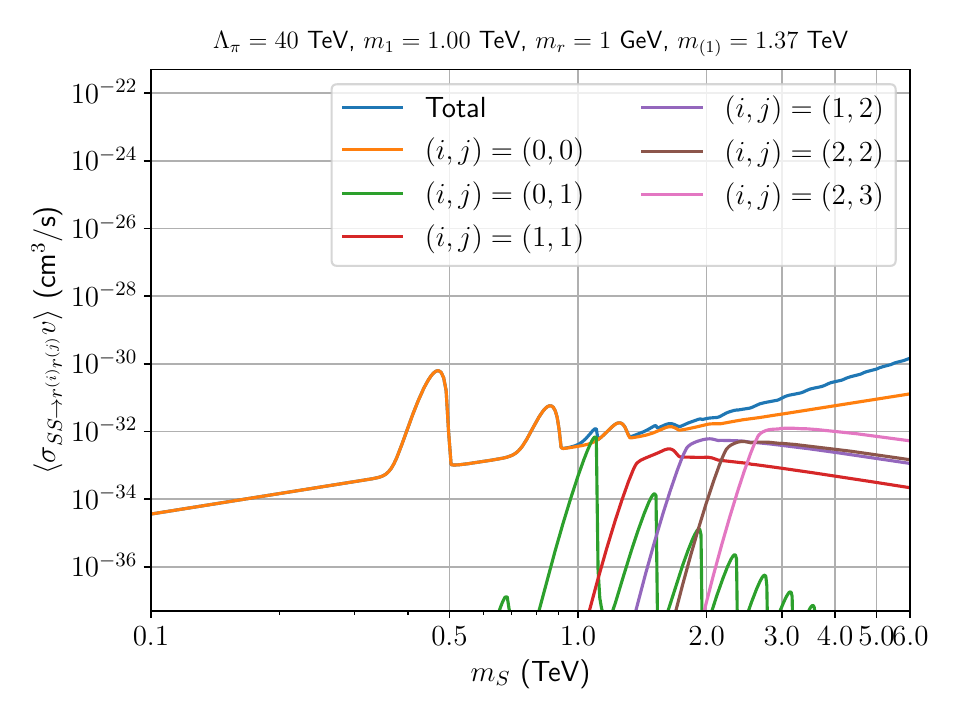}
    \caption{Cross-section (left panel) and velocity averaged cross section (right panel) for the process $S S \rightarrow r^{ \left( i \right)} r^{ \left( j \right)}$ for the mass of the brane scalar of $0.5$ TeV. The cross-sections are computed by summing over the truncated KK tower of $30$ internal spin-$0$ and $30$ internal spin-$2$ KK modes. \label{fig:CrossSectsSSRR} } 
\end{figure}

The resulting cross-sections for the processes in question, obtained numerically by summing over the truncated spin-$0$ and spin-$2$ towers, and are presented in Fig.~\ref{fig:CrossSectsSSRR}, with the corresponding slice of the velocity averaged cross-section at the typical freeze-out temperatures depicted on Fig.~\ref{fig:CrossSectsSSRR}. We have checked that including up to 30 KK states are numerically sufficient to accurately compute the relevant cross-sections for our purposes.

Repeating the same process for the case of brane localized vector boson, we expand the $S$-matrix element in terms of COM energy as 
\begin{equation}
        \mathcal{M}_{\lambda_1, \lambda_2} = \sum_{p \in \mathbb{Z}} \mathcal{M}^{\left( p / 2 \right)}_{\lambda_1, \lambda_2} s^{p/2}, 
\end{equation}
where $\lambda_{\{1, 2 \}}$ are the polarizations of the incoming vector bosons. The corresponding non-zero leading order terms are
\begin{align}
    \mathcal{M}_{\pm,\mp}^{\left( 1 \right)} =~& - \frac{i \kappa_1}{8} \left( 1 - \cos 2 \theta \right) \sum_{k=0}^\infty \kappa_k \left( d_{k \left( i \right) \left( j \right)} + 2 \tilde{d}_{k \left( i \right) \left( j \right)} \right) = - \frac{i \kappa_{\left( i \right)} \kappa_{\left( j \right)}}{8} \left( 1 - \cos 2 \theta \right), \\
    \mathcal{M}_{0,0}^{\left( 1 \right)} =~&  \frac{i \kappa_1 }{6} \left[2 \sum_{n=0}^\infty \kappa_{\left( n \right)} e_{\left( i \right) \left( j \right) \left( n \right)} - \frac{ \kappa_{\left( i \right)} \kappa_{\left( j \right)} }{\kappa_1} - \kappa_0 \left( d_{0 \left( i \right) \left( j \right)} - 2 d_{0 \left( i \right) \left( j \right)} \right) \right.  \nonumber \\
    & + \left.  \left( m_{\left( i \right)}^2 + m_{\left( j \right)}^2 \right) \sum_{k=1}^\infty \frac{ \kappa_k \left( d_{k \left( i \right) \left( j \right)} + 2 \tilde{d}_{k \left( i \right) \left( j \right)} \right)}{m_k^2} + 4 \sum_{k=1}^\infty \frac{\kappa_k \bar{d}_{k \left( i \right) \left( j \right)} }{m_k^2} \right]  \nonumber \\
    & - \frac{i \kappa_{\left( i \right)} \kappa_{\left( j \right)} }{24} \left( 1+ 3   \cos 2 \theta \right) \nonumber\\
    =&~ - \frac{i \kappa_{\left( i \right)} \kappa_{\left( j \right)} }{24} \left( 1+ 3   \cos 2 \theta \right),
\end{align}
where we have applied Eqs.~(\ref{eqn:SumRule7}),~(\ref{eqn:SumRule8}),~and~(\ref{eqn:SumRuleRR}).

\begin{figure}[t]
    \centering
    \includegraphics[width=0.49\linewidth]{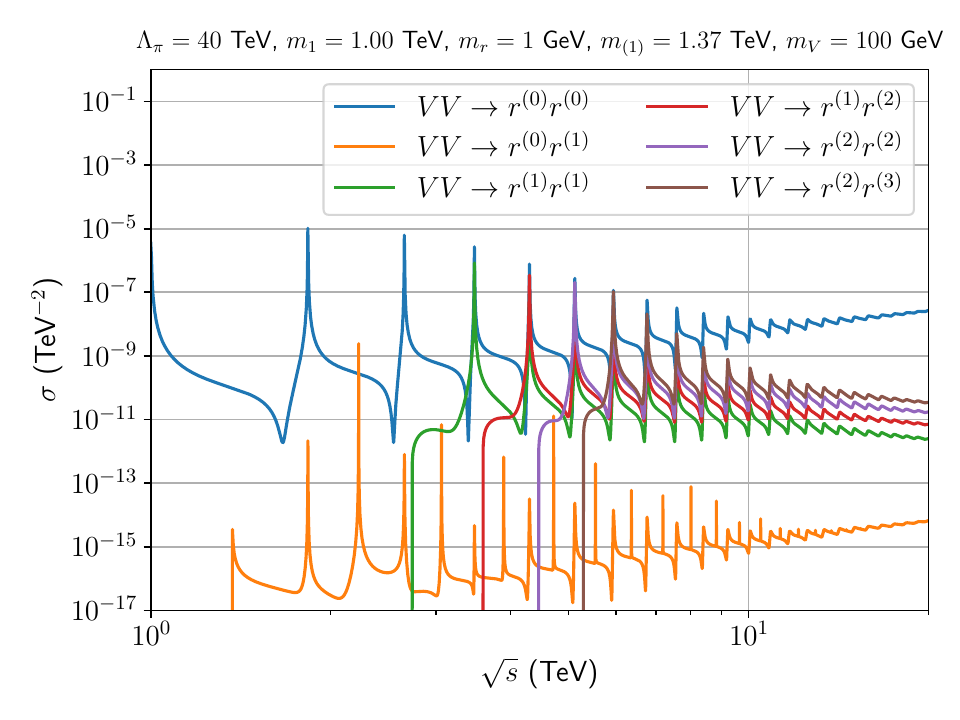}
    \includegraphics[width=0.49\linewidth]{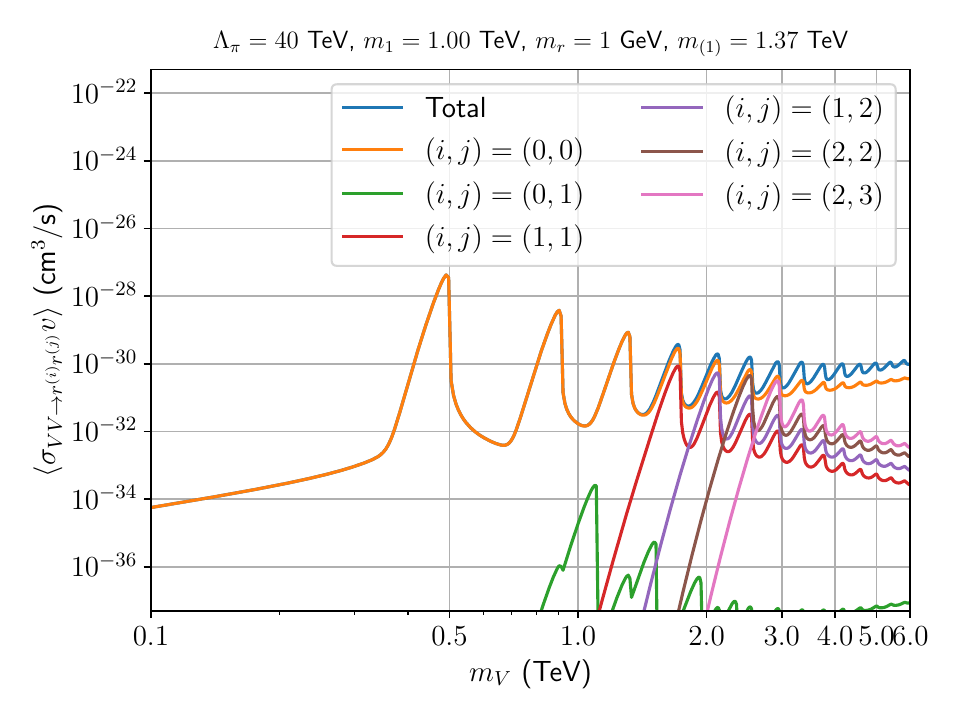}
    \caption{Cross-section (left panel) and velocity averaged cross section (right panel) for the process $V V \rightarrow r^{ \left( i \right)} r^{ \left( j \right)}$ for the mass of the brane scalar of $0.5$ TeV. The cross-sections are computed by summing over the truncated KK tower of $30$ internal spin-$0$ and $30$ internal spin-$2$ KK modes. \label{fig:CrossSectsAARR} } 
\end{figure}

The resulting cross-sections for the processes in question, obtained numerically by summing over the truncated spin-$0$ and spin-$2$ towers, are presented in the left panel of Fig.~\ref{fig:CrossSectsAARR}, with the corresponding slice of the velocity averaged cross-section at the typical freeze-out temperatures depicted on the right panel of this figure.

 Finally, for the case of brane localised fermion, we analogously expand the $S$-matrix element in terms of COM energy as 
\begin{equation}
        \mathcal{M}_{s_1, s_2} = \sum_{p \in \mathbb{Z}} \mathcal{M}^{\left( p / 2 \right)}_{s_1, s_2} s^{p/2}, 
\end{equation}
where $s_{\{1, 2 \}}$ are the spins of the incoming fermions. The corresponding non-zero leading order terms are
\begin{equation}
    \mathcal{M}_{\pm,\mp}^{\left( 1 \right)} = - \frac{i \kappa_1}{8} \sin 2 \theta  \sum_{k=0}^\infty \kappa_k \left( d_{k \left( i \right) \left( j \right) } + 2 \tilde{d}_{k \left( i \right) \left( j \right)} \right) = - \frac{i \kappa_{\left( i \right)} \kappa_{\left( j \right)}}{8} \sin 2 \theta,
\end{equation}
where we've applied the requisite 
sum rules.

\begin{figure}[t]
    \centering
    \includegraphics[width=0.49\linewidth]{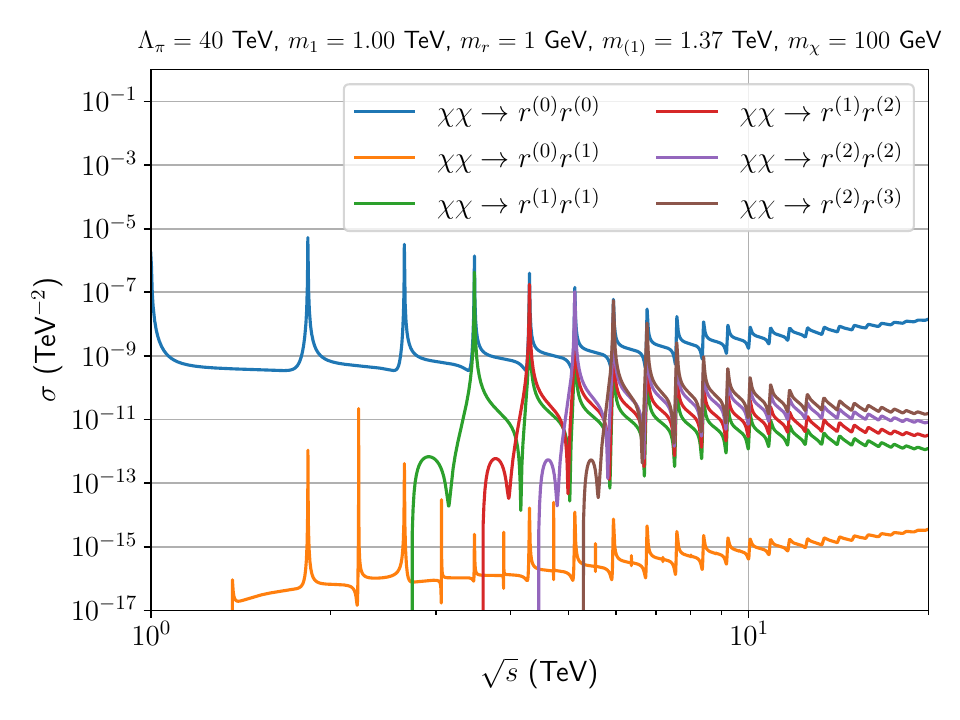}
    \includegraphics[width=0.49\linewidth]{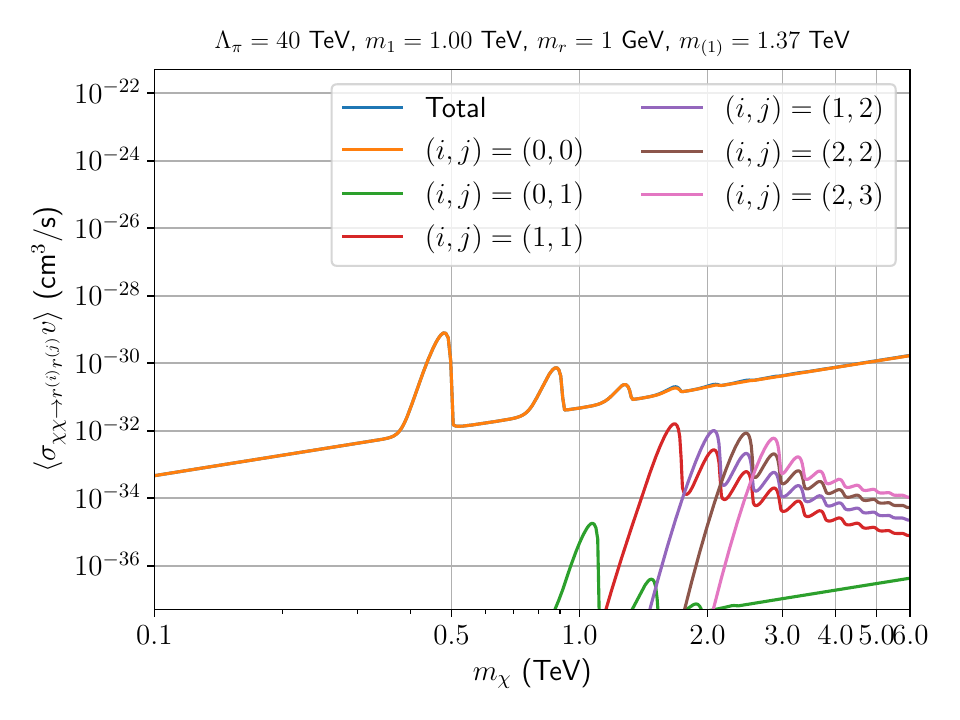}
    \caption{Cross-section (left panel) and velocity averaged cross section (right panel) for the process $\mathcal{\chi} \mathcal{\chi} \rightarrow r^{\left( i \right)} r^{\left( j \right)}$ for the mass of the brane fermion of $0.5$ TeV. The cross-sections are computed by summing over the truncated KK tower of $30$ internal spin-$0$ and $30$ internal spin-$2$ KK modes. \label{fig:CrossSectsFFRR} } 
\end{figure}

The resulting cross-sections for the processes in question, obtained numerically by summing over the truncated spin-$0$ and spin-$2$ towers, are presented in Fig.~\ref{fig:CrossSectsFFRR}, with the corresponding slice of the velocity averaged cross-section at the typical freeze-out temperatures depicted on the left panel of Fig.~\ref{fig:CrossSectsFFRR}.

\newpage

\bibliographystyle{apsrev4-1.bst}

\bibliography{main}

\end{document}